\documentclass[aip,cha,amsmath,amssymb,reprint,natbib]{revtex4-1}
\usepackage{epsf}

\usepackage{graphics}
\usepackage{color}
\usepackage{subcaption}
\usepackage{graphicx}
\usepackage{float}
\usepackage{rotfloat}
\usepackage{tabularx}
\usepackage{placeins} 
\floatstyle{boxed} 
\usepackage{dcolumn}
\usepackage{bm}
\usepackage[shortlabels]{enumitem}
\usepackage{soul}
\usepackage{newtxtext}      
\usepackage[utf8]{inputenc}
\usepackage[T1]{fontenc}
\usepackage{amsthm}

\usepackage{sidecap}
\usepackage{grffile}
\usepackage{scalerel}
\usepackage[draft]{changes}
\usepackage{tikz}
\usetikzlibrary{shapes.geometric, arrows, positioning}
\usetikzlibrary{calc}
\usepackage{xcolor}
\usepackage{afterpage}

\usepackage{tikz}
\usetikzlibrary{arrows.meta}

\usepackage{newtxtext}
\usepackage[colorlinks=true, pdfstartview=FitV, linkcolor=blue,
            citecolor=blue, urlcolor=blue]{hyperref}

\usepackage{xcolor}

\definecolor{darkgreen}{rgb}{0.0, 0.5, 0.0}
\definecolor{applegreen}{rgb}{0.55, 0.71, 0.0}
\definecolor{bostonuniversityred}{rgb}{0.8, 0.0, 0.0}
\definecolor{capri}{rgb}{0.0, 0.75, 1.0}
\definecolor{cyan}{rgb}{0.0, 1.0, 1.0}
\definecolor{darkred}{rgb}{0.55, 0.0, 0.0}
\definecolor{electricultramarine}{rgb}{0.25, 0.0, 1.0}
\definecolor{green(colorwheel)(x11green)}{rgb}{0.0, 1.0, 0.0}
\definecolor{malachite}{rgb}{0.04, 0.85, 0.32}
\definecolor{ao}{rgb}{0.0, 0.0, 1.0}

\def\bea{\begin{equation} \begin{aligned}}
\def\eea{\end{aligned} \end{equation}}
\def\beas{\begin{equation*} \begin{aligned}}
\def\eeas{\end{aligned} \end{equation*}}
\def\bes{\begin{equation*}}
\def\ees{\end{equation*}}

\def\be{\begin{equation}}
\def\ee{\end{equation}}

\begin{document}

\preprint{AIP/123-QED}

\title{Topological modes of variability  of the wind-driven ocean circulation}
\author{Gisela D. Charó}
\email{gisela.charo@cima.fcen.uba.ar}
\affiliation{CONICET – Universidad de Buenos Aires. Centro de Investigaciones
del Mar y la Atmósfera (CIMA), C1428EGA  Ciudad Autónoma de Buenos Aires, Argentina.}
\affiliation{CNRS – IRD – CONICET – UBA. Institut Franco-Argentin d'Études sur le Climat et ses Impacts (IRL 3351 IFAECI), C1428EGA Ciudad Autónoma de Buenos Aires, Argentina.}
\affiliation{Laboratoire des Sciences du Climat et de l’Environnement, CEA Saclay l’Orme des Merisiers, UMR 8212 CEA-CNRS-UVSQ, Université Paris-Saclay \& IPSL, 91191, Gif-sur-Yvette, France.}

\author{Denisse Sciamarella}
\email{denisse.sciamarella@cnrs.fr}
\affiliation{CNRS – IRD – CONICET – UBA. Institut Franco-Argentin d'Études sur le Climat et ses Impacts (IRL 3351 IFAECI), C1428EGA Ciudad Autónoma de Buenos Aires, Argentina.}
\affiliation{CNRS – Centre National de la Recherche Scientifique, 75016 Paris, France.}%

\author{Juan Ruiz}
\email{jruiz@cima.fcen.uba.ar}
\affiliation{CNRS – IRD – CONICET – UBA. Institut Franco-Argentin d'Études sur le Climat et ses Impacts (IRL 3351 IFAECI), C1428EGA Ciudad Autónoma de Buenos Aires, Argentina.}
\affiliation{Departamento de Ciencias de la Atmósfera y los Océanos, Facultad de Ciencias Exactas y Naturales, Universidad de Buenos Aires, C1428EGA  CABA, Argentina.}

\author{Stefano Pierini}
\email{stefano.pierini@collaboratore.uniparthenope.it}
\affiliation{Department of Science and Technology, Parthenope University of Naples, Naples, Italy.}

\author{Michael Ghil}
\email{ghil@atmos.ucla.edu}
\affiliation{Geosciences Department and Laboratoire de M\'et\'eorologie Dynamique (CNRS and IPSL), \'Ecole Normale Sup\'erieure and PSL University, 75231 Paris Cedex 05, France.}
\affiliation{Department of Atmospheric \& Oceanic Sciences, University of California, Los Angeles, CA 90095-1565, USA.}
\affiliation{Department of Mathematics, Imperial College London, London SW7 2BX, UK.}

\date{\today}

\begin{abstract}
Templexes are topological objects that encode the branching organization of a flow in phase space. We build on these objects to introduce the concept of \textit{topological modes of variability} (TMVs). TMVs are defined as dynamical manifestations of algebraically defined cycles, called generatexes, in the templex; they provide a concrete link between abstract topological invariants and time-dependent behavior in a model or in observations. 
We apply this approach to a low-order model of the wind-driven ocean circulation, subject to both periodic and aperiodic forcing, and show how TMVs emerge or vanish over time in nonautonomous settings. The analysis reveals that TMVs allow for a qualitatively new understanding of variability in complex systems where linear modes fail to describe the nonlinear dynamics.
\end{abstract}

\maketitle

\begin{quotation}
The theory of nonautonomous dynamical systems (NDSs) provides an appropriate framework for studying how a system's internal variability responds to changes in external forcing, while algebraic topology provides key elements for the description and understanding of nonautonomous dynamics \cite{ghil2023dynamical}.
“Topological modes” appear in various fields of science \cite{wan2024nonlinear,zhang2025time}, but their application to nonlinear flows in phase space remains largely unexplored. In this study, we introduce the concept of \emph{topological modes of variability} (TMVs), defined as entities that characterize specific manifestations of nonlinear behavior. TMVs represent the metric counterpart of the topological properties computed with the help of a templex\cite{charo2022templex,charo2023random}. Particular templex components, called {\em generatexes}, help identify topologically distinct paths in phase space, and allow one to dissect a time series into sequentially activated, distinct modes. These concepts and methods are illustrated using a conceptual model of the wind-driven ocean circulation \cite{Pierini.ea.2016, Pierini.ea.2018}.
In this context, it is the external forcing that activates specific TMVs of the ocean's internal variability. If the forcing is periodic, the sequence of TMVs repeats from period to period, although the order in which they appear may vary. In contrast, under aperiodic forcing, the description is never complete: different TMVs can emerge or vanish. These modes offer a genuinely nonlinear alternative to traditional modal decompositions.
\end{quotation}

\section{Introduction and motivation}\label{sec:intro}

Studying intrinsic low-frequency variability of the mid-latitude double-gyre circulation is crucial for understanding ocean dynamics and climate variability in general. \cite{Dijkstra.Ghil.2005} Simplified, conceptual models play an important role in identifying the nonlinear processes that are responsible for distinct flow regimes and the transitions between them. The effects of time-dependent forcing on intrinsic ocean variability have been investigated using a full hierarchy of oceanic and coupled ocean-atmosphere models, from the simplest conceptual ones to the highest resolution ones \cite{Ghil.Luc.2020}.

\subsection{Simple models of the wind-driven circulation}
\label{ssec:WDG}

Idealized models of the wind-driven circulation that emphasize nonlinear effects and the bifurcations they give rise to include the four-dimensional (4-D), spectral quasi-geostrophic (QG) model of \citet{veronis1963analysis}, who found periodic solutions, and the spatially two-dimensional (2-D) shallow-water model of \citet{jiang1995multiple}, who studied a finite-difference version thereof that had intermediate horizontal resolution, and found multiple steady states, as well as limit cycles and chaotic solutions. The latter authors also investigated a QG, spectrally 2-D version of their model, in which the western boundary currents \cite{Ghil.Chil.1987}, absent in earlier work, were introduced by an exponential term in the basis functions.

\citet{simonnet2005homoclinic} and \citet{pierini2011low} both studied 4-D spectral QG models using the westward intensification of the flow introduced in \citet{jiang1995multiple}, but differed in the selection of their basis functions: the former used the tensor product of one longitudinal mode with four meridional ones, while the latter used the tensor product of two longitudinal and two meridional modes. \citet{simonnet2005homoclinic} emphasized the role of a homoclinic bifurcation in finding two types of chaos, of \citet{lorenz1963lorenz} type and of \citet{Shilnikov.1965} type, while \citet{pierini2011low} emphasized the role relaxation oscillations play in giving rise to chaos, while \citet{Pierini.ea.2016} went further in exploring the role of time-dependent forcing in affecting the model's intrinsic chaos. Furthermore, \citet{Pier.Ghil.2021} analyzed the tipping points that arise in the same model \cite{pierini2011low} under the action of a parameter drift.

The model's four ordinary differential equations form a nonlinear dynamical system for the amplitudes of the flow's streamfunction modes. Reasonably realistic values for the spatial scales and the wind stress curl can be used, allowing for the expected bifurcation scenario to occur in spite of the severe truncation. The autonomous case was studied for increasing values of the wind stress, yielding steady states, periodic solutions and chaotic ones \cite{pierini2011low, Pierini.ea.2016}, including Shilnikov type chaos, in which the homoclinic orbit spirals back to the unstable fixed point. \citet{Pierini.ea.2016} considered the nonautonomous case for an aperiodic forcing with a decadal characteristic time, while \citet{pierini2014ensemble} and \citet{Pierini.ea.2018} studied the model's response to periodic forcing by metric methods that did not appeal to algebraic topology.  

The purpose of the present paper is to bring to bear on this simple oceanic model recent  concepts and methods from the theory of nonautonomous dynamical systems (NDSs) and algebraic topology \cite{ghil2023dynamical}, leading up to the introduction of Topological Modes of Variability (TMVs). These TMVs will be shown to be totally different from various other data-driven decomposition modes that are either purely statistical or borrow only moderately from the underlying dynamics, such as empirical orthogonal functions and principal component analysis \cite{Preisendorfer.1988, Jolliffe.2002} or proper orthogonal decomposition \cite{Berkooz.ea.1993}.

\subsection{Nonautonomous dynamical systems}
\label{ssec:NDS}

NDS theory deals with the ways that dynamical systems with time-dependent forcing or coefficients can be handled as completely as autonomous systems in which no time dependence is explicitly built in. For the latter ones, solutions $x(t; t_0)$ depend only on the elapsed time $t - t_0$ between the fixed initial time $t_0$ and the time of observation $t$. For NDSs, solutions depend separately, as a two-parameter family, on the initial time usually denoted by $s$ and the time of observation $t$. 

For autonomous systems, the only attractors are forward attractors, obtained as $t \to + \infty$. For NDSs, one has either {\em pullback attractors (PBAs)}, obtained for a given $t$ as $s \to - \infty$, or {\em forward attractors}, which we will label as {\em FWAs}, obtained for a given $s$ as $t \to + \infty$. The latter though are distinct from the forward attractors of autonomous systems, since the conditions for the existence and uniqueness of an FWA are more complicated \cite{Car.Han.2017, Kloeden.Yang.2020}. In the mathematical literature, PBAs were introduced first for random dynamical systems by \citet{crauel1994attractors} and by  \citet{arnold1998random}; they were applied to climate problems first by \citet{Ghil.ea.2008} and by \citet{chekroun2011stochastic}.

At roughly the same time, in the physical literature, attractors for NDSs were proposed under the name of {\em snapshot attractors} by \citet{namenson1996fractal}. While these attractors were defined somewhat more loosely than the FWAs in the mathematical literature, the two are similar in spirit and \citet{Bodai.Tel.2012} and \citet{Tel.ea2020}, for instance, used the snapshot attractor name in climate applications. In recent papers by \citet{herein2017theory, janosi2024overview}, 
the authors referred to their methodology as {\em parallel climate realizations}, a terminology which corresponds roughly to the ensemble simulations that the Intergovernmental Panel on Climate Change (IPCC) uses in its successive assessment reports of past, present and future climate change. \cite{IPCC.AR4.2007}

\citet{Maraldi.ea.2025} have recently discussed in this journal the relations between PBAs, FWAs and {\em uniform attractors} \cite{Haraux.1991, Vishik.1992}; see, in particular, their Appendix~A. Since the model treated herein is dissipative and the forcing in both the periodic and aperiodic case is bounded, we have reasons to believe that the attractors we compute exist in both the forward and pullback sense, and that they are therewith actually uniform \cite{Car.Han.2017,Kloeden.Yang.2020}. In the meantime, we will use the term FWAs for the attractors we are computing and leave the mathematically rigorous treatment of the matter for future work.

\subsection{Algebraic topology and dynamics: the cell complex}
\label{ssec:AlgTop}

As Henri Poincaré observed in 1892 \cite{poincare1893methodes}, perturbation theory fails to account for the mechanisms of stretching and squeezing of the flow in phase space under the action of deterministic governing equations. This observation led him to lay the foundations of {\em Analysis Situs} \cite{poincare1895analysis}, relying on earlier results of Enrico Betti, Bernhard Riemann and James Joseph Sylvester. This branch of mathematics, now called algebraic topology, has become central to identifying the essential features that distinguish different types of dynamics in phase space. The computation of topological invariants in phase space relies in practice on the analysis of point clouds derived from numerical simulations of dynamical models as well as from time series measured in the lab or observed in nature. 
 
Recent work demonstrates that algebraic topology can characterize chaotic attractors in both deterministic and stochastic settings \cite{charo2021noise,ghil2023dynamical}, by using {\em Branched Manifold Analysis through Homologies} to construct a {\sc BraMAH} cell complex \cite{sciamarella1999topological,sciamarella2001unveiling} in the deterministic case or families of indexed cell complexes in the random one \cite{charo2021noise,charo2023random}. In algebraic topology, a cell complex is a structure composed of cells of various dimensions --- e.g., points are 0-cells, segments are 1-cells, polygons are 2-cells, polyhedra are 3-cells, and so on --- that can be built from a point cloud to approximate its shape. 

Once the cell complex is constructed, it is analyzed algebraically, focusing not on the specific coordinates of the 0-cells that anchor the complex to the phase space but on the way the cells of the complex are glued together \cite{munkres2018elements}. The algebraic analysis of a {\sc BraMAH} complex is independent of the particular cell complex used to approximate the point cloud: the cell complex acts as a Wittgenstein ladder \cite{Wittgenstein.1922} that is discarded once the topological properties are identified. These topological properties are invariant and they are not affected by the number, distribution, or size of the cells used to approximate the manifold into which the dataset is embedded.

\subsection{Flow on the cell complex and the templex}
\label{ssec:Templex}

The point cloud being analyzed may be derived from embedding a time series in phase space or from a trajectory in phase space. In either case, the cells of the complex are visited by the trajectory in an order determined by the flow. This order is entirely different from the orientation of the cells as defined in algebraic topology. Cell orientation merely serves to define an algebra of chains, enabling the computation of homology groups, which represent holes of various dimensions.  

In contrast, the sequence in which a trajectory associated with a dynamical system visits the cells is limited to the highest-dimensional cells of the complex, and the possibility of flowing from one cell to another is related to causality. This information can be incorporated using a directed graph (digraph), where the nodes correspond to the highest-dimensional cells of the complex and the edges stand for the connections between cells established by the flow. Together, the cell complex and this digraph form what \citet{charo2022templex} called a {\em templex}.  A templex encodes the topological structure underlying the point cloud, along with the organization of the flow within that structure.

\subsection{Constructing a templex for the problem at hand}
\label{ssec:Tplx_WDG}

When a system is subjected to a time-dependent driving force, one seeks to understand how this forcing alters the flow's topological structure under the resulting nonautonomous governing equations.  How is the topology affected? Does forcing only alter the properties of the digraph, or does it also induce changes in the cell complex?  How do the ensemble simulations that correspond to the FWA map onto the templex? To address these questions, we construct herein templexes for both a periodically and an aperiodically forced model of low-frequency variability in the wind-driven ocean circulation.

\subsection{Topological modes}
\label{ssec:Dtm}

The notion of a mode arises in many areas of mathematics and the sciences, often with slightly different interpretations depending on the area under study and on the analytical framework. Broadly speaking, a mode is a fundamental building block or characteristic pattern that helps describe a system's behavior. A summary of commonly used linear modal decomposition methods, along with remarks on their inherent linear assumptions, is presented in Table~\ref{tab:lin_modes}.
The essential distinction between TMVs and the methods in this table is that the TMVs are concatenated in time, rather than superposed, like the modes in the table, for all times.

\begin{table*}[htpb]
\centering
\renewcommand{\arraystretch}{1.3}
\begin{tabularx}{\textwidth}{|>{\raggedright\arraybackslash}p{4cm}|X|}
\hline
\textbf{Method} & \textbf{Linear nature} \\
\hline
Fourier Analysis (sines/cosines) 
& Assumes linear superposition \& stationarity. Cannot capture nonlinear features in phase space. \cite{Alessio.2015}\\
\hline
PCA / EOF \& Linear combinations of variables 
& Yields directions of maximal variance, based on the covariance matrix. Fails to capture nonlinear manifolds. \cite{Jolliffe.2002} \\
\hline
SVD \& Linear decomposition (rank-1 components) 
& Decomposes any matrix into a sum of orthogonal linear components. Foundation of PCA and SSA. \cite{Golub.Loan.2013} \\
\hline
DMD \& Best-fit linear operator for dynamics 
& The Koopman operator is linear and operates on observables of nonlinear dynamics. DMD analyzes a finite-dimensional approximation of the Koopman operator. \cite{Schmid.2010} \\
\hline
SSA \& Linear decomposition after time-delay embedding 
& Uses SVD on a trajectory matrix after time-delay embedding. The latter captures some nonlinearity through delay vectors, but decomposition remains linear. \cite{Ghil.ea.2002} \\
\hline
\end{tabularx}
\caption{Intrinsic linearity of common modal decomposition methods. PCA: Principal Component Analysis; EOF: Empirical Orthogonal Functions; SVD: Singular Value Decomposition; DMD: Dynamic Mode Decomposition; SSA: Singular Spectrum Analysis. General references are provided for each method.} 
\label{tab:lin_modes}
\end{table*}

Modes of a topological nature have been introduced in the field of condensed matter physics, for instance \cite{wan2024nonlinear,zhang2025time}. In these studies, topological modes refer to spatially localized excitations. The systems under consideration operate in physical space—such as positions on a lattice or corners of a metamaterial—and, in some cases, involve temporal or frequency aspects through Floquet theory. They do not, however, address phase space, understood here as the space of all possible states of a dynamical system.

This paper focuses on the application of the templex framework to study the TMVs associated with the wind-driven ocean circulation, showing how  different forms of external forcing affect these inherently nonlinear modes. Beyond this specific application, TMVs seem well adapted to the study of truly nonlinear aspects of evolution problems, be it in the climate sciences or elsewhere.

The article is organized as follows. Section~\ref{sec:templex_def} introduces the main concepts and tools of chaos topology, to prepare the model's topological analysis using the templex approach.  Section~\ref{section:TMV} introduces the TMV concept.
Section~\ref{sec:autonomous_part} presents the templex analysis of the four-dimensional autonomous system associated with our low-order spectral QG model. Section~\ref{sec:nonautonomous_part} examines the nonautonomous cases, where first periodic and then aperiodic forcing is applied to the model, and the corresponding templex analyses are conducted. In Section~\ref{sec:snapshots}, we compute the FWAs for the nonautonomous scenarios and explore the relationship between these attractors and the TMVs. The conclusions of the study are summarized in the final section.

\section{Chaos topology and the templex approach}
\label{sec:templex_def}
The structure of a dynamical system's flow in phase space emerges from a series of processes --- stretching, folding, tearing, and squeezing --- that are driven by the governing equations and repeatedly occur for any initial condition \cite{gilmore2013topology}. This section presents newly developed tools that characterize the topological structure of a deterministic flow in phase space \cite{sciamarella2024new}.

\subsection{The cell complex and homology groups} 
\label{ssec:Bramah}

The analysis starts with a point cloud in the full phase space of dimension $n$ or in a lower-dimensional phase space, as long as the subspace is free of false neighbors so as to  preserve the topology. To compute the topology of the manifold in which the points lie, this point cloud must be (i) sufficiently well sampled, and (ii) have a long enough time span for the attractor to be explored to a satisfactory exten. Grouping subsets of points into cells is the first step in the construction of a cell complex. The subset of points in a cell should be locally homeomorphic to either a full or a half disk in $d \leq n$ dimensions, where $d$ is the local dimension of the subset. The cell representing such a subset is termed a $d$-cell. 

Let $\bm{\phi}$ represent a sufficiently well sampled flow with an underlying $d$-dimensional manifold in $\mathbb{R}^n$. A {\sc BraMAH} complex of dimension $d$ consists of $k$-cells with $k \leq d$ such that:
\begin{enumerate}[(i), nosep]
\item The $0$-cells form a sparse subset of the original point cloud in phase space; 
\item $d\leq n$ denotes the local dimension of the branched manifold; and  
\item Each $d$-cell is defined so that $d$ of the singular values that characterize the distribution of points around the barycenter of the $d$-cell scale linearly with the number of points within the cell. 
\end{enumerate}

The third property fails to hold when the local curvature of the branched manifold becomes significant. In this case, singular values that grow with higher powers reflect the extent to which the manifold deviates from its tangent space. Depending on the chosen tolerance to curvature during the computation, various {\sc BraMAH} complexes can be constructed from a point cloud. Despite these variations, all properly constructed complexes share the key property of providing an accurate skeleton that represents the underlying structure supporting the flow. 

The topological properties of a cell complex $K$ — such as homology groups, torsions, and weak boundaries — are computed following standard procedures in algebraic topology \cite{Hatcher2002}. The homology groups $H_k(K)$ of $K$ capture the nontrivial loops or $k$-holes in a topological space for different dimensions. A $k$-hole in $K$ is a $k$-cycle that is not the boundary of any ($k+1$)-chain \cite{kinsey2012topology}. We provide in Appendix~\ref{ap:BraMAH} a brief account of the computation of homology groups on a toroidal surface and a Klein bottle.

\begin{figure}[ht]
\centering
\begin{subfigure}[b]{0.49\textwidth}
\includegraphics[width=\textwidth]{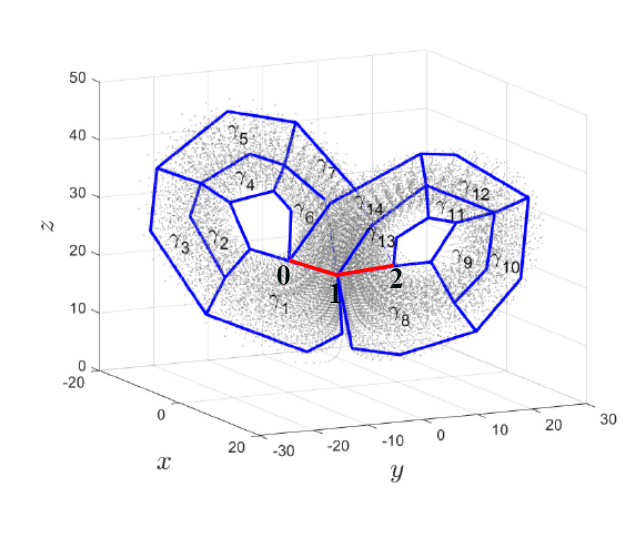}
\caption{$K(L)$}
	\end{subfigure}
~	
\begin{subfigure}[b]{0.4\textwidth}
	\includegraphics[width=\textwidth]{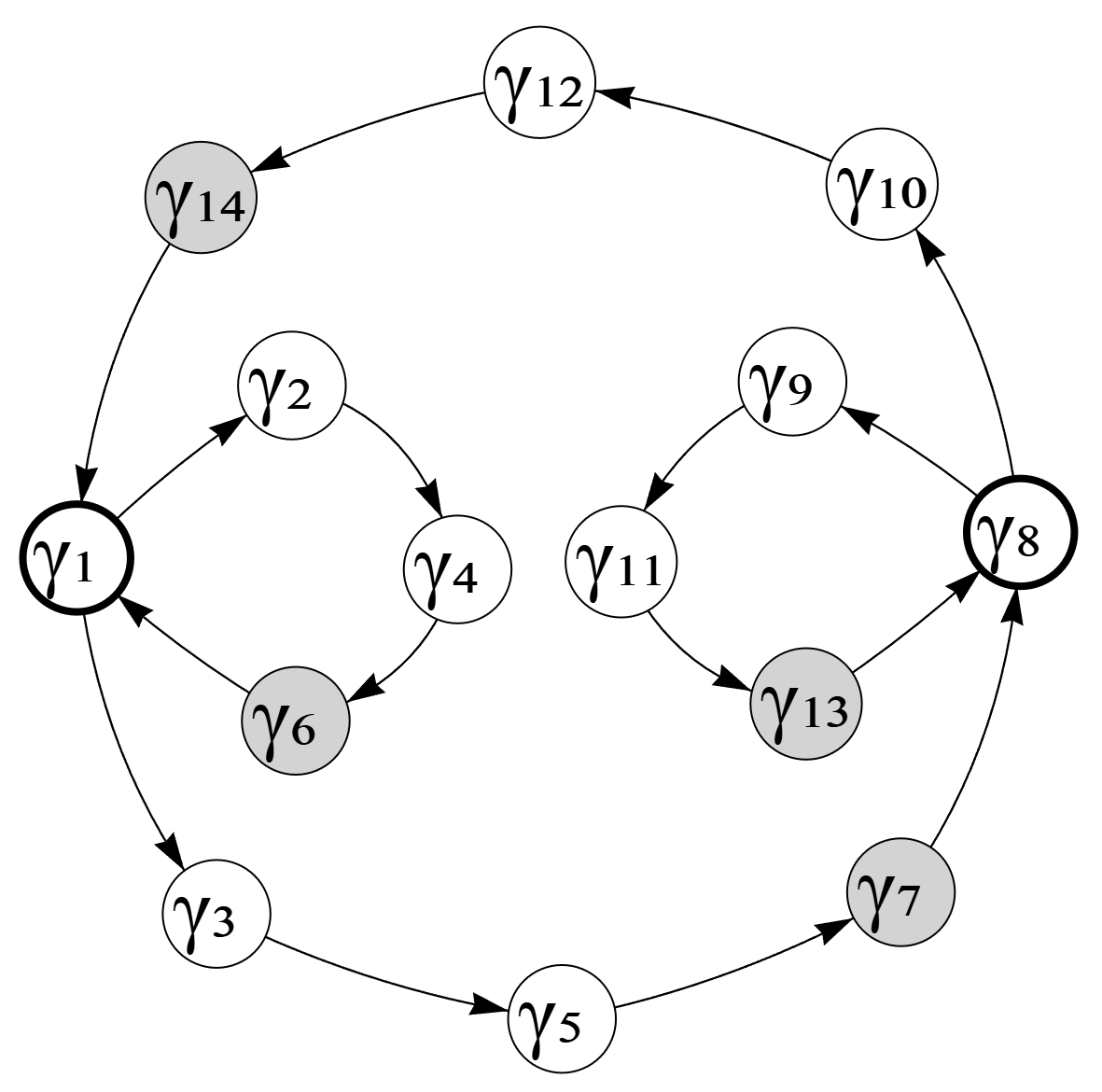}
	\caption{$G(L)$}
	\end{subfigure}
\caption{A templex $T(L)=(K(L),G(L))$ for a three-dimensional point cloud $L$ corresponding to the Lorenz attractor \cite{lorenz1963lorenz}. (a) The {\sc BraMAH} cell complex $K(L)$; the two $1$-cells colored in red constitute the joining line $J=\langle 0,1 \rangle \cup \langle 1,2 \rangle$. (b) The digraph $G(L)$: the ingoing nodes $\gamma_6,\gamma_7, \gamma_{13}$ and $\gamma_{14}$ are shaded, and the outgoing nodes $\gamma_1$ and $\gamma_8$ have a thicker border. Note that the digraph $G(L)$ has three nonequivalent cycles: an exterior one and two interior ones; that latter two connect to the exterior one at the nodes 
$\gamma_1$ and $\gamma_8$, respectively.
}
\label{fig:L63}       
\end{figure}

An example of a {\sc BraMAH} cell complex $K(L)$ for the Lorenz attractor\cite{lorenz1963lorenz} is provided in Figure~\ref{fig:L63}(a). This cell complex is characterized by a key feature, namely the "sowing together" of the two wings of the butterfly by their interconnection through the two $1$-cells that are highlighted in color. This structural element assumes a pivotal role within the cell complex and the subsequent tracking of the flow supported by the complex.

\subsection{The joining locus and the digraph} 
\label{ssec:digraph}

In general, any {\sc BraMAH} complex $K$ of dimension $d$ is said to have a {\it joining locus} if there are $(d-1)$-cells shared by at least three distinct $d$-cells. If $d=2$, the joining locus is called a joining line, as it is composed of $1$-cells. Joining lines can be seen as the analogs of the components of a Poincaré section. The Lorenz attractor is thus found to have a two-component joining locus, formed by the $1$-cells $\langle0,1\rangle$ and $\langle1,2\rangle$.

Let us now consider the underlying flow across the joining locus as depicted in Figure~\ref{fig:L63}(b). The flow goes from the $2$-cells $\gamma_6$ and $\gamma_{14}$ into $\gamma_1$ through $\langle0,1\rangle$, and it goes from $\gamma_{7}$ and $\gamma_{13}$ into $\gamma_8$ through $\langle1,2\rangle$.  We discriminate between these 2-cells according to the sense of the flow: $\gamma_6$, $\gamma_7$, $\gamma_{13}$ and $\gamma_{14}$ are the ingoing cells; and  $\gamma_1$, $\gamma_8$ are the outgoing cells. 

Defining flow paths on the cell complex is straightforward: it can be specified not only for the cells abutting the joining locus but also for the rest of the $2$-cells, leading to a digraph whose nodes coincide with the $2$-cells and whose edges determine whether the nodes are connected by the flow. As previously stated in Sect.~\ref{sec:intro}, we call the mathematical object formed by the cell complex $K$ and the associated digraph $G$ a templex.

Homologies provide the proof that nontrivial loops in the cell complex are fundamental for distinguishing different topological structures. Can we find nontrivial loops in the digraph that capture distinct essential properties of the flow along the structure?  In fact, we can introduce a few definitions to capture such properties, just as we have retained the essential properties of the structure itself through the computation of the homology group generators. 

A nontrivial loop in the digraph corresponds to a \textit{generatex}. More precisely, a \textit{generatex} is a subtemplex whose subgraph corresponds to a cycle in the digraph and the subcomplex is the set of cells corresponding to that cycle. As the number of cells in a cell complex is arbitrary, we define an equivalence relation to obtain a single representative for each group of equivalent generatexes, much like what is done with $k$-generators in homology group theory, leaving us with the minimal set of representative generatexes for a given point cloud.

Two generatexes are considered equivalent if they share the same set of ingoing and outgoing nodes, and thus correspond to a single cycle with no other equivalents in the digraph. The subset of generatexes associated with nonequivalent cycles will be referred to as the \textit{representative generatexes}. A representative generatex is said to be of order~$p$, with $p \in \mathbb{N}$ and $p \geq 1$, if its cycle contains $p$ distinct ingoing nodes. Hereafter, for simplicity, we use the term generatex to refer to a representative generatex.

Last but not least, just as torsion groups can be computed alongside homology groups—enabling, for instance, the distinction between a regular strip and a Möbius strip — it is also possible to detect local twists in a templex.

Figure~\ref{fig:L63}(b) shows the digraph $G(L)$. The generatex set is formed by three elements:
\begin{subequations}
\begin{align}
    \mathcal{G}_1(L) & = \{\gamma_1 \rightarrow \gamma_2 \rightarrow \gamma_4 \rightarrow \gamma_6 \rightarrow \gamma_1 \}, \\
    \mathcal{G}_2(L) & = \{\gamma_8 \rightarrow \gamma_9 \rightarrow \gamma_{11} \rightarrow \gamma_{13} \rightarrow \gamma_8 \}, \\
    \mathcal{G}_3(L) & = \{\gamma_1 \rightarrow \gamma_3 \rightarrow \gamma_5 \rightarrow \gamma_7 \rightarrow \gamma_8 \notag \\
    & \hspace{1.5cm} \rightarrow \gamma_{10} \rightarrow \gamma_{12} \rightarrow \gamma_{14} \rightarrow \gamma_1 \}.
\end{align}
\end{subequations}
The first two generatexes have order 1, whereas the third one has order 2, and none of them exhibits any torsion.

Identifying a generatex in the phase space is simple in the case of a templex, as the coordinates of the $0$-cells of the {\sc BraMAH} complex are provided directly by the construction.

With this set of concepts, we can describe in detail not only the structure associated with a deterministic attractor in phase space but also the representative flow paths around the structure. In the absence of joining loci no alternative paths — and, therefore, no chaos — would be possible.

\section{Topological Modes of Variability (TMV{\sc s})} \label{section:TMV}

The topology of chaos is known to serve various purposes, from unraveling how chaotic behavior arises to informing the modeling of system dynamics directly from data. In this work, we go a step further by building a modal decomposition of system evolution in time upon the generatexes. While generatexes, like holes, arise from purely topological, and thus metric-free, considerations, we focus here on how they manifest themselves within the phase space's metrics. This leads to a new use of generatexes, namely defining the system’s TMVs.

In this section, we define the system’s TMVs as state sequences in phase space that each correspond to a generatex in the templex. The introduction of TMVs allows us to translate topological properties of the system's dynamics into the features of a time series that the system generates. The typical duration of each TMV is defined as the time it takes for a trajectory to loop once around its corresponding generatex. This value should not be confused with the variable amount of time the system actually spends within the region of phase space associated with a specific TMV during each of its visits.

\begin{figure*}[t]
\centering
\begin{subfigure}[b]{0.7\textwidth}
	\includegraphics[width=\textwidth]{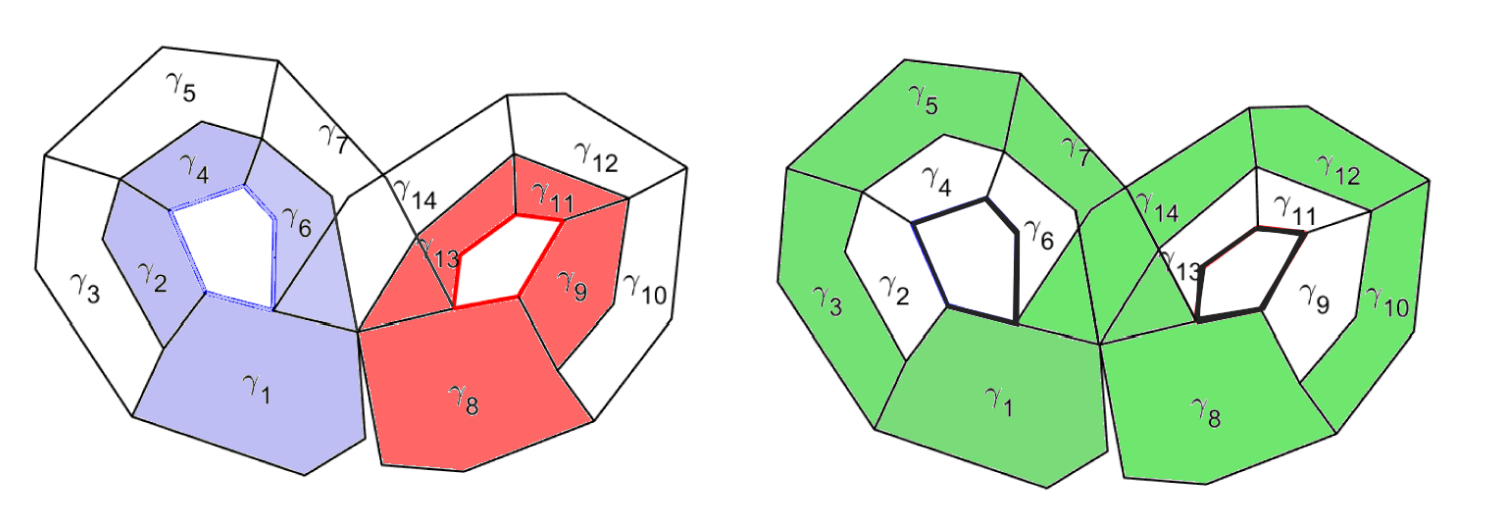}
	\end{subfigure}	
\begin{subfigure}[b]{0.95\textwidth}
	\includegraphics[width=\textwidth]{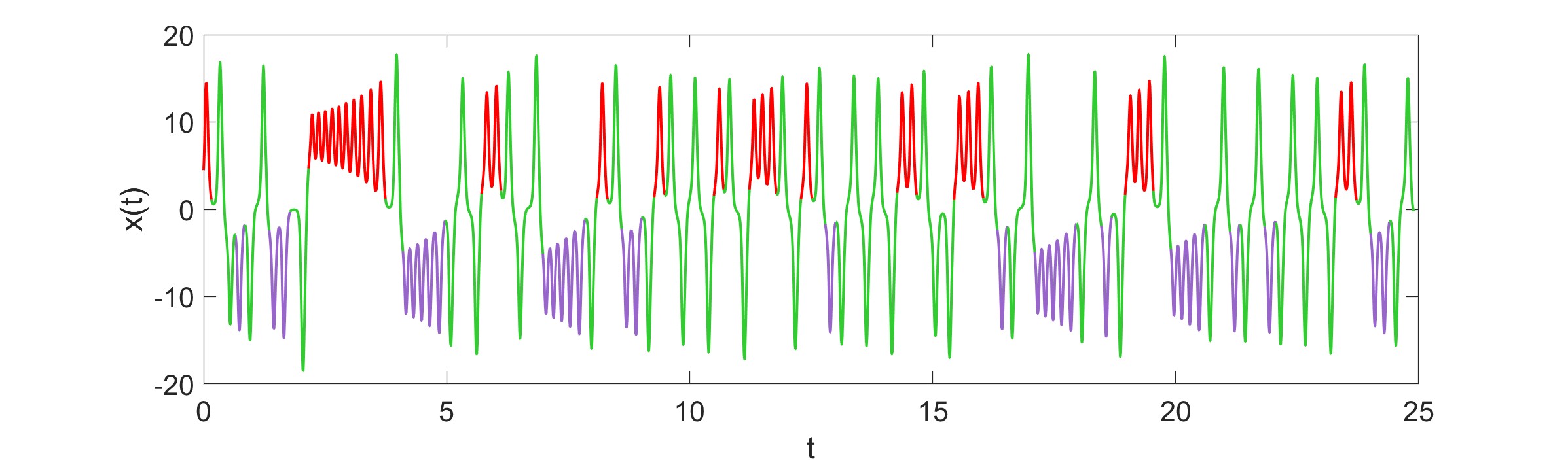}
	\end{subfigure}
\caption{Generatexes and TMVs for the classical Lorenz strange attractor \cite{lorenz1963lorenz}. (Top) Set of the three generatexes: $\mathcal{G}_1(L)$ (purple), $\mathcal{G}_2(L)$ (red), and $\mathcal{G}_3(L)$ (green) that are illustrated in Fig.~\ref{fig:L63}(b) and correspond to TMV-1, TMV-2, and TMV-3, respectively. (Bottom) A segment of a numerical solution $x(t)$ of the Lorenz model \cite{lorenz1963lorenz}, colored according to the decomposition into TMVs.}
\label{fig:tvm_lor}
\end{figure*}

In the classical Lorenz case \cite{lorenz1963lorenz}, Figure~\ref{fig:tvm_lor} (top) displays three colored generatexes in the cell complex: $\mathcal{G}_1(L)$ (purple), $\mathcal{G}_2(L)$ (red), and $\mathcal{G}_3(L)$ (green). Each generatex $\mathcal{G}_i(L)$ corresponds to TMV-$i$, for $i = 1, 2, 3$. Figure~\ref{fig:tvm_lor} (bottom) shows a segment of the $x(t)$ variable of a model solution, with the signal colored according to its decomposition into TMVs. Note that when a generatex is of order $p \ge 2$, the signal may only partially visit 
it before transitioning to another one. In contrast, when $p = 1$, the signal fully traverses the generatex before moving on.

\section{A low-order model of the wind-driven ocean circulation and its TMV{\sc s}} 
\label{sec:WDG}

Following \citet{pierini2011low} and \citet{Pierini.ea.2016}, we investigate here a 4-D spectral QG model of the wind-driven, double-gyre ocean circulation in midlatitudes. The full model's streamfunction is expanded in a rectangular domain according to
\begin{equation} \label{eq:streamfcn}
    \psi(x,y, t) = \sum_{i=1}^{4} \Psi_i(t) \lvert i \rangle. 
\end{equation}
\noindent
The orthonormal basis $\lvert i \rangle$ is defined by:
\begin{subequations} \label{eq:basis}
	\begin{align}
   & \lvert 1 \rangle = e^{-\alpha x} \sin x \sin y, \\
  & \lvert 2 \rangle  = e^{-\alpha x} \sin x \sin 2y, \\
    & \lvert 3 \rangle = e^{-\alpha x} \sin 2x \sin y, \\
    & \lvert 4 \rangle = e^{-\alpha x} \sin 2x \sin 2y,
\end{align}
\end{subequations}
with a constant value for $\alpha > 0$. This basis satisfies the free-slip boundary conditions along the boundaries of the rectangular domain and it also accounts for the oceanic flow's western boundary layer, often referred to in physical oceanography as westward intensification \cite{Gill.1982, Ghil.Chil.1987}. 

The set of four nonlinear coupled ordinary differential equations governing the evolution of the state vector $\mathbf{\Psi}(t) = (\Psi_1(t), \Psi_2(t), \Psi_3(t), \Psi_4(t))^{\rm T}$ can be expressed in vector-matrix notation as:
\begin{subequations} \label{eq:ODEs}
	\begin{align}
   &  \frac{d\mathbf{\Psi}}{dt} + \mathbf{\Psi} \mathbf{J} \mathbf{\Psi} + \mathbf{L} \mathbf{\Psi} = \bm{W}(\bm{x}, t), \label{eq:ppal} \\
   & \bm{W}(\bm{x}, t) = \gamma [1 + \varepsilon f(t)] \bm{w}(\bm{x}). \label{eq:forcing}
\end{align}
\end{subequations}
\noindent
where $\varepsilon$ and $\gamma$ are real dimensionless constants.

In the forcing term on the right-hand side of Eq.~\eqref{eq:ppal}, $\bm{w}$ gives the spatial structure of the wind stress curl and $1 + \varepsilon f(t)$ defines the time dependence of its intensity; see \citet{pierini2011low} for the definitions of $\mathbf{J}$, $\mathbf{L}$, and $\bm{w}$.

In the autonomous case of $\varepsilon=0$, the forcing term is constant in time, $\bm{W}(\bm{x}) = \gamma ~\bm{w}(\bm{x})$, and self-sustained relaxation oscillations occur for $\gamma > 1$. These equations are integrated numerically using the Bulirsch–Stoer method \cite{press1992numerical}. Each integration is carried out for a duration that (a) sufficiently exceeds the spin-up time in a given scenario; and (b) lasts for long enough thereafter to permit observing and characterizing the attractor.

\begin{figure*}[t]
	\centering
		\includegraphics[width=0.7\textwidth]{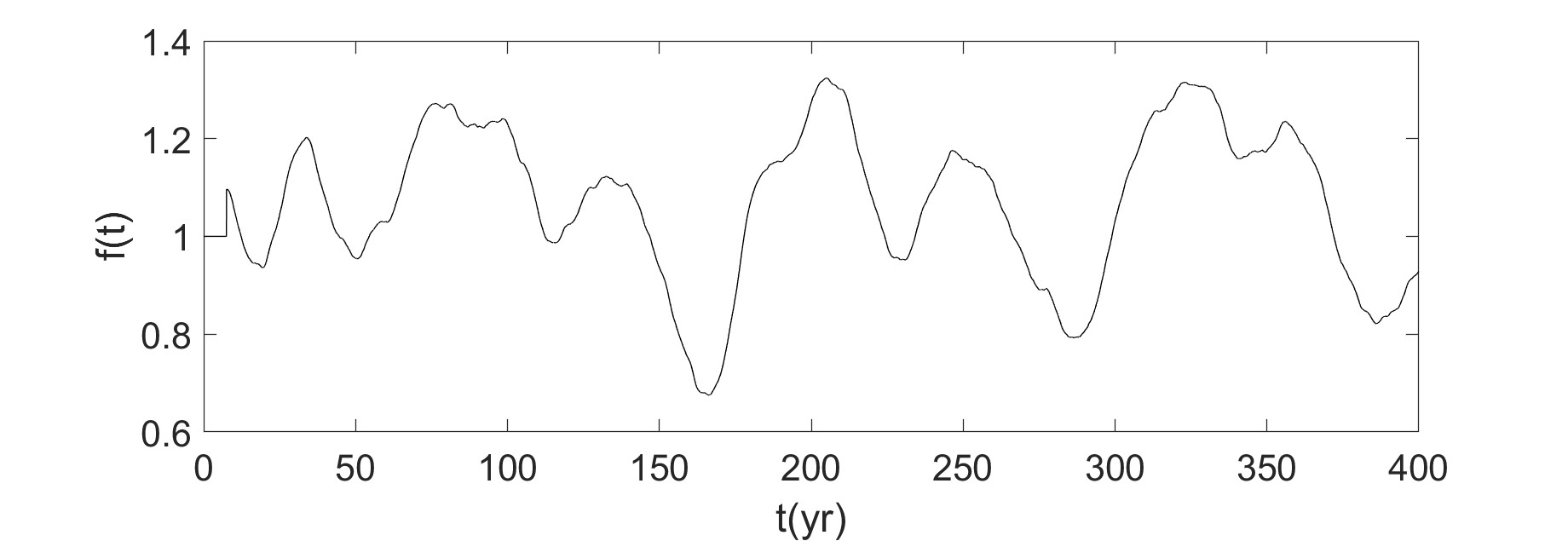}
\caption{Aperiodic forcing $f(t) = F_{T_f} [\zeta(t)]$, $T_f=15$.}
		\label{fig:aperiodic_forcing}
	\end{figure*}
    
In this work, we will study two types of time-dependent forcing. The first one is periodic, with $ f(t) = \sin(\omega t)$ \cite{pierini2014ensemble,Pierini.ea.2018}. The second one involves an aperiodic function  $f(t) = F_{T_f} [\zeta(t)]$ \cite{Pierini.ea.2016}, where $\zeta(t)$ is a fixed realization of an Ornstein-Uhlenbeck process. This process is normalized to have unit variance and autocorrelation time $T_a$. The realization is then smoothed using a sliding window operator $F_{T_f}$ with width $T_f$ \cite{Pierini.ea.2016}, and it is plotted in Fig.~\ref{fig:aperiodic_forcing}.

 In the periodic case, given the dissipative character of the model, an FWA is known to exist. In the aperiodic case, the forcing is bounded, cf.~Fig.~\ref{fig:aperiodic_forcing}, and the existence of an FWA and, hence, of a uniform attractor, is still quite plausible. Please see the discussion of these matters in \citet{Maraldi.ea.2025}, especially Appendices A, B, and D, and references therein \cite{Car.Han.2017,Haraux.1991,Kloeden.Yang.2020,Vishik.1992}, as well as \citet{Wang.ea.2004} in connection with inflated PBAs.

\begin{figure*}[t]
	\centering
		\includegraphics[width=0.8\textwidth]{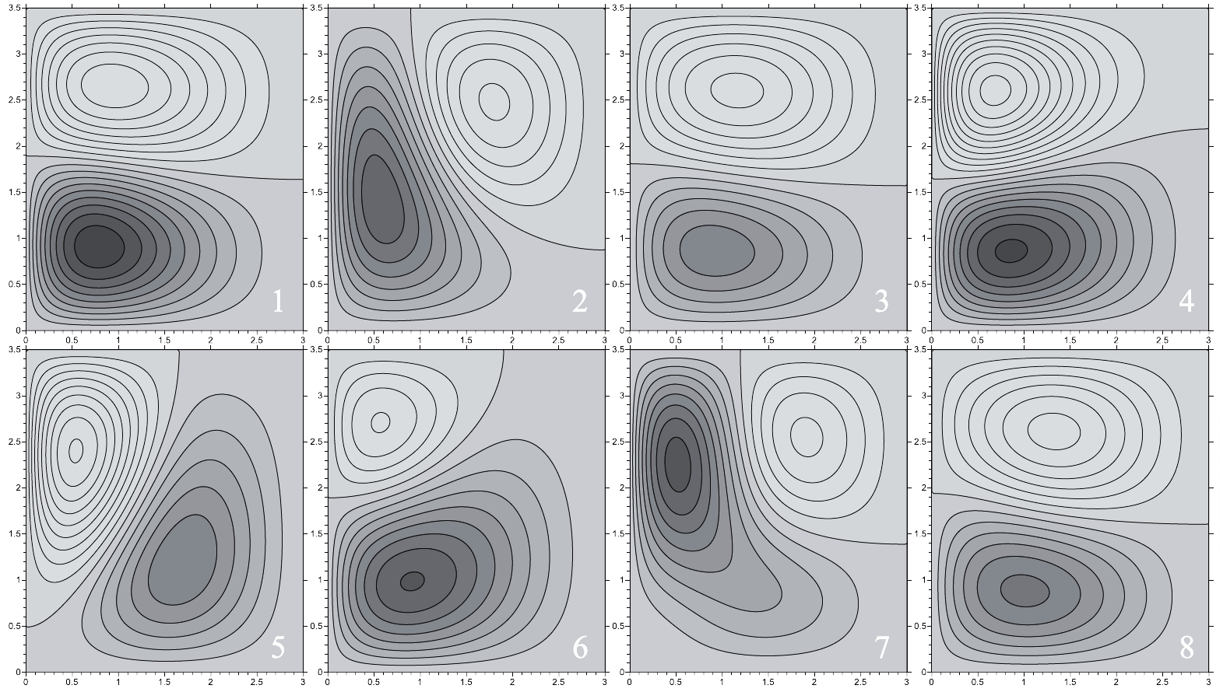}
\caption{Streamfunction evolution for a relaxation oscillation of the idealized ocean model in the autonomous case, with $\gamma = 1.043$. The white numerals indicate the order of the panels in their lower-right corner. Adapted with permission from S. Pierini, J. Phys. Oceanogr., 41, 1585–1604 (2011). Copyright 2011 American Meteorological Society.}
		\label{fig:RO}
	\end{figure*}

\subsection{Templex analysis for the autonomous case}
\label{sec:autonomous_part} 

In this section, we examine the autonomous case for $\varepsilon=0$, where the forcing $\bm{W}(\bm{x}) = \gamma ~\bm{w}(\bm{x})$ is constant in time and the phase space is spanned by $(\Psi_1, \Psi_2, \Psi_3, \Psi_4)^{\rm T}$, cf. Eqs.~\eqref{eq:streamfcn}--\eqref{eq:ODEs}. Templex analysis will be conducted to elucidate the intrinsic variability of the strange attractor that emerges when the parameter value is $\gamma = 1.35$. The bifurcation diagram for the autonomous case is shown in Fig.~1 of \citet{Pierini.ea.2018} and at this value of $\gamma$, large-amplitude relaxation oscillations characterize the model's autonomous behavior. One period of such an oscillation is shown in Fig.~\ref{fig:RO}.

\begin{figure}[h]
	\centering
	\includegraphics[width=0.45\textwidth]{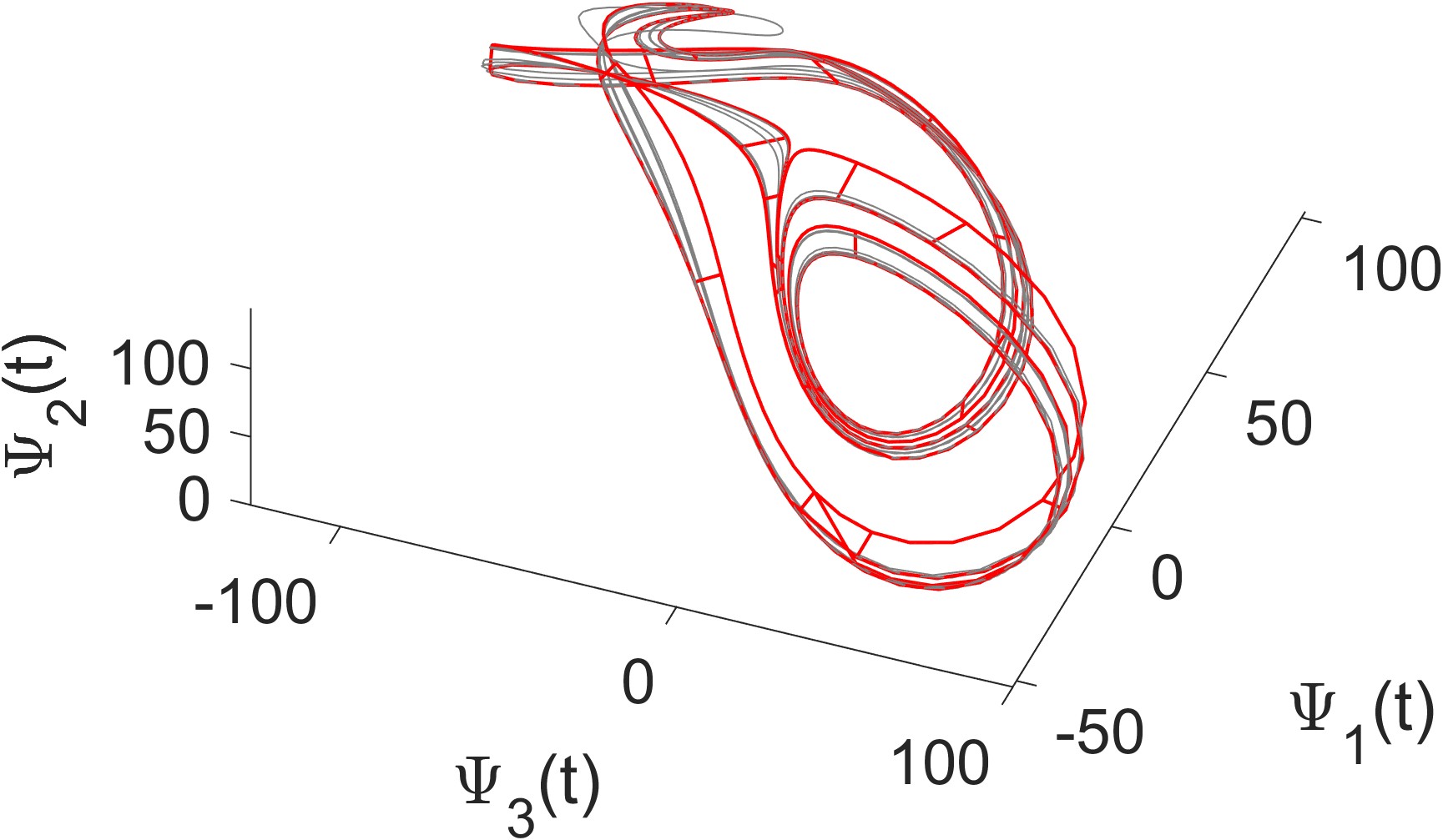} \\[-0.3cm]
\caption{ {\sc BraMAH} cell complex $K_1$ (red) for the autonomous model version's solutions at 
	$\gamma = 1.35$. The point cloud (gray dots) yielding $K_1$ is projected onto the $(\psi_1, \psi_2, \psi_3)$ subspace; the solutions at this $\gamma$-value are relaxation oscillations.}
	\label{fig:autonomous_complex}
\end{figure}

Figure~\ref{fig:autonomous_complex} shows the {\sc BraMAH} cell complex $K_1$, which approximates the four-dimensional point cloud projected onto the subspace \( (\Psi_1, \Psi_2, \Psi_3) \). To improve clarity, the cell complex $K_1$ was simplified by merging some of its 2-cells to yield $\bar{K}_1$, as shown in Figure \ref{fig:autonomous_templex}, where the heavy pink line segment indicates the joining locus $\langle 0, 1 \rangle$. This simplification does not alter the topological structure of the original complex. 

The 1-cell $\langle 2, 3, 4 \rangle$ is a splitting locus, where the $0$-cell $\langle 2 \rangle$ is a tearing point and is called a {\it splitting} 0-cell. The flow that emanates from $\sigma_{12}$ bifurcates into two branches, $\sigma_1$ and $\sigma_2$, due to the {\it splitting} 0-cell. We name the templex of this case $T_1=(\bar{K}_1,G_1)$.

The digraph $G_1$ is shown in Figure~\ref{fig:autonomous_templex}~(b). The ingoing nodes $\sigma_6$ and $\sigma_{11}$ are shaded in gray, and the outgoing node $\sigma_8$ is highlighted by a heavier border. This digraph contains two nonequivalent cycles that correspond to two generatexes of order~1:
\begin{figure*}[t]
\centering
\begin{subfigure}{0.44\textwidth}
\includegraphics{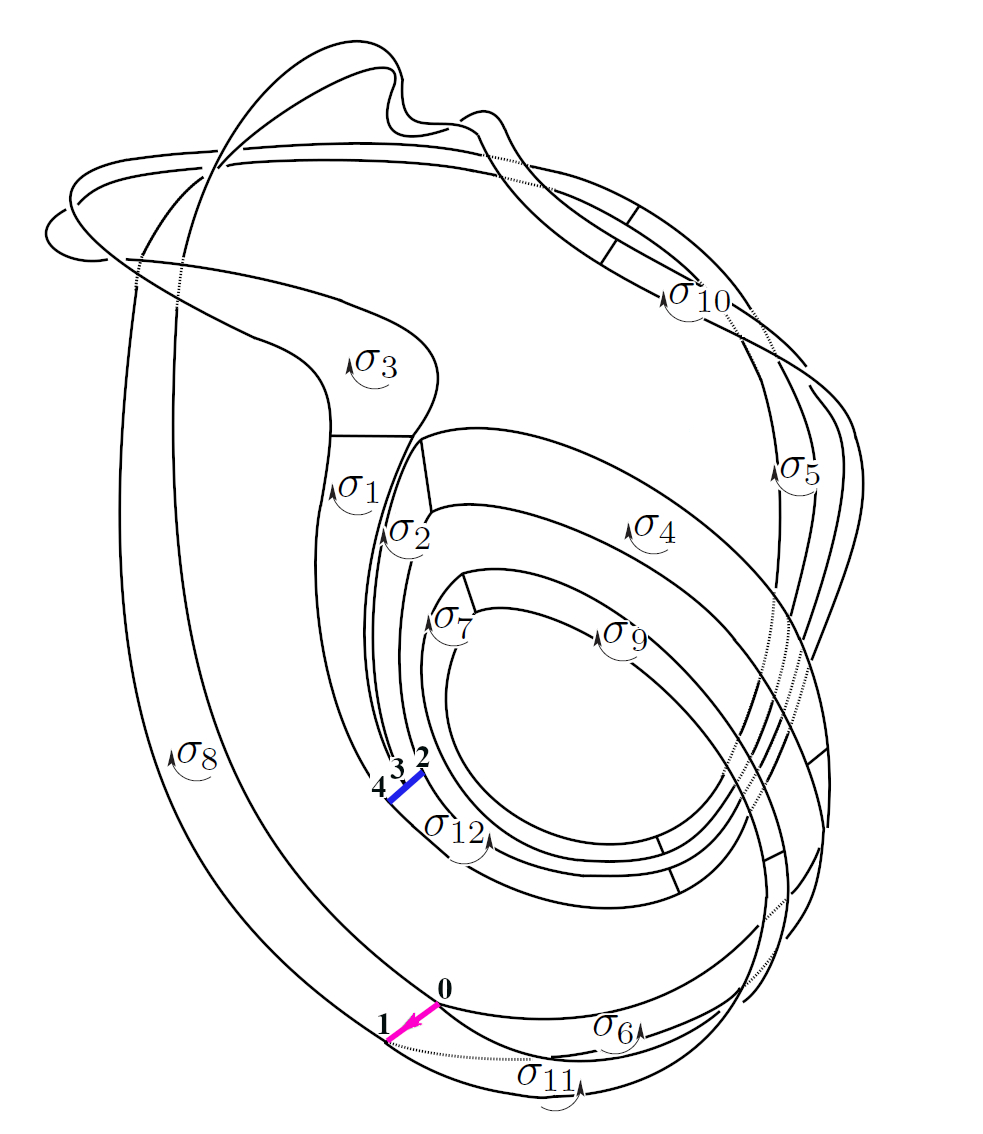}
	\caption{Simplified complex $\bar{K}_1$}
	\end{subfigure}
    \begin{subfigure}{0.38\textwidth}
\includegraphics{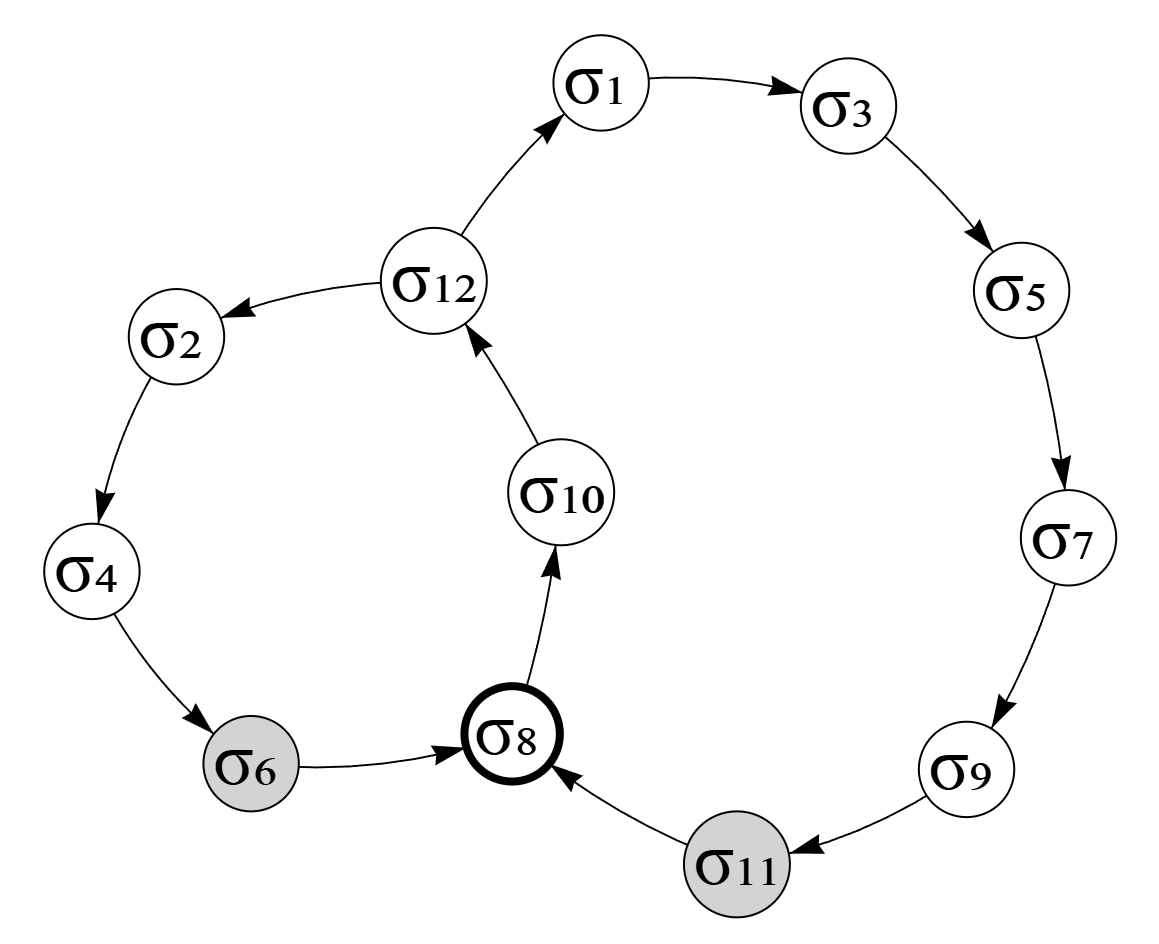}
\caption{Digraph $G_1$}
	\end{subfigure}
\caption{ Templex $T_1=( \bar{K}_1,G_1)$  for the autonomous case in Fig.~\ref{fig:autonomous_complex}. (a) Diagram of the simplified cell complex $\bar{K}_1$. The heavy pink line indicates the joining line $\langle 0,1 \rangle$, and the blue segment correspond to the splitting line  $\langle 2,3,4 \rangle$. (b) Digraph $G_1$: the ingoing cells $\sigma_6$ and $\sigma_{11}$ are shaded, and the outgoing cell $\sigma_8$ has a heavier border; this digraph has 2 nonequivalent cycles.}
\label{fig:autonomous_templex}       
\end{figure*}
\begin{subequations}
\begin{align}
    \mathcal{G}_1(T_1) & = \{\sigma_8 \rightarrow \sigma_{10} \rightarrow \sigma_{12} \rightarrow \sigma_2 \rightarrow \sigma_4 \notag \\ 
    & \qquad \rightarrow \sigma_6 \rightarrow \sigma_8 \}, \\
    \mathcal{G}_2(T_1) & = \{\sigma_8 \rightarrow \sigma_{10} \rightarrow \sigma_{12} \rightarrow \sigma_1 \rightarrow \sigma_3 \rightarrow \sigma_5 \rightarrow \sigma_7 \notag \\
    & \qquad \rightarrow \sigma_{9} \rightarrow \sigma_{11} \rightarrow \sigma_8 \}.
\end{align}
\end{subequations}

\begin{figure*}[!ht]
	\centering
	\begin{subfigure}[b]{0.35\textwidth}
		\includegraphics[width=\textwidth]{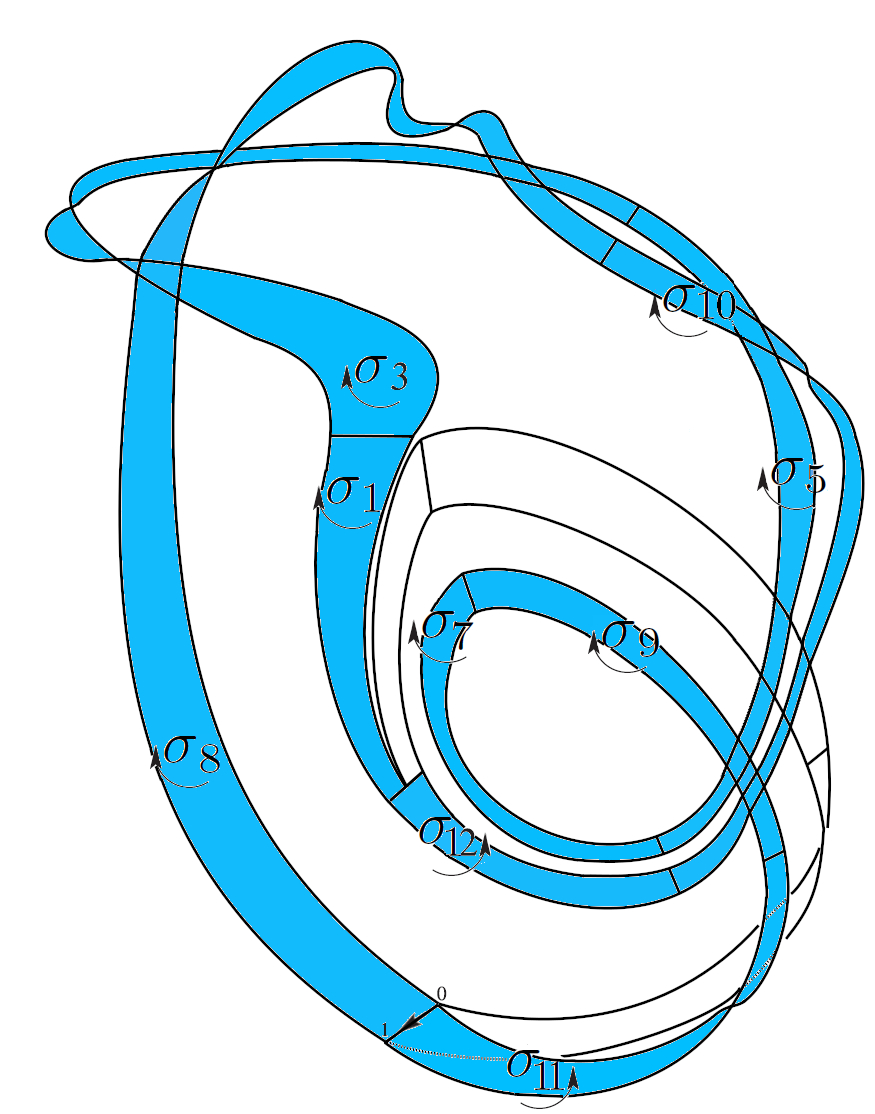}
		\caption*{(a) $\mathcal{G}_1(T_1)$}
	\end{subfigure}
	~
	\begin{subfigure}[b]{0.35\textwidth}
		\includegraphics[width=\textwidth]{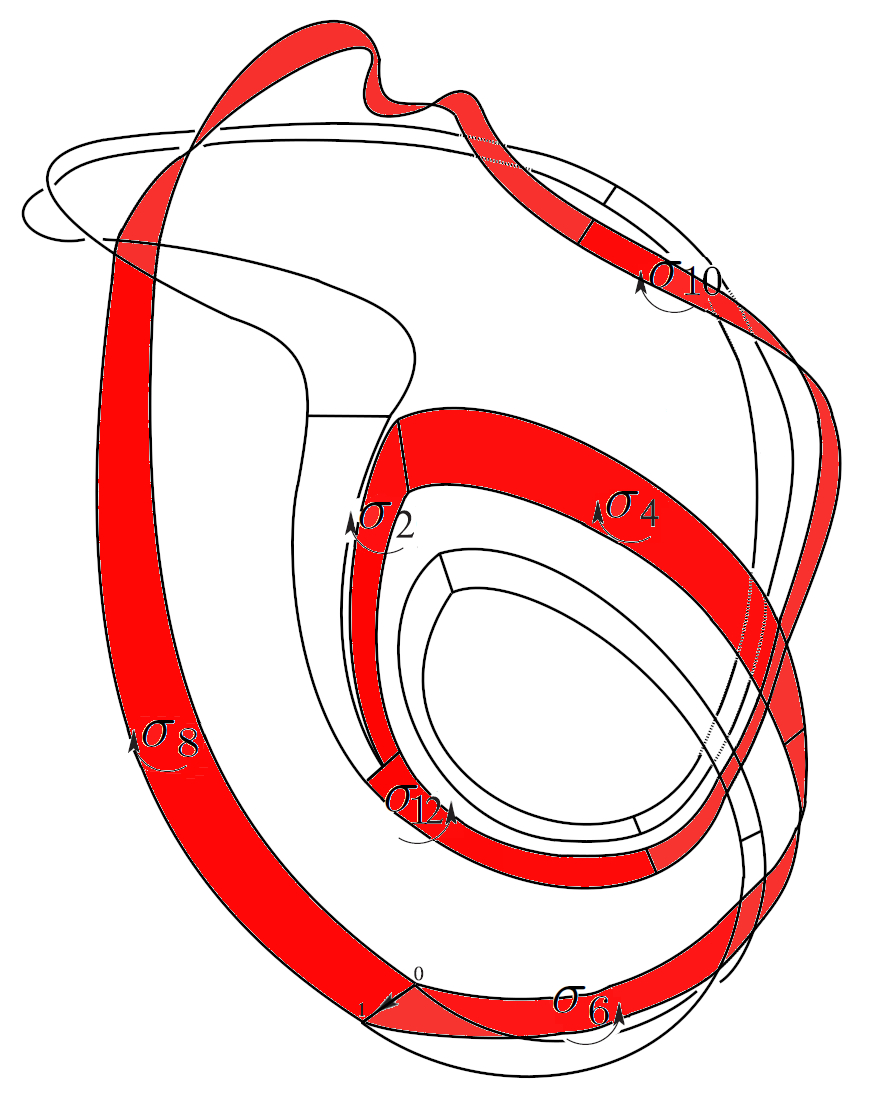}
		\caption*{(b) $\mathcal{G}_2(T_1)$}
	\end{subfigure}

	\vspace{0.5em} 
	\begin{subfigure}[b]{0.9\textwidth}
		\centering
		\includegraphics[width=\textwidth]{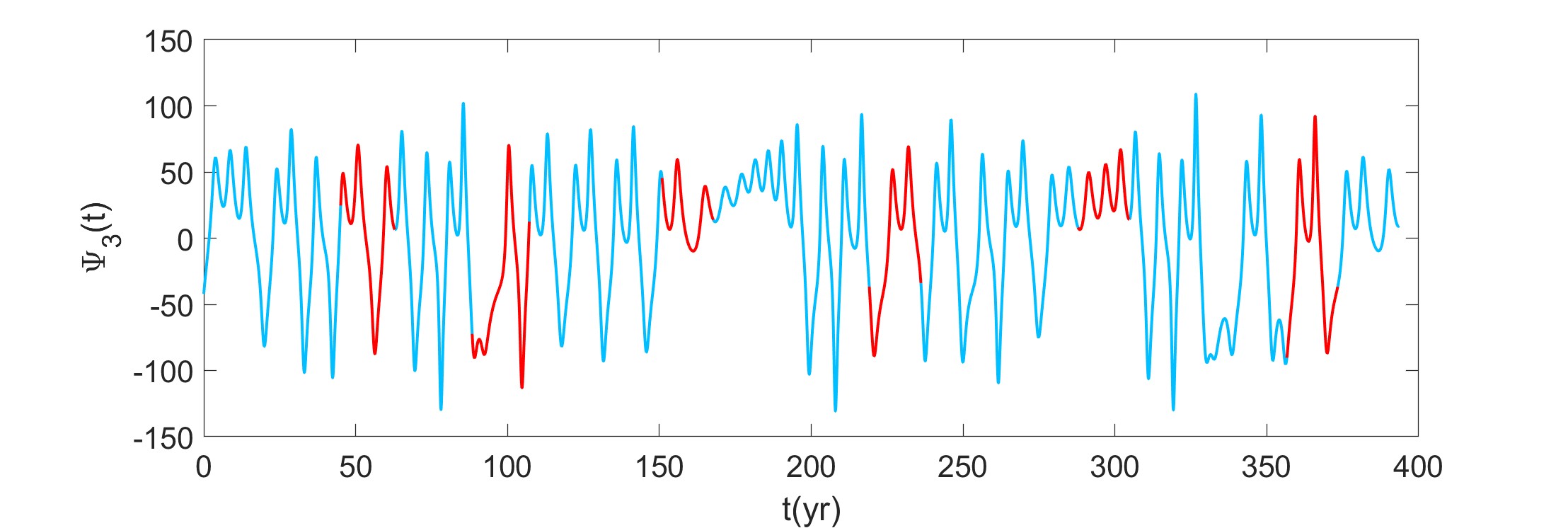}
		\caption*{(c)}
	\end{subfigure}
	\caption{The set of the two generatexes: $\mathcal{G}_1(T_1)$ (a), and $\mathcal{G}_2(T_1)$ (b) which corresponds to TMV-1 and TMV-2, respectively for the autonomous case in Fig.~\ref{fig:autonomous_templex}. (c) The time series $\Psi_3(t)$ of a solution is colored according to the generatex through which the solution passes.} 
	\label{fig:autonomous_generatexes}       
\end{figure*}

Figure~\ref{fig:autonomous_generatexes} presents $\Psi_3(t)$ for a solution, along with the two generatexes $\mathcal{G}_1(T_1)$ and $\mathcal{G}_2(T_1)$, shown within the cell complex $\bar{K}_1$. Segments of $\Psi_3(t)$ are colored according to the generatex through which the solution passes. These two generatexes define two distinct TMVs. The typical durations of the TMVs, in years, are  $25.4$ for TMV-1 and $16.7$ for TMV-2.

\begin{figure*}[!ht]
\centering
\begin{subfigure}[b]{0.44\textwidth}
\includegraphics[width=\textwidth]{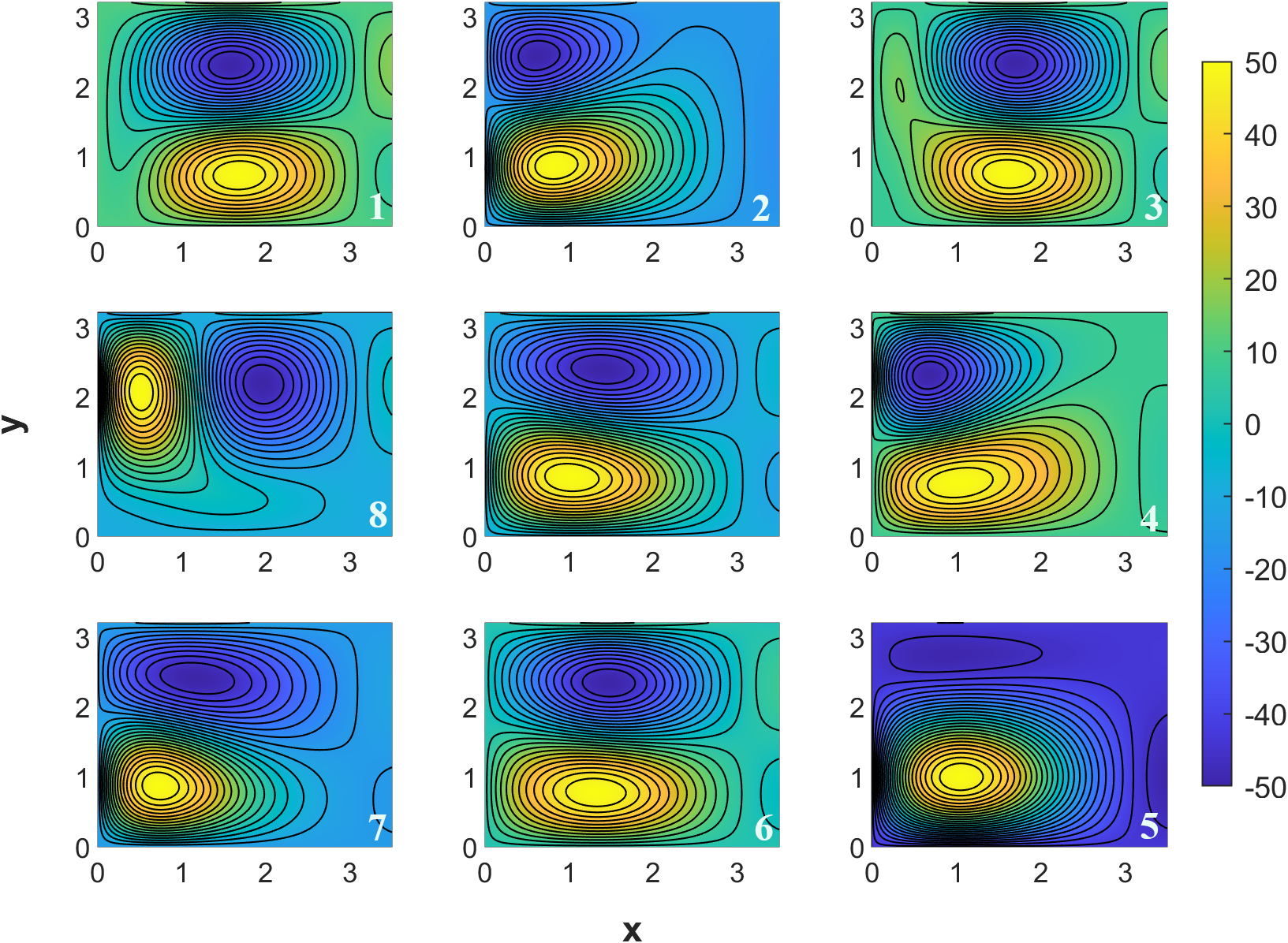}
	\caption{TMV-1}
	\end{subfigure}
~	
\begin{subfigure}[b]{0.44\textwidth}
	\includegraphics[width=\textwidth]{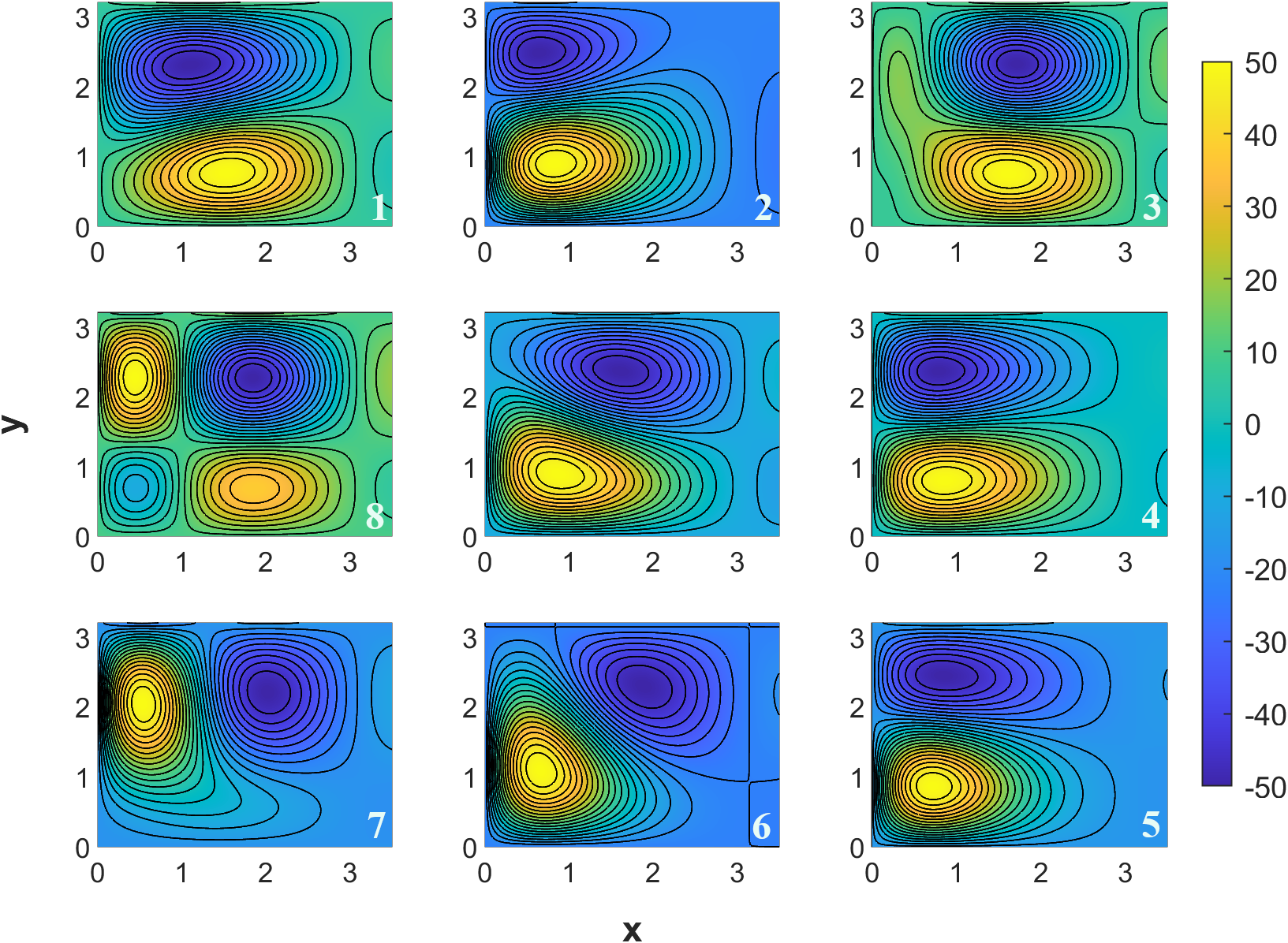}
	\caption{TMV-2}
	\end{subfigure}
\caption{TMV plots in physical space for the autonomous system: for each TMV, a sequence of eight snapshots (1–8) of $\psi = \psi(x, y, t)$ is displayed in a circular arrangement around a central subpanel, which shows the mean field $\overline{\psi(x, y)}$ associated with the corresponding TMV. TMV-$i$ in phase space corresponds to the generatex $\mathcal{G}_i(T_1)$, for $i = 1,2$. The snapshot labels (1–8) appear in white in the lower-right corner of each subpanel, over a dark background. The correspondence between the numbering of the subpanels and the cells of the generatex is as follows:(a) TMV-1: $\{1\} \to \sigma_8$, $\{2\} \to \sigma_{10}$, $\{3\} \to \gamma_{12}$, $\{4\} \to \sigma_3$, $\{5\} \to \sigma_5$, $\{6\} \to \sigma_7$, $\{7\} \to \sigma_9$, and $\{8\} \to \sigma_{11}$. (b) TMV-2: $\{1\} \to\sigma_8$, $\{2\} \to \sigma_{10}$, $\{3\} \to \sigma_{12}$, $\{4\} \to \sigma_2$, $\{5,6\} \to \sigma_4$, and $\{7,8\} \to \sigma_6$.}
  \label{fig:tmv_autonomo}
\end{figure*}

In Figure~\ref{fig:tmv_autonomo}, the TMVs are represented in physical space. Each sequence of snapshots $\{\psi (x, y, t_k) : k = 1, \ldots, 8 \}$ corresponds to a trajectory in phase space looping around one of the two generatexes or the other. The figure caption indicates the correspondence between the subpanels in this figure and the specific cells in the corresponding generatex in Fig.~\ref{fig:autonomous_generatexes}.
	
A cautionary note is in order, since similar fields may arise from trajectories that pass through a set of 2-cells that are shared by two or more generatexes. The features that distinguish two TMVs from each other stem from those portions of one TMV that fall within the cells of a given generatex but are not shared with another generatex.

\subsection{Templex analysis for the nonautonomous cases} \label{sec:nonautonomous_part} 

We now turn to the analysis of nonautonomous cases, where the system is subjected to time-dependent forcing. In this context, it is important to distinguish between periodic and aperiodic forcing, as they give rise to different features of the corresponding TMVs.

When the global attracting set of a dynamical system contains multiple attractors, as suggested by the numerical experiments of \citet{Pierini.ea.2016}, a single trajectory may not suffice to capture the full structure associated with all possible initial conditions. In our numerical experiments, however, we did not observe such multiplicity. Should multiple attractors be present, the templex can still be constructed from several long trajectories. This is still possible because time is not treated as a metric variable in our framework, and the directionality of the semi-flow is encoded in the constituent digraph of the templex. As a result, data obtained from different initial conditions can  still be integrated into a single coherent structure. 

Given the unique attractor present in the systems under study herein — whether periodically or aperiodically forced — we therefore construct the templex using a point cloud obtained from a single long trajectory after the transient has decayed. In all cases, the resulting templex structure proves robust with respect to the choice of initial condition, including in the aperiodically forced regime.

\subsubsection{Periodically forced case}
\label{subsec:periodic}

In this section, we analyze the periodically forced case, as before, in the phase space spanned by $(\Psi_1, \Psi_2, \Psi_3, \Psi_4)^{\rm T}$, 
given that the forcing produces a negligible number of false neighbors, which do not affect the construction of the cell complex. The parameter values are $\gamma = 1.10$ and $\varepsilon = 0.20$, and the forcing period is $\bar T_p = 2\pi/\omega = 30$~yr. 

\begin{figure}[ht]
  \centering
  \includegraphics[width=0.42\textwidth]{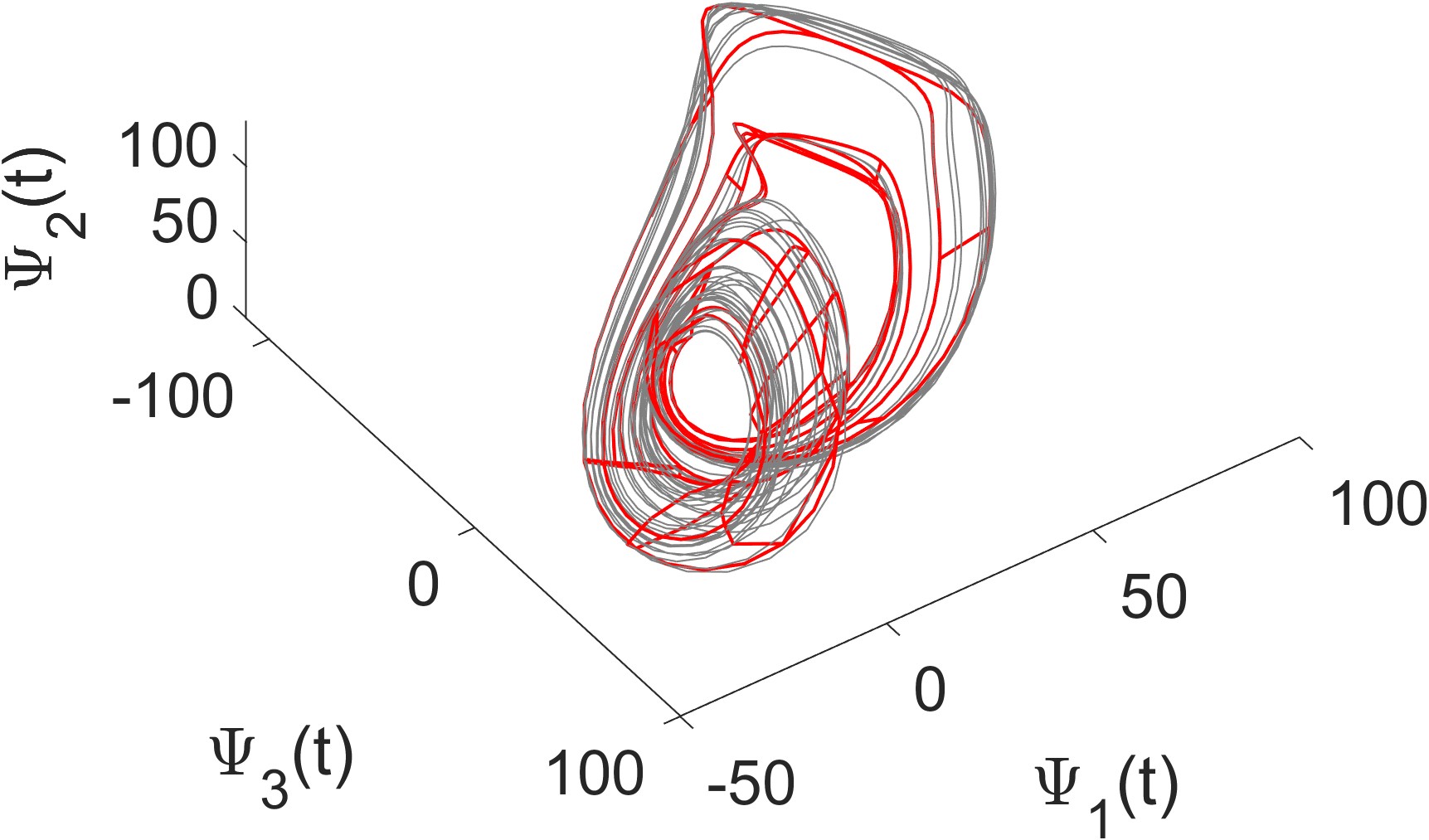} \\[-0.3cm]
  \caption{Point cloud (gray dots) juxtaposed  over the  BraMAH 
   cell complex $K_2$ (in red) in a $(\psi_1,\psi_2,\psi_3)$ projection for the periodic-forcing case with $\gamma = 1.10, \varepsilon = 0.20$ and forcing period $\bar T_p = 30$~yr.}
  \label{fig:periodic_complex}
\end{figure}

\begin{figure*}[ht]
\centering
\begin{subfigure}[b]{0.44\textwidth}
\includegraphics[width=\textwidth]{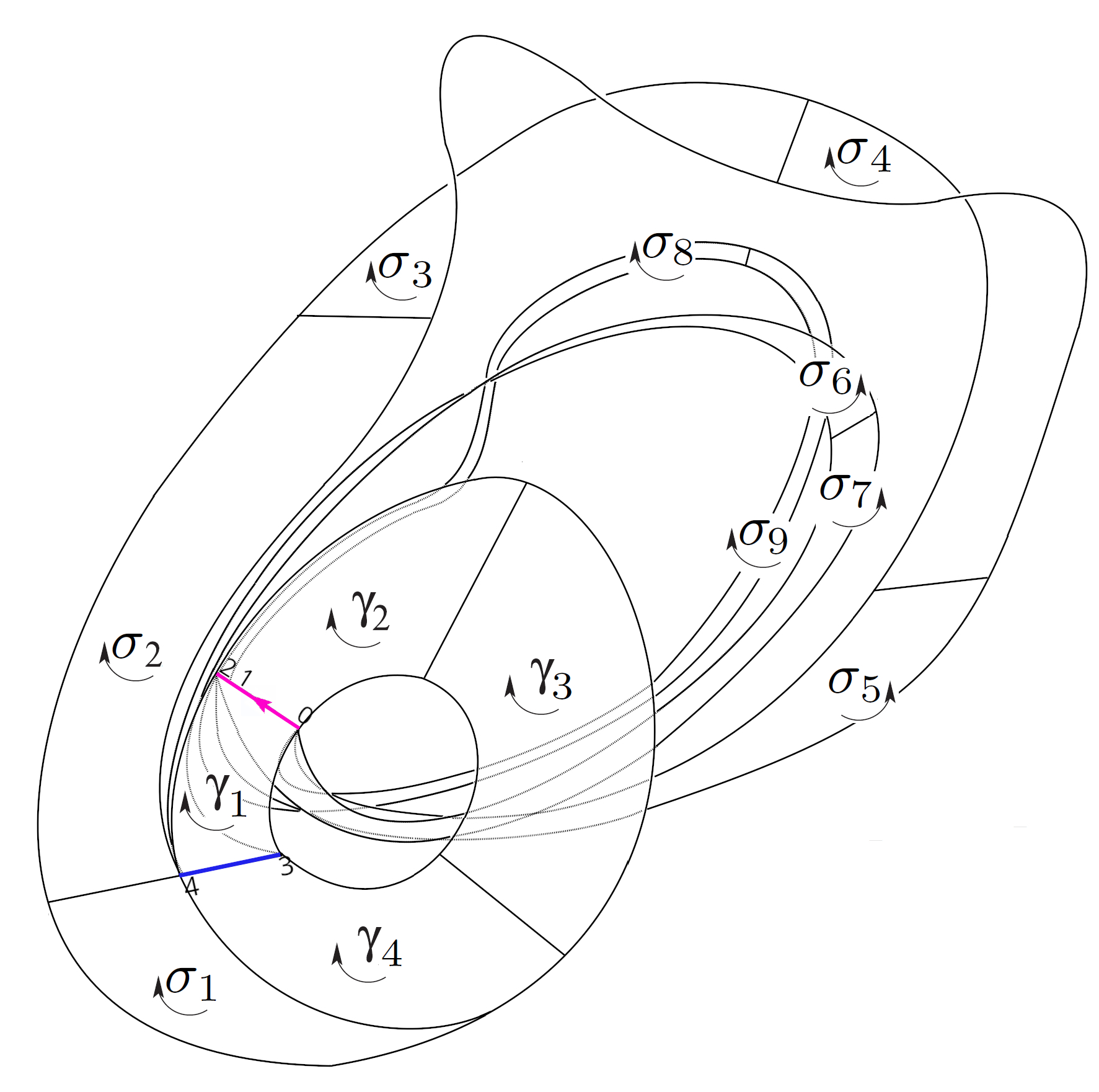}
	\caption{Simplified complex $\bar{K}_2$}
	\end{subfigure}
~
    \begin{subfigure}[b]{0.45\textwidth}
\includegraphics[width=\textwidth]{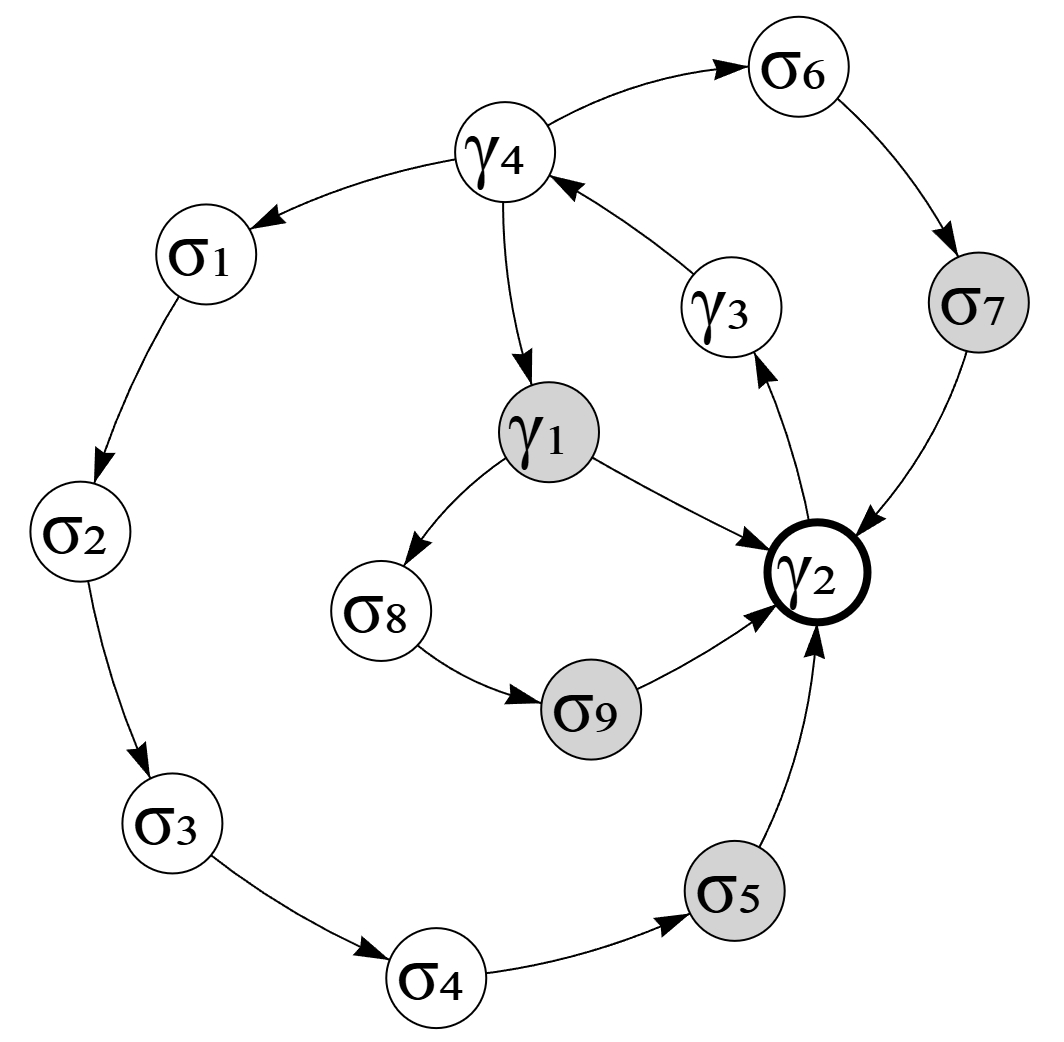}
\caption{Digraph $G_2$}
	\end{subfigure}
\caption{ Templex $T_2=( \bar{K}_2,G_2)$ for the periodic-forcing case 
		in Fig.~\ref{fig:periodic_complex}.
	(a) Diagram of the simplified cell complex $ \bar{K}_2$, where $\{\gamma_i : i= 1, \ldots, 4$\} are 3-cells and $\{\sigma_j : j = 1, \ldots, 9 \}$ are 2-cells. The heavy pink line segment indicates the joining locus at $\langle 0,1,2 \rangle$, and the blue segment corresponds to the  splitting locus at $\langle 3,4 \rangle$.
	(b) Digraph $G_2$: the ingoing cells $\gamma_1,\sigma_5,\sigma_7$ and $\sigma_{9}$ are shaded, and the outgoing cell $\gamma_2$  is heavy solid; this digraph has 4 nonequivalent cycles.}
\label{fig:periodic_templex}
\end{figure*}

The cell complex $K_2$ and its simplified form $\bar{K}_2$ are shown in Figures~\ref{fig:periodic_complex} and~\ref{fig:periodic_templex}~(a).  The presence of time-dependent forcing transforms several 2-cells from the autonomous complex ($\sigma_1$, $\sigma_2$, $\sigma_4$, $\sigma_6$, $\sigma_7$, $\sigma_9$, $\sigma_{11}$) into 3-cells ($\gamma_1$, $\gamma_2$, $\gamma_3$, $\gamma_4$) in the nonautonomous case: this leads to the emergence of a three-dimensional filled torus in the complex, defined by the chain of 3-cells $\gamma_1$, $\gamma_2$, $\gamma_3$, and $\gamma_4$. This feature has also been observed in the analysis of templex properties in an idealized model of the Atlantic Meridional Overturning Circulation~\cite{mosto2025templex}.

In the cell complex $\bar{K}_2$, two prominent lines stand out: the joining locus $\langle 0,1,2 \rangle$ and the splitting locus $\langle 4,3 \rangle$. At the splitting locus, trajectories may follow two distinct paths—either from $\gamma_4$ to $\gamma_1$, or from $\gamma_4$ to $\sigma_6$. 

In the digraph $G_2$ shown in Figure~\ref{fig:periodic_templex} (b), ingoing cells are represented as shaded nodes, while the outgoing node is marked with a thicker border. Together, the cell complex $\bar{K}_2$ and its corresponding digraph $G_2$ define the templex $T_2 = (\bar{K}_2, G_2)$. The presence of the filled torus gives rise to multiple branches emerging at the joining locus $\langle 0,1,2 \rangle$, resulting in two additional generatexes compared to the autonomous case.

\begin{figure*}[t]
\centering
\begin{subfigure}[b]{0.35\textwidth}
\includegraphics[width=\textwidth]{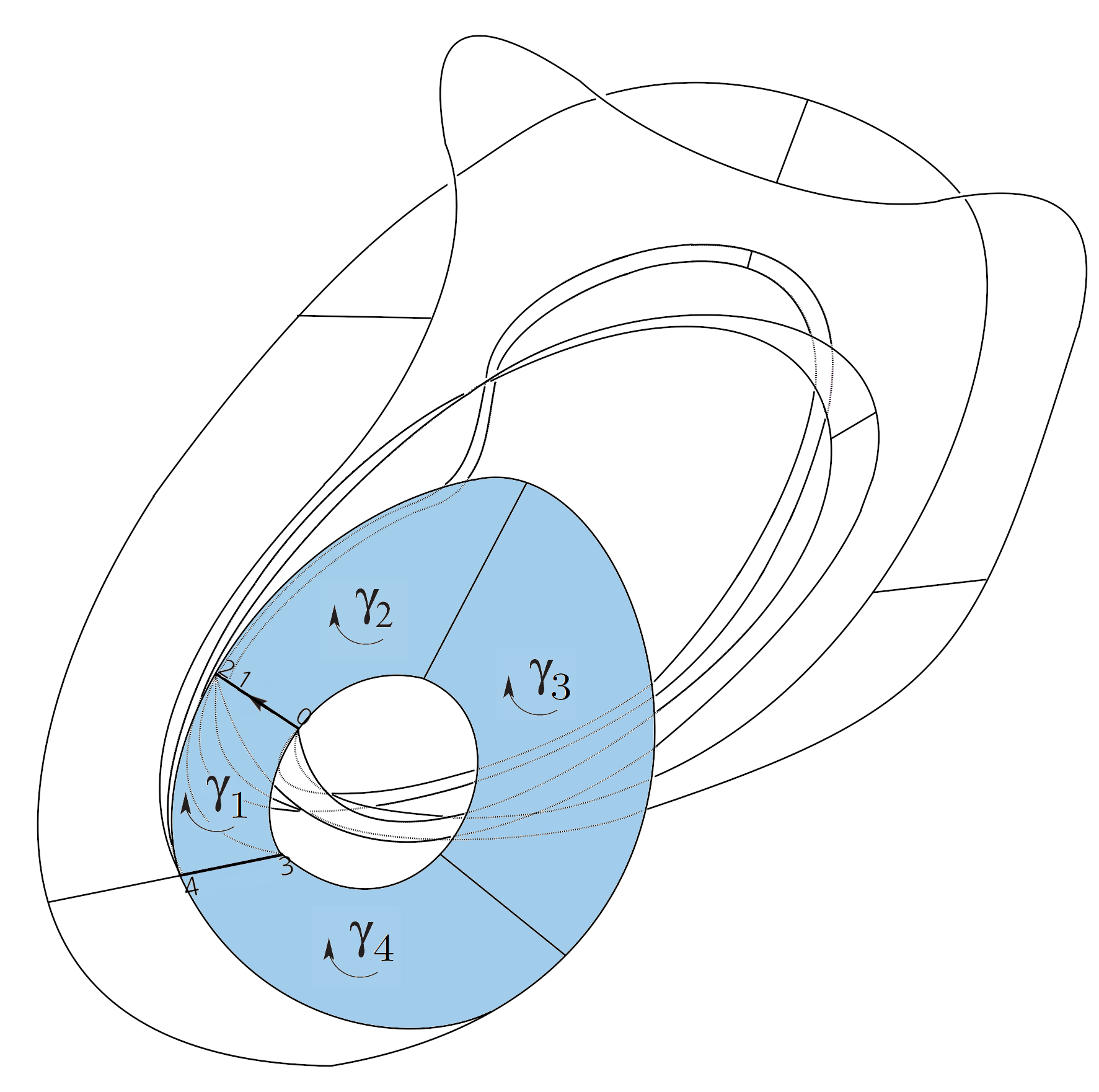}
	\caption*{$\mathcal{G}_1(T_2)$}
	\end{subfigure}
~	
\begin{subfigure}[b]{0.35\textwidth}
	\includegraphics[width=\textwidth]{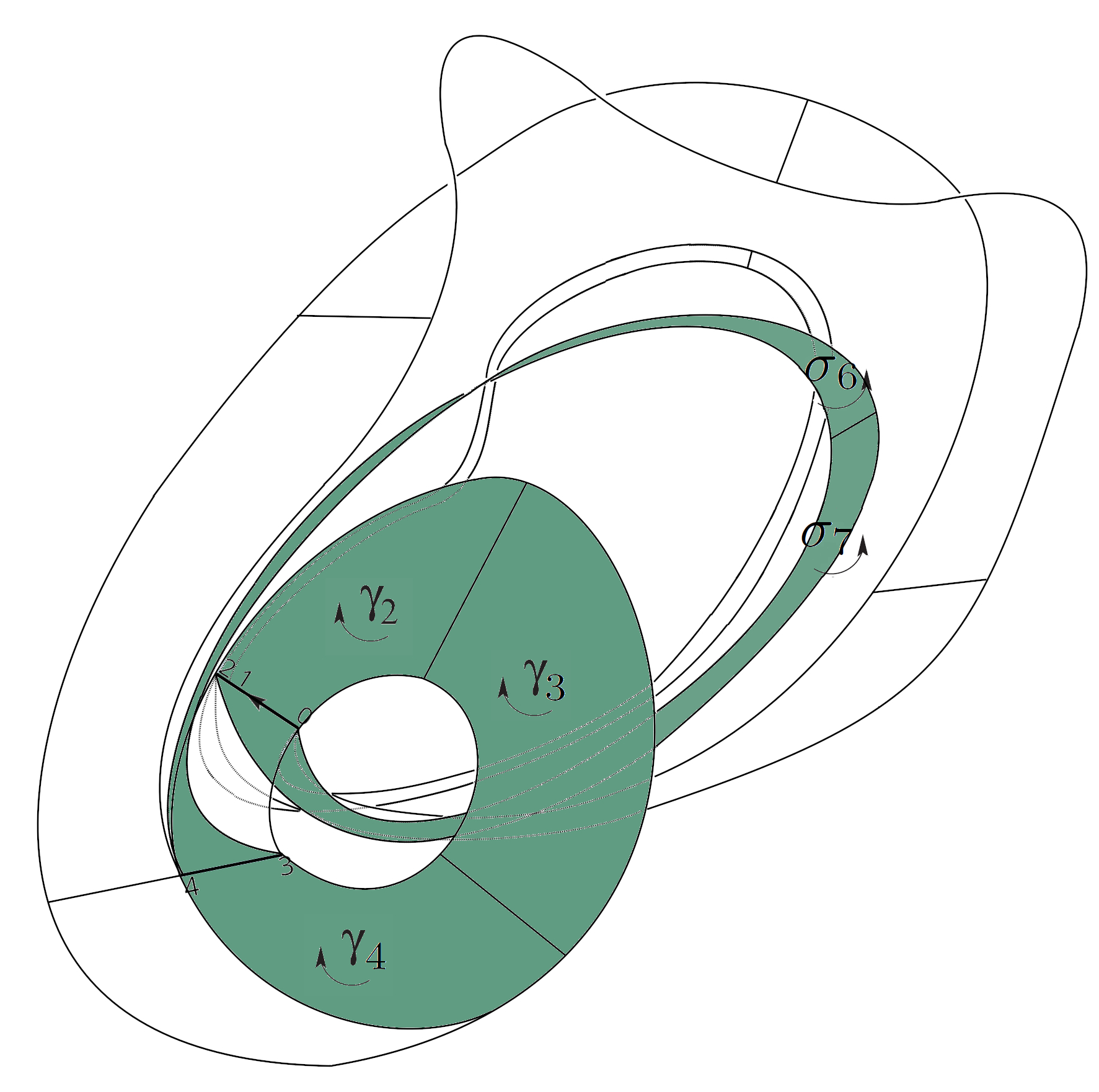}
	\caption*{$\mathcal{G}_2(T_2)$}
	\end{subfigure}

\begin{subfigure}[b]{0.8\textwidth}
	\includegraphics[width=\textwidth]{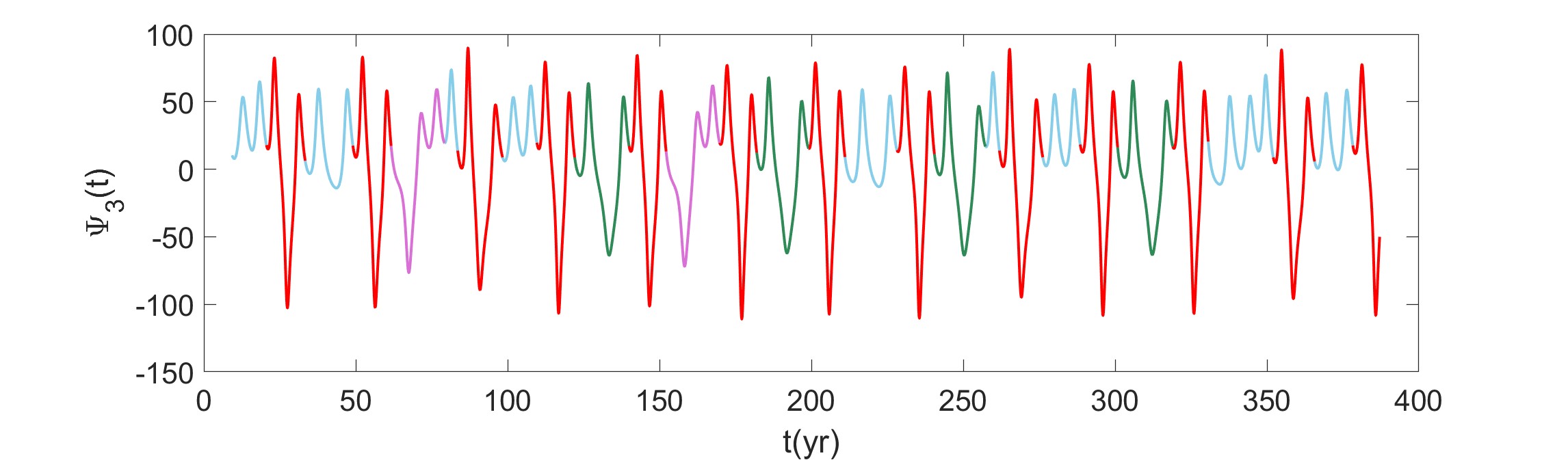}
	\caption*{}
	\end{subfigure}
\begin{subfigure}[b]{0.35\textwidth}
\includegraphics[width=\textwidth]{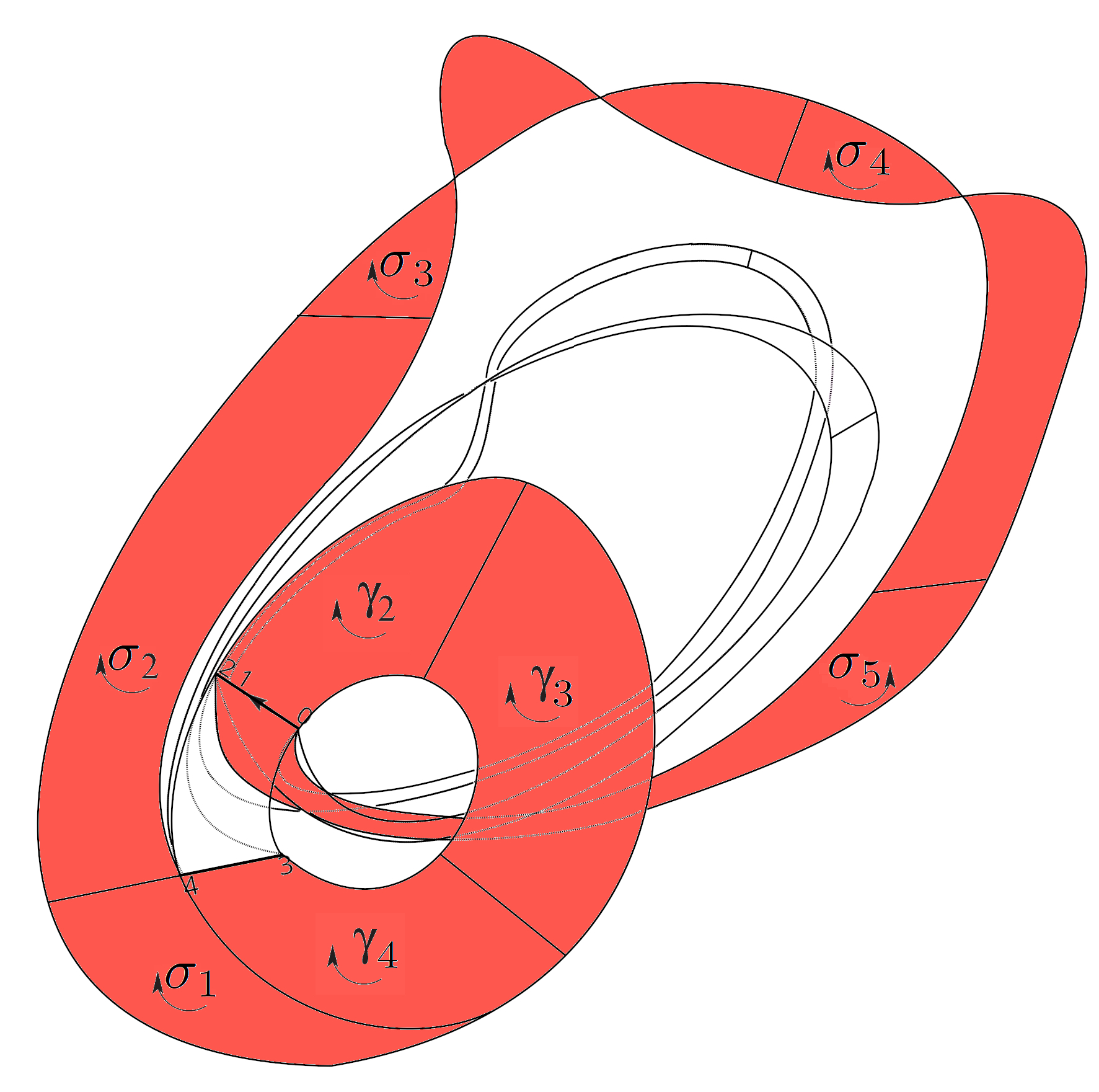}
	\caption*{$\mathcal{G}_3(T_2)$}
	\end{subfigure}
~	
\begin{subfigure}[b]{0.35\textwidth}
	\includegraphics[width=\textwidth]{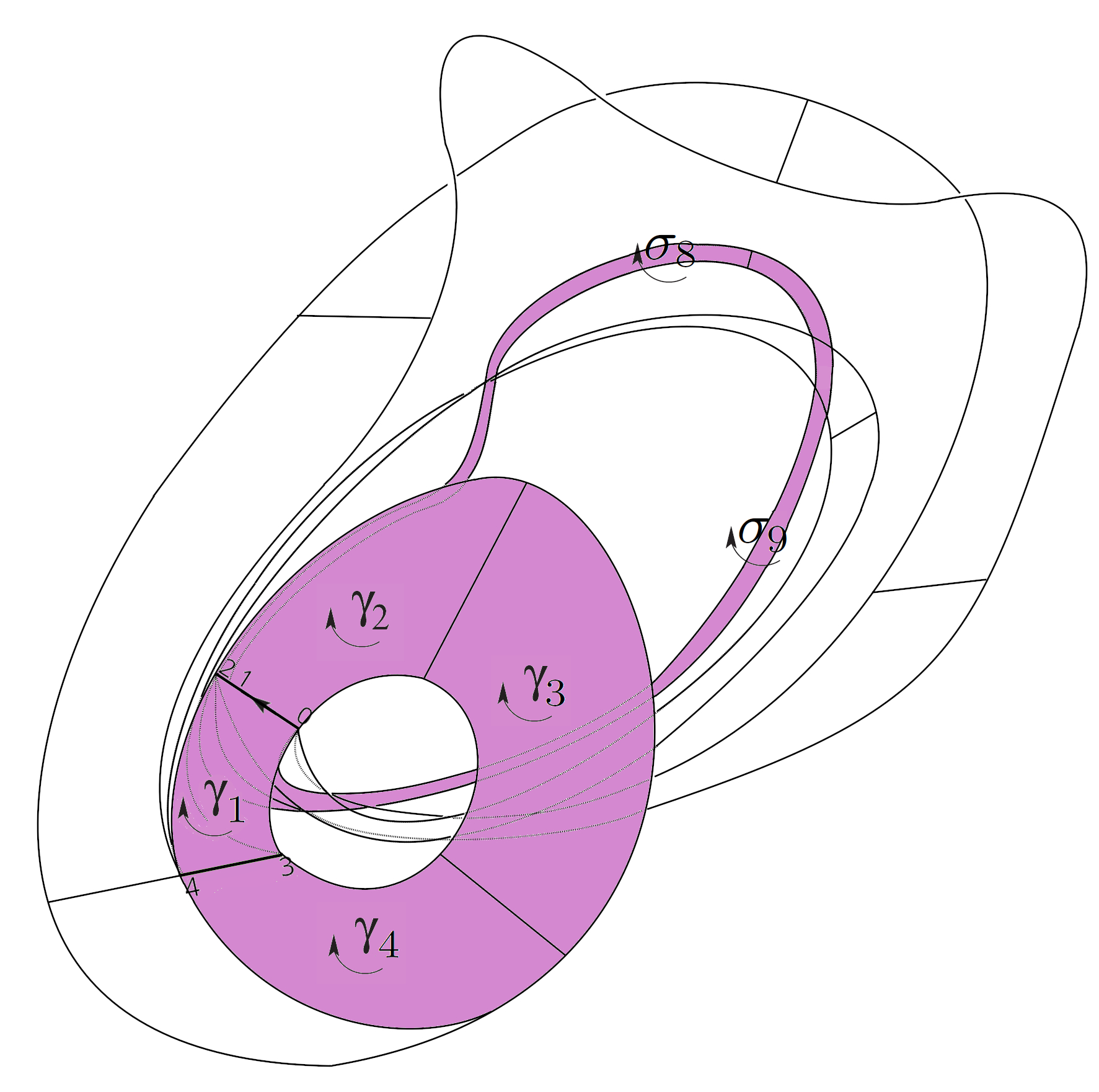}
	\caption*{$\mathcal{G}_4(T_2)$}
	\end{subfigure}     
\caption{Same as Fig.~\ref{fig:autonomous_generatexes}  for the periodic-forcing case 
		in Fig.~\ref{fig:periodic_complex}. Time series $\Psi_3(t)$ of an individual solution within the FWA and the four generatexes $\mathcal{G}_1, \mathcal{G}_2, \mathcal{G}_3,$ and $\mathcal{G}_4$ of the templex $T_2$, as seen within the cell complex $\bar{K}_2$. Segments of the time series $\Psi_3(t)$ are colored in accordance with the generatex through which the model's trajectories pass.}
\label{fig:nonautonomous_generatexes}
\end{figure*}

The digraph $G_2$ contains four nonequivalent cycles, which correspond to the new generatex set composed of four elements of order 1. Among them, $\mathcal{G}_2(T_2)$ and $\mathcal{G}_3(T_2)$ exhibit a twist;  see Fig.~\ref{fig:nonautonomous_generatexes}. The set of generatexes is listed below:
\begin{subequations}
\begin{align}
    \mathcal{G}_1(T_2) & = \{\gamma_2 \rightarrow \gamma_3 \rightarrow \gamma_4 \rightarrow \gamma_1 \rightarrow \gamma_2 \}, \\
    \mathcal{G}_2(T_2) & = \{\gamma_2 \rightarrow \gamma_{3} \rightarrow \gamma_{4} \rightarrow \sigma_6 \rightarrow \sigma_7 \rightarrow \gamma_{2} \}, \\
    \mathcal{G}_3(T_2) & = \{\gamma_2 \rightarrow \gamma_{3} \rightarrow \gamma_{4} \rightarrow \sigma_{1} \rightarrow \sigma_{2} \rightarrow \sigma_3 \notag \\
    & \hspace{1.5cm} \rightarrow \sigma_4 \rightarrow \sigma_5 \rightarrow \gamma_{2} \}, \\
    \mathcal{G}_4(T_2) & = \{\gamma_2 \rightarrow \gamma_{3} \rightarrow \gamma_{4} \rightarrow \gamma_{1} \rightarrow \sigma_{8} \rightarrow \sigma_{9} \rightarrow \gamma_2 \}.
\end{align}
\end{subequations}

Figure~\ref{fig:nonautonomous_generatexes} displays the time series $\Psi_3(t)$ of an individual solution, along with the four generatexes $\mathcal{G}_1(T_2)$, $\mathcal{G}_2(T_2)$, $\mathcal{G}_3(T_2)$, and $\mathcal{G}_4(T_2)$, represented in distinct colors within the cell complex $\bar{K}_2$. The same colors are used to identify the corresponding segments of $\Psi_3(t)$ during which the solution follows each generatex, i.e. remains in a certain TMV. Although each generatex corresponds to a single loop in the complex, the associated TMV in the time series may extend over multiple passages. The typical duration of a TMV is estimated as the time taken by the trajectory to complete one pass along its generatex. The resulting durations, in years, are 6.5 for TMV-1,  17.9 for TMV-2, 12.6 for TMV-3, and 17.6 for TMV-4.

\begin{figure*}[t]
\centering
\begin{subfigure}[b]{0.44\textwidth}
\includegraphics[width=\textwidth]{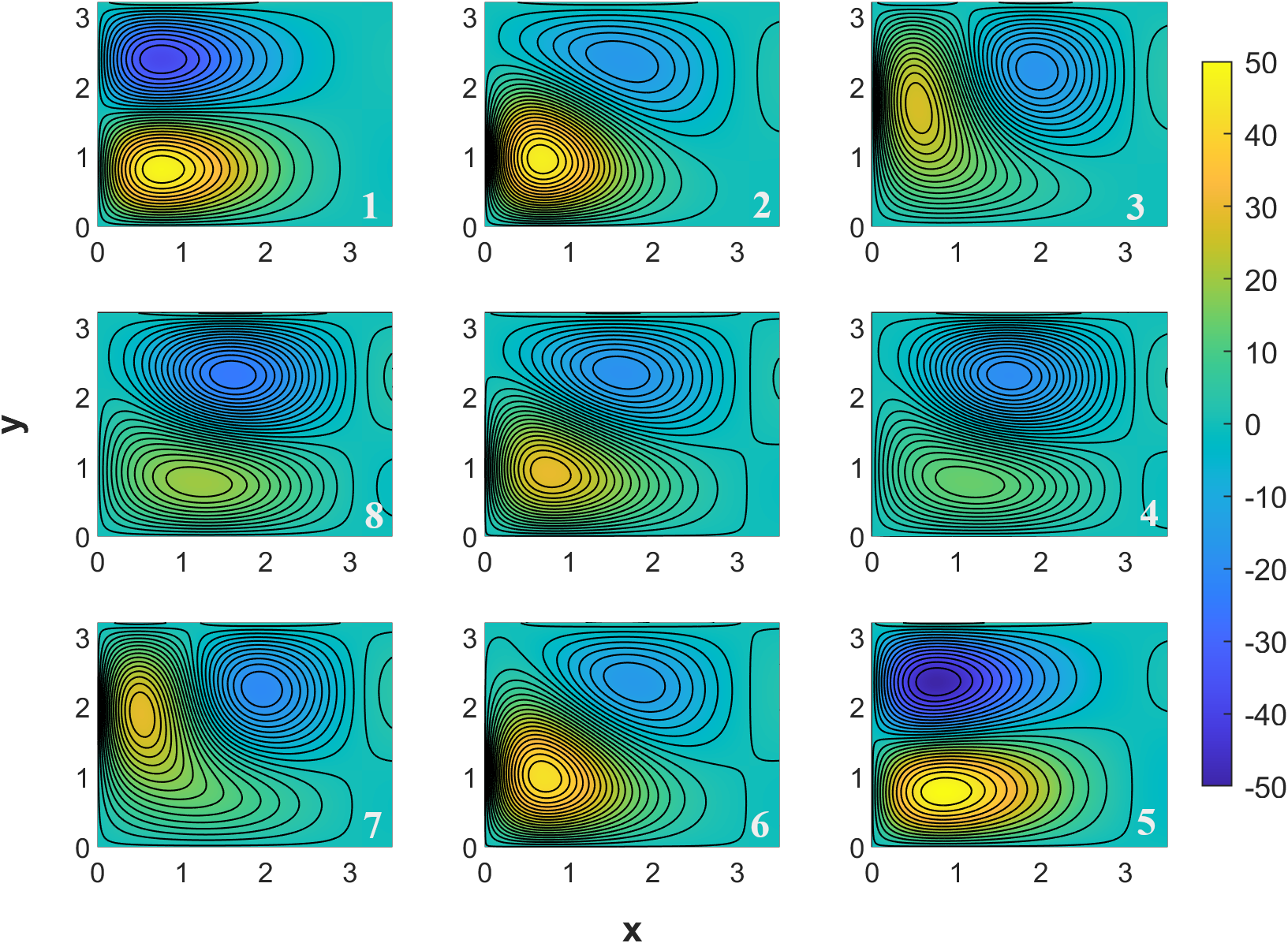}
	\caption{TMV-1}
	\end{subfigure}
~	
\begin{subfigure}[b]{0.44\textwidth}
	\includegraphics[width=\textwidth]{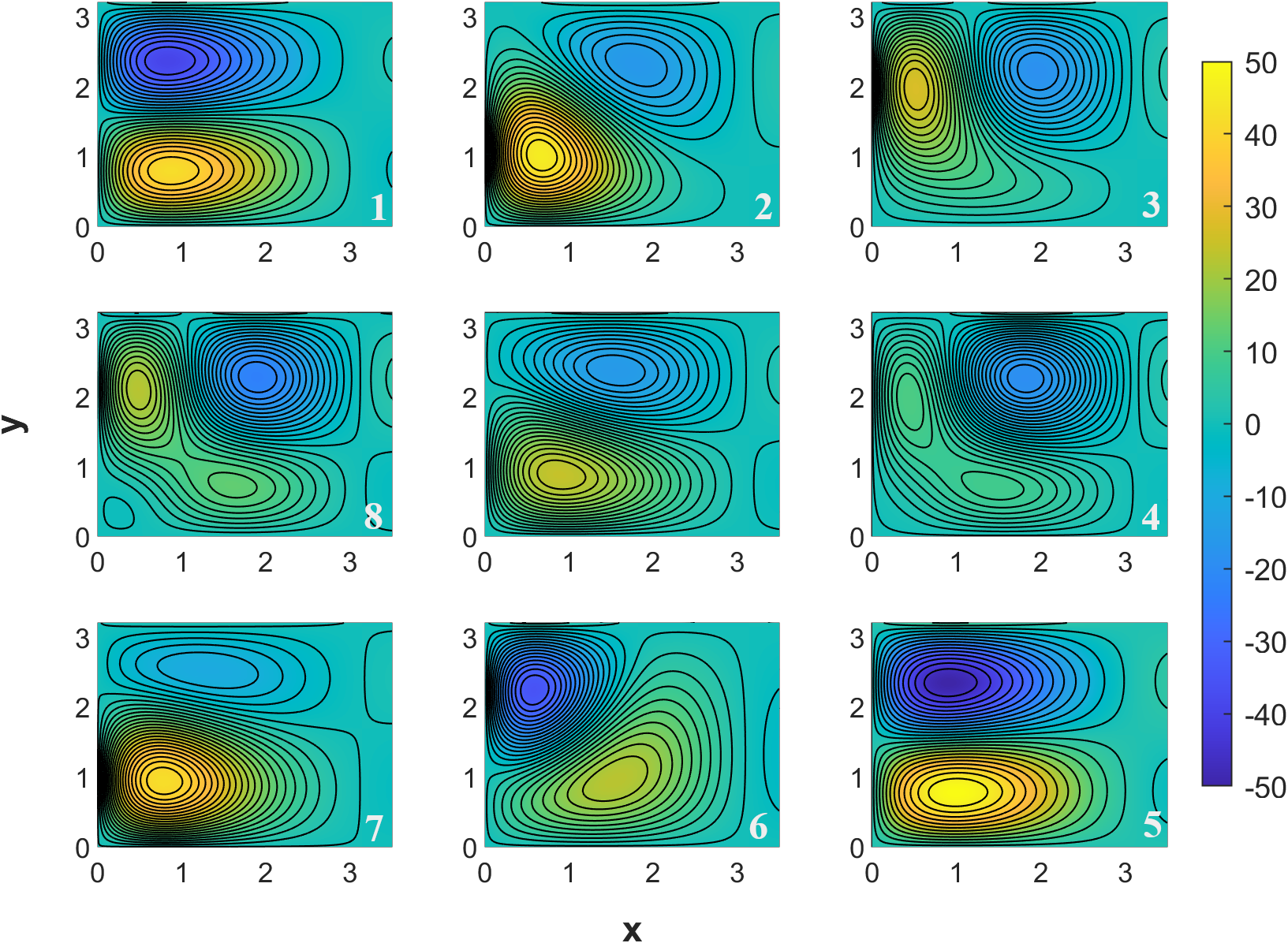}
	\caption{TMV-2}
	\end{subfigure}

\begin{subfigure}[b]{0.44\textwidth}
\includegraphics[width=\textwidth]{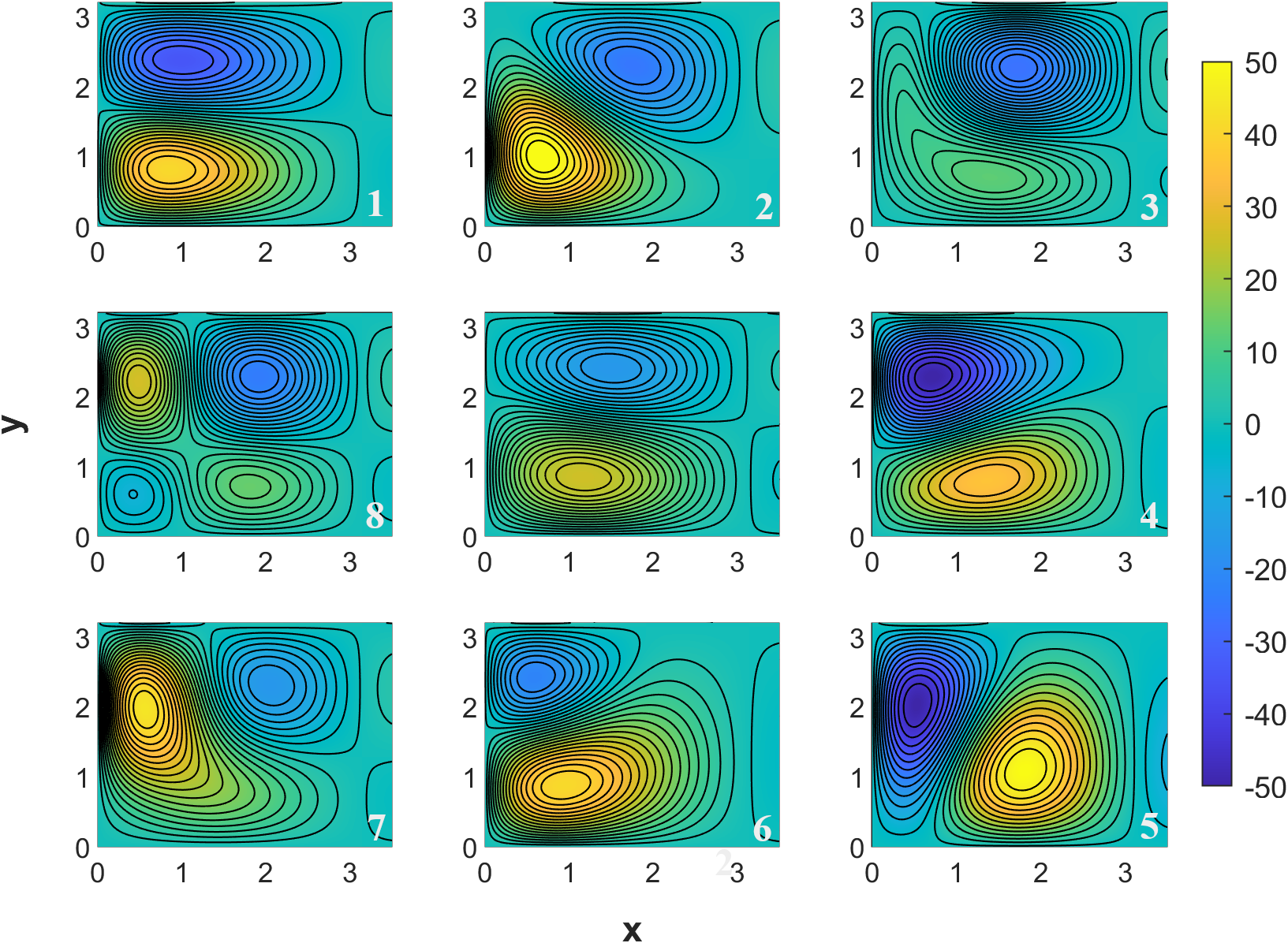}
	\caption{TMV-3}
	\end{subfigure}
~	
\begin{subfigure}[b]{0.44\textwidth}
	\includegraphics[width=\textwidth]{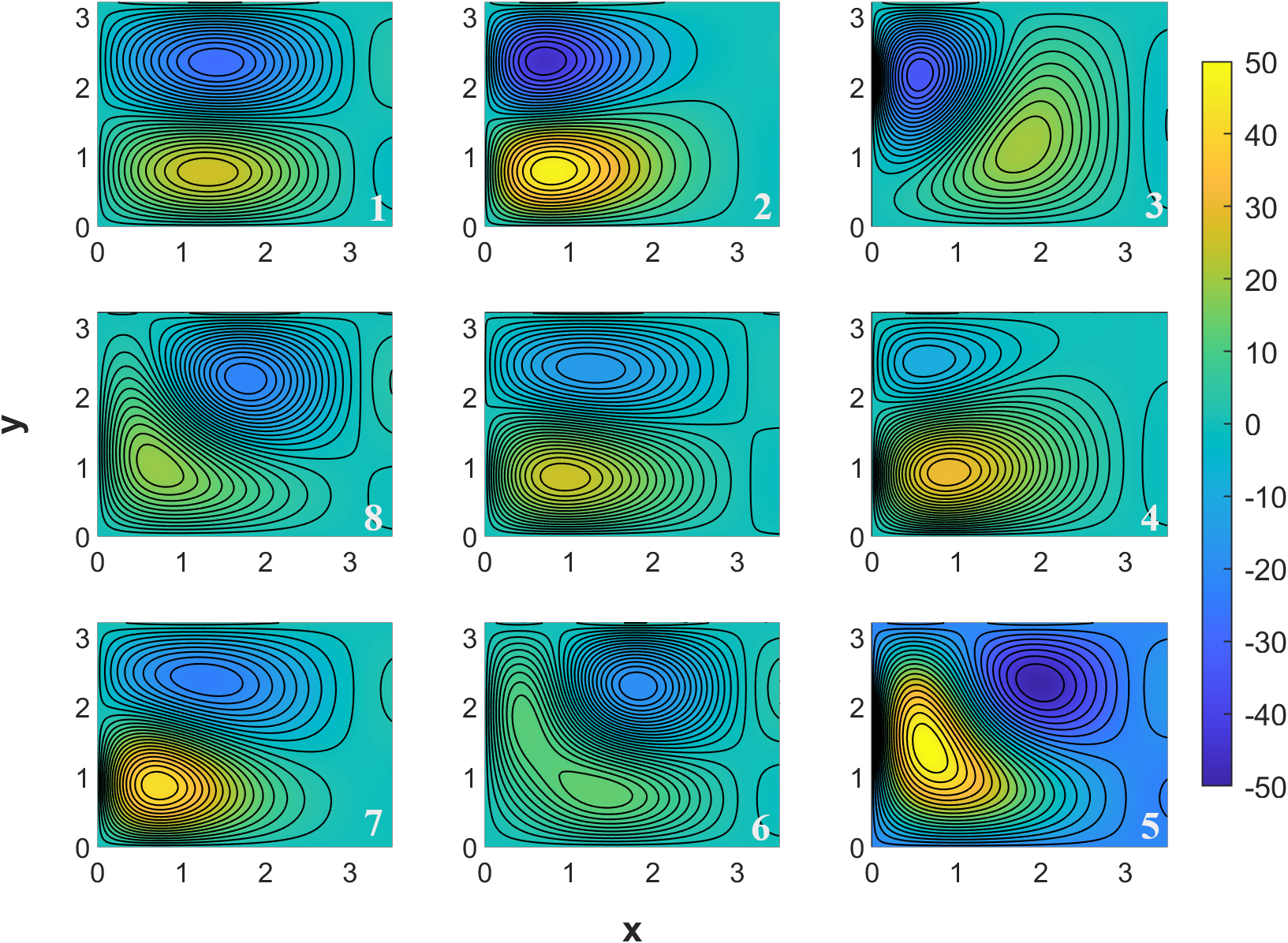}
	\caption{TMV-4}
	\end{subfigure}    
\caption{TMVs plots in physical space for the periodically forced model. Same arrangement and
		labeling of subpanels as in Fig.~\ref{fig:tmv_autonomo}. TMV-$i$ in phase space corresponds to the generatex $\mathcal{G}_i(T_2)$, for $i = 1, \ldots, 4$. The correspondence between the subpanel numbering and the cells of the generatex is as follows: (a) TMV-1: $\{1,5\} \to\gamma_2$, $\{2,6\} \to \gamma_3$, $\{3,7\} \to \gamma_4$, and $\{4,8\} \to \gamma_1$. (b) TMV-2: $\{1\} \to\gamma_2$, $\{2\} \to \gamma_3$, $\{3\} \to \gamma_4$, $\{4,5,6\} \to \sigma_6$, and $\{7,8\} \to \sigma_7$. (c) TMV-3:$\{1\} \to\gamma_2$, $\{2\} \to \gamma_3$, $\{3\} \to \sigma_1$, $\{4 \} \to \sigma_2$, $\{5\} \to \sigma_3$, $\{6\} \to \sigma_4$, and $\{7,8\} \to \sigma_5$.(d) TMV-4: $\{1,2,3\} \to\sigma_8$, $\{4,5,6\} \to \sigma_9$, $\{7\} \to \gamma_3$, and $\{8\} \to \gamma_4$.
	 }
  \label{fig:tmv_periodic}
\end{figure*}

In Fig.~\ref{fig:tmv_periodic}, the TMVs are plotted in physical space. Note that the set of 3-cells $\{\gamma_2, \gamma_3, \gamma_4\}$ is shared by all four generatexes. For instance, in panels~(a) and~(b), snapshots 1, 2, 3, and 4 of the streamfunction field $\psi(x, t)$ appear pairwise fairly similar, as they correspond to trajectories passing through these shared 3-cells in both cases. In contrast, snapshots 5 through 8 reveal marked differences between TMV-1 and TMV-2, reflecting the distinct subentities of the templex that these trajectories explore.

This analysis shows that, under periodic forcing, the TMVs reflect the topological organization encoded in the templex, and the generatex set can be regarded as complete, even though the time series used to build it is finite. The system evolves within a fixed topological scaffold, while the ordering and activation time of each TMV are variable.

As we shall see in the next section, aperiodic forcing leads to a genuinely nonautonomous regime, where the forcing remains exogenous and cannot be absorbed into an autonomous-type description. New generatexes may emerge or vanish beyond the time  interval used to build the templex rendering the topological analysis necessarily partial and dependent on both the initial and final times.

\subsubsection{Aperiodically forced case}
\label{sec:aper}

In this section, we analyze the aperiodically forced case with parameter values $\gamma = 1.10, \varepsilon = 0.20$, and $T_f = 15$; the time interval studied is the 400-year span of forcing illustrated in Fig.~\ref{fig:aperiodic_forcing}. As in the periodic case, the number of false nearest neighbors in four dimensions remains small and does not affect the templex construction.

\begin{figure}[ht]
  \centering
  \includegraphics[width=0.49\textwidth]{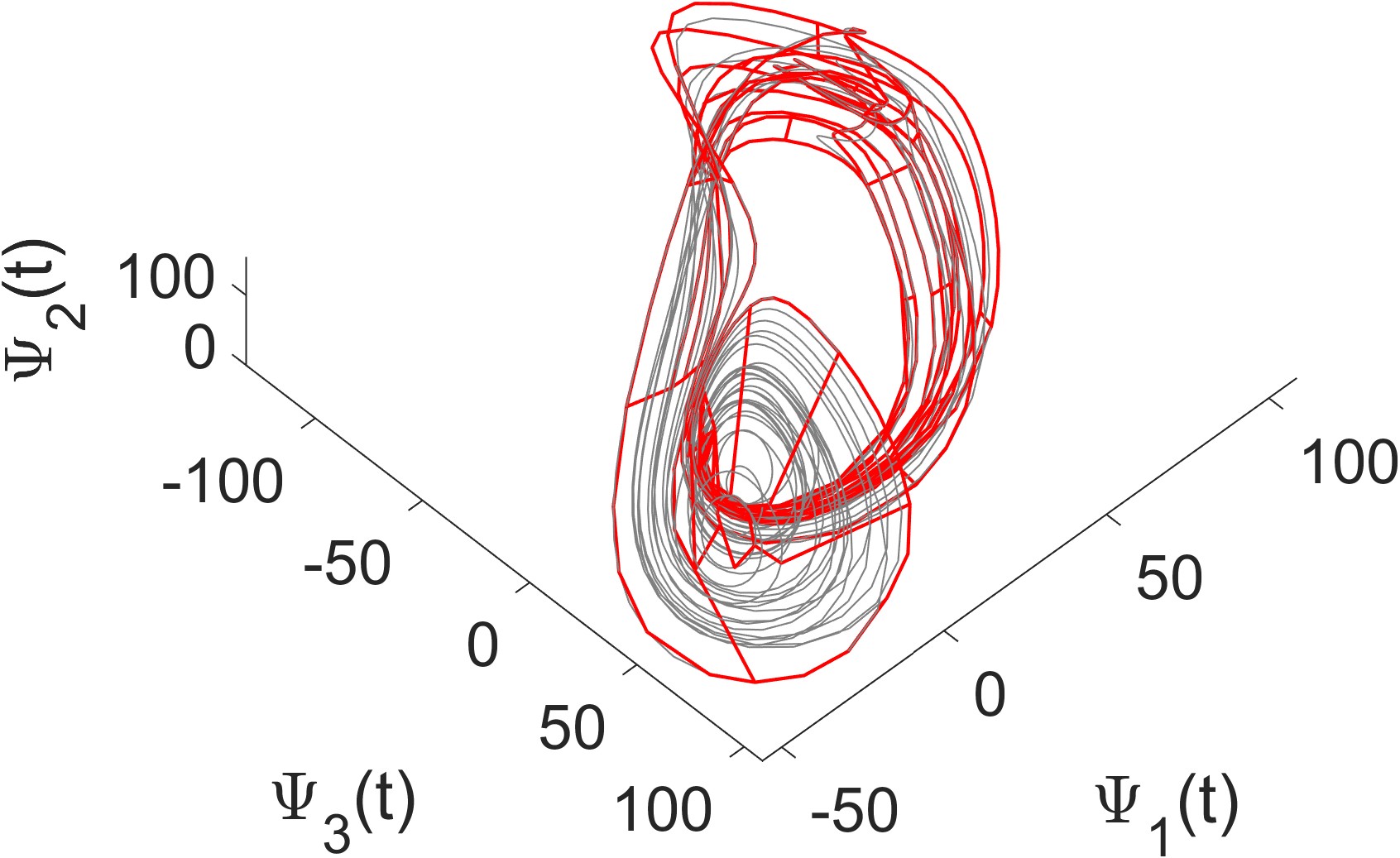}
   \caption{Plot of the point cloud (gray dots) juxtaposed over the {\sc BraMAH} cell complex $K_3$ (in red) for the aperiodic forcing illustrated in Fig.~\ref{fig:aperiodic_forcing}; the parameter values $\gamma = 1.10, \varepsilon = 0.20$ are the same as for the periodic forcing in Fig.~\ref{fig:periodic_complex} and the same projection is used.}
  \label{fig:complex_aperiodic}
\end{figure}

\begin{figure*}[ht]
\centering
\begin{subfigure}[b]{0.4\textwidth}
	\includegraphics[width=\textwidth]{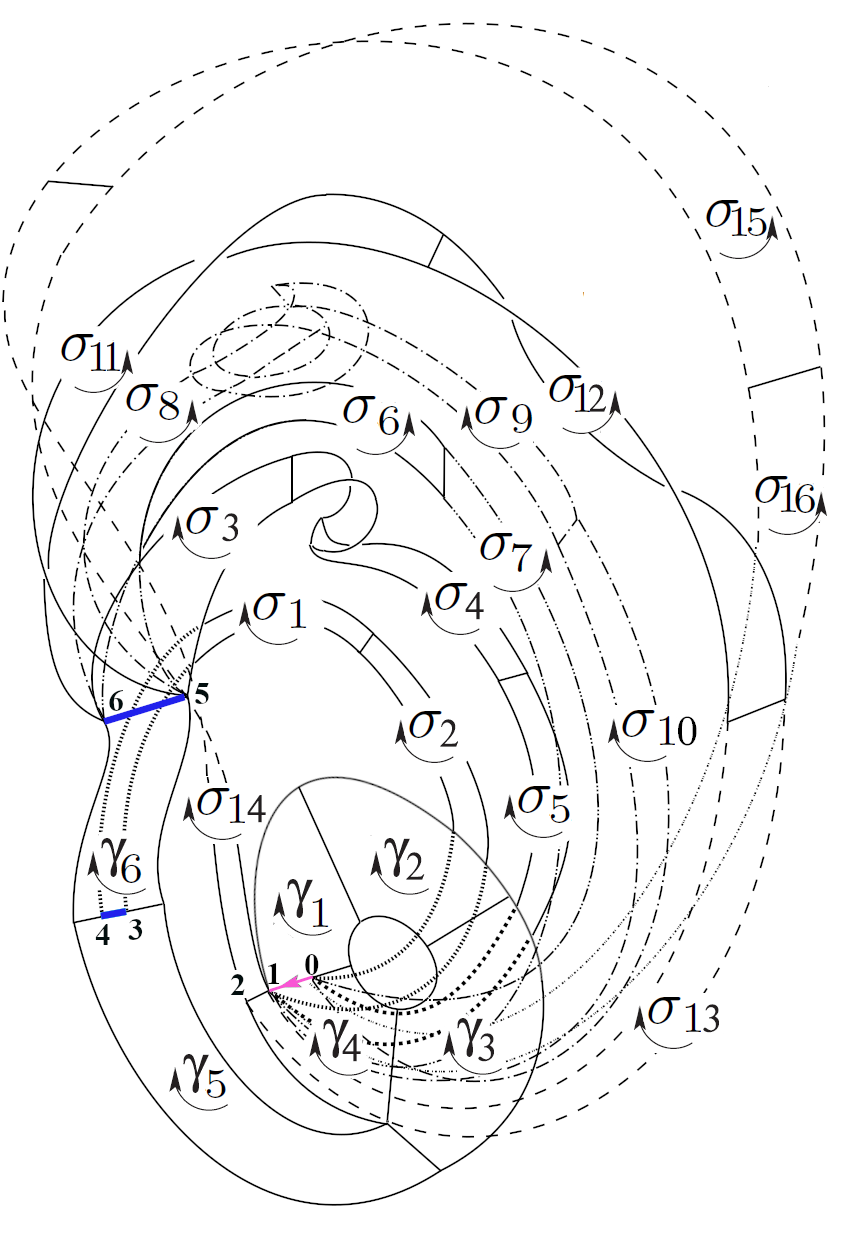}
	\caption{Simplified complex $\bar{K}_3$}
	\end{subfigure}
    ~
    \begin{subfigure}[b]{0.45\textwidth}
\includegraphics[width=\textwidth]
{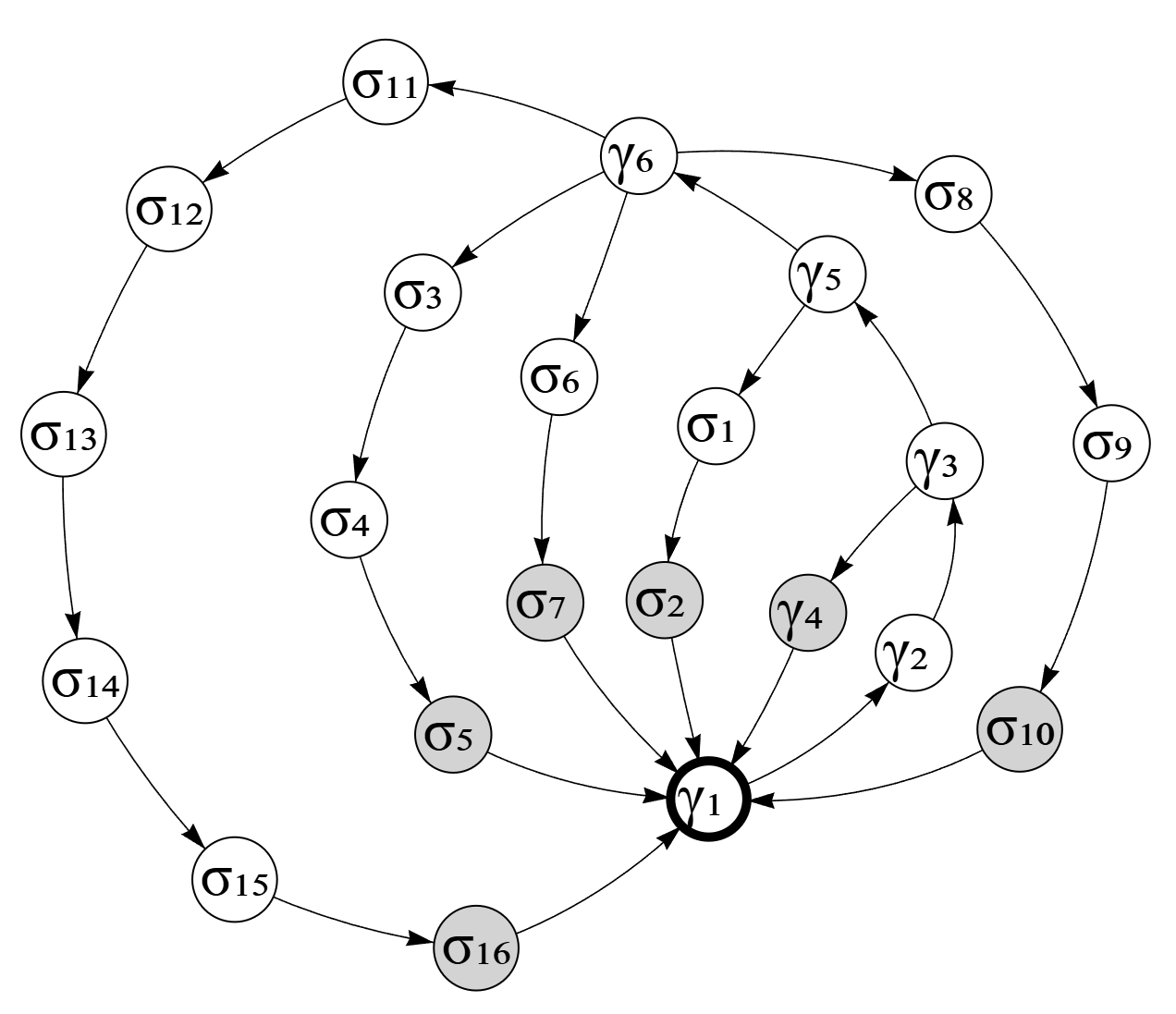}
\caption{Digraph $G_3$}
\end{subfigure} \\[-0.3cm]
\caption{Templex $T_3 = (\bar{K}_3,G_3)$ for the aperiodic forcing case. (a) Diagram of the simplified cell complex $ \bar{K}_3$, where $\{\gamma_i : i= 1, \ldots, 6\}$ are 3-cells and $\{\sigma_j : j = 1, \ldots, 16 \}$ are 2-cells. The heavy pink line segment indicates the joining locus at $\langle 0,1 \rangle$, while the two blue segments mark the splitting loci at $\langle 3,4 \rangle$ and $\langle 5,6 \rangle$, respectively. In the digraph $G_3$ the  ingoing cells $\gamma_4,\sigma_2,\sigma_5,\sigma_7, \sigma_{10}$ and $\sigma_{16}$ are shaded, and the one of the outgoing cell $\gamma_1$ is heavy solid. This digraph has 6 nonequivalent cycles.}
\label{fig:aperiodic_templex}
\end{figure*}

\begin{figure*}
\centering
\begin{subfigure}[b]{0.25\textwidth}
\includegraphics[width=\textwidth]{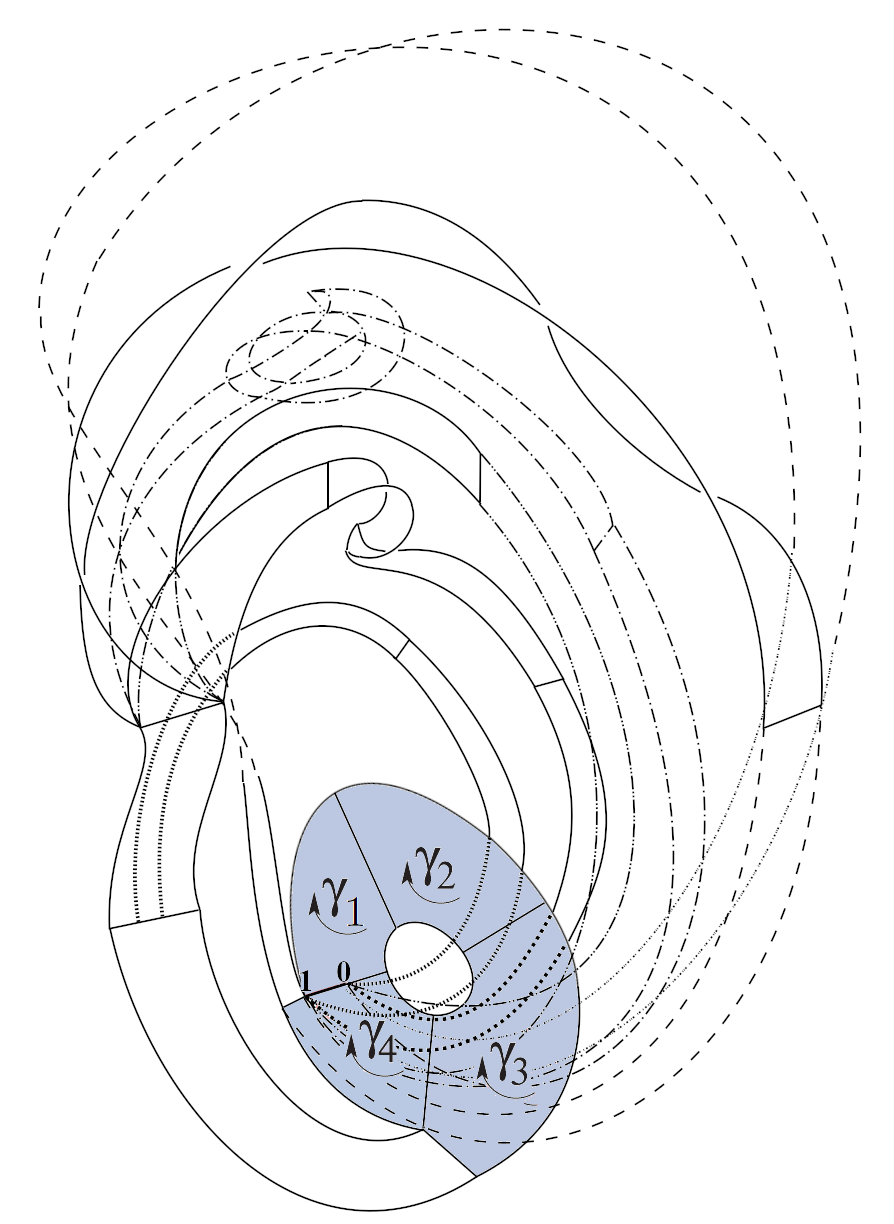}
	\caption*{$\mathcal{G}_1(T_3)$}
	\end{subfigure}
~	
\begin{subfigure}[b]{0.25\textwidth}
	\includegraphics[width=\textwidth]{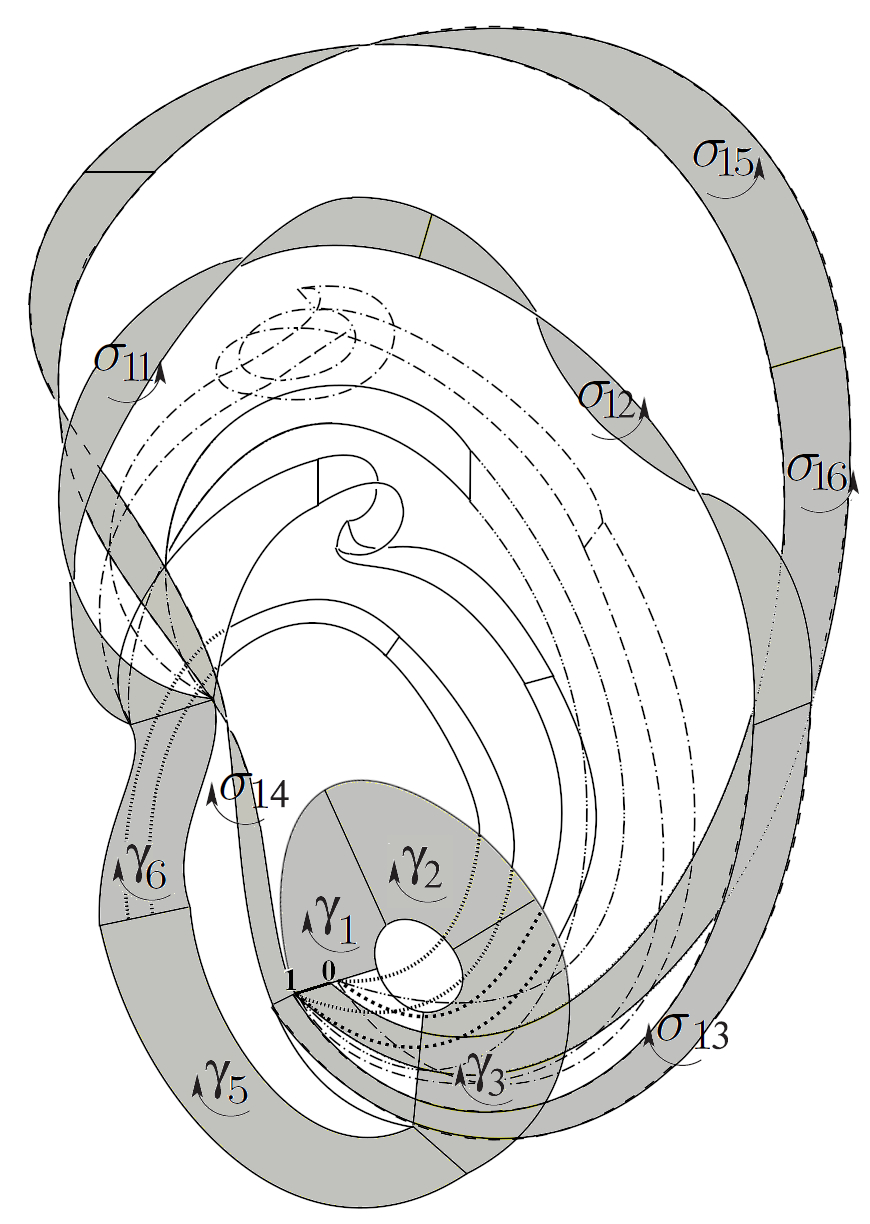}
	\caption*{$\mathcal{G}_2(T_3)$}
	\end{subfigure}
    ~	
\begin{subfigure}[b]{0.25\textwidth}
	\includegraphics[width=\textwidth]{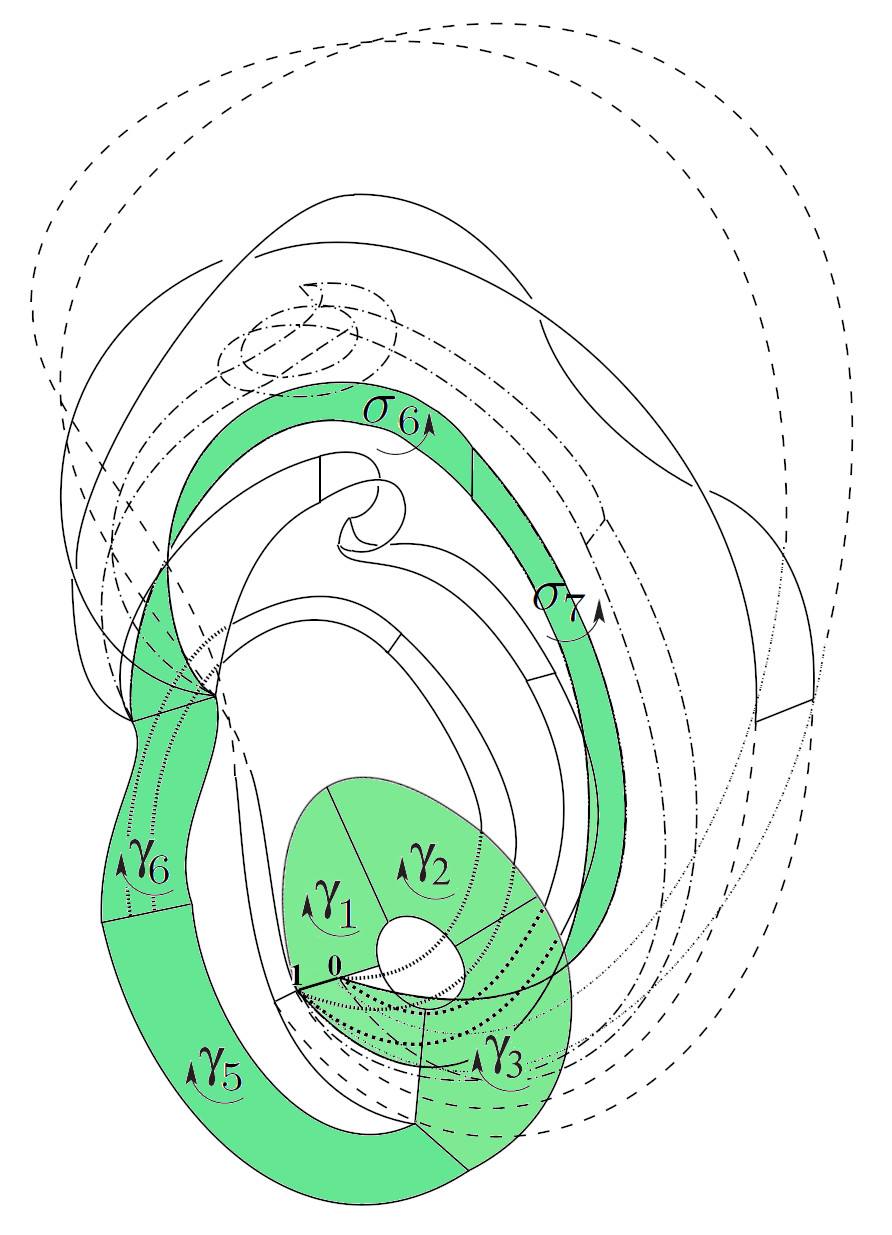}
	\caption*{$\mathcal{G}_3(T_3)$}
	\end{subfigure}

\begin{subfigure}[b]{0.9\textwidth}
	\includegraphics[width=\textwidth]{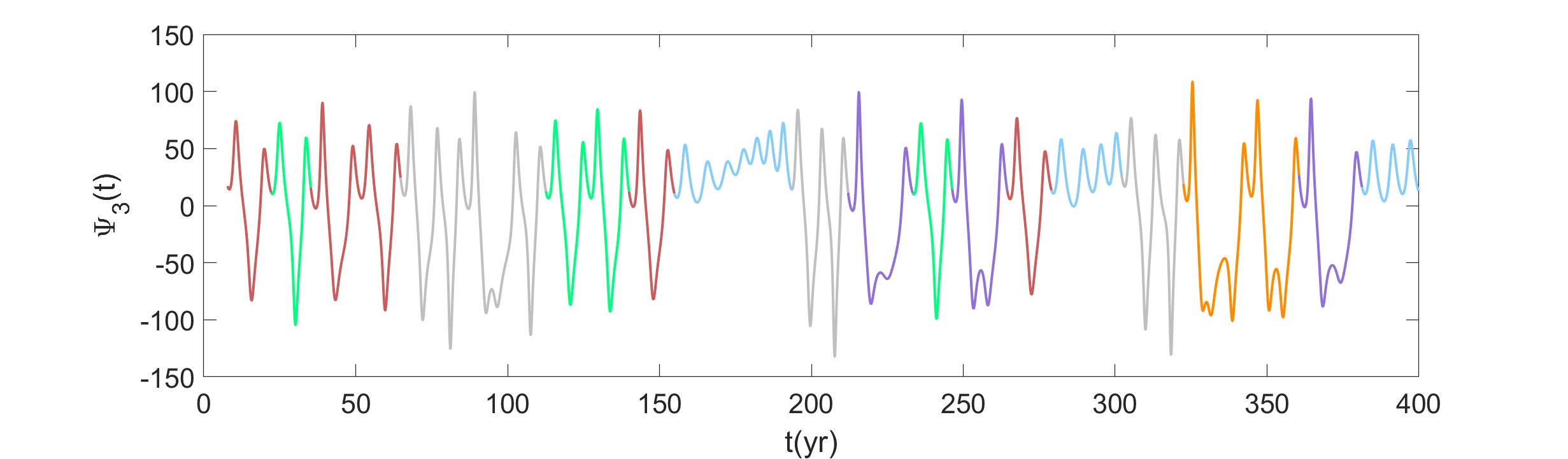}
	\caption*{}
	\end{subfigure}
    
\begin{subfigure}[b]{0.25\textwidth}
\includegraphics[width=\textwidth]{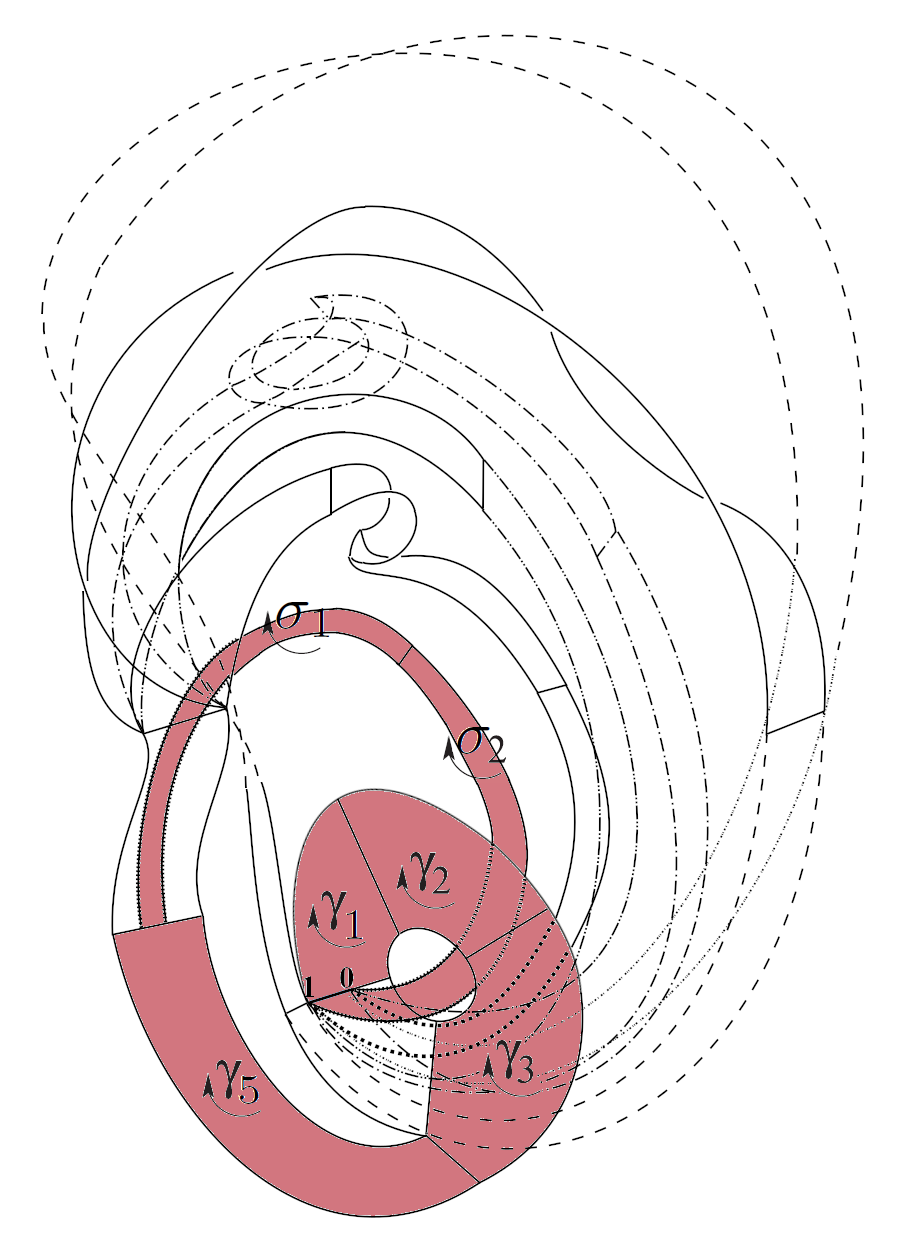}
	\caption*{$\mathcal{G}_4(T_3)$}
	\end{subfigure}
    ~
\begin{subfigure}[b]{0.25\textwidth}
\includegraphics[width=\textwidth]{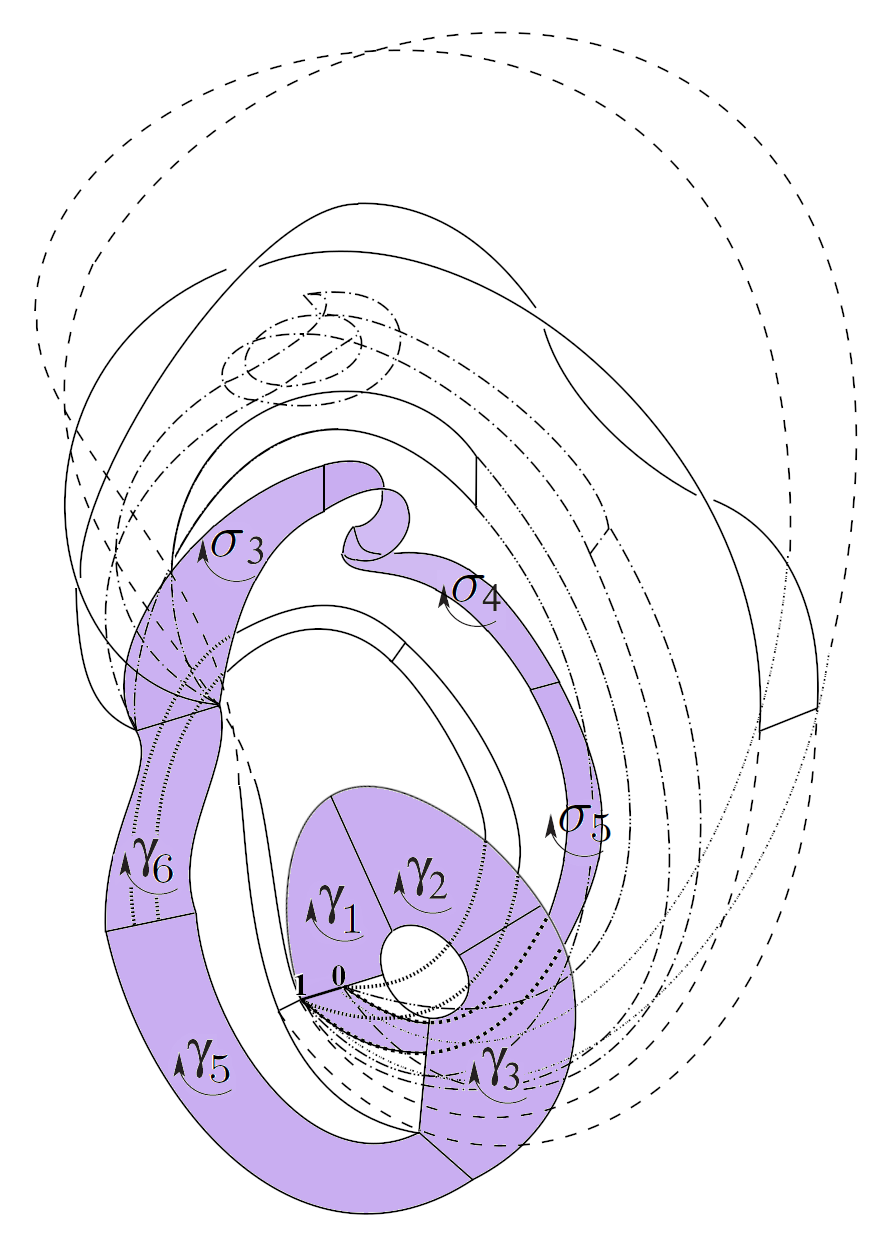}
	\caption*{$\mathcal{G}_5(T_3)$}
	\end{subfigure}
~	
\begin{subfigure}[b]{0.25\textwidth}
	\includegraphics[width=\textwidth]{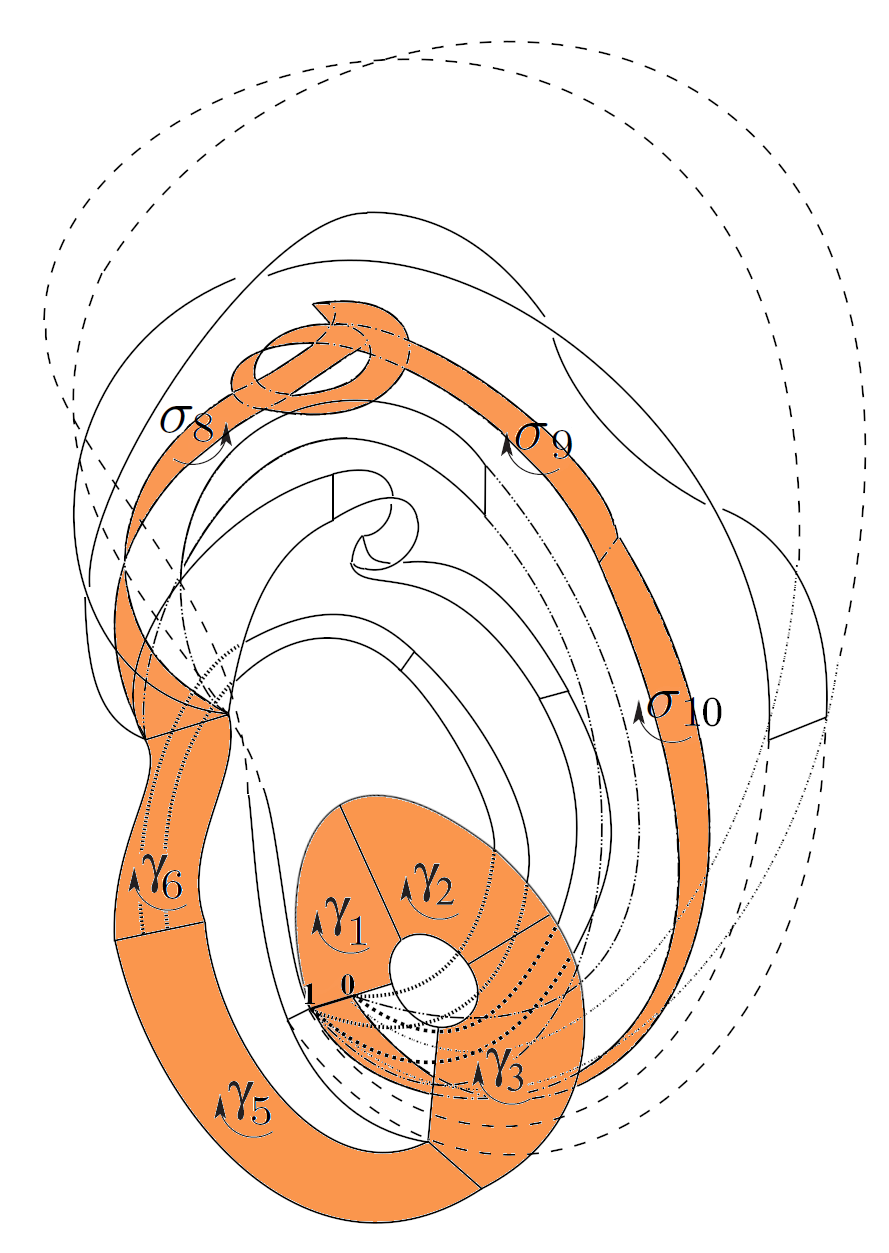}
	\caption*{$\mathcal{G}_6(T_3)$}
	\end{subfigure}    	 
\caption{Time series $\Psi_3(t)$ of an individual solution within the FWA, and the six generatexes $\{\mathcal{G}_i : i=1 \cdots 6\}$ in the aperiodic-forcing templex $T_3$, as seen within the cell complex $\bar{K}_3$. Segments of the time series $\Psi_3(t)$ are colored in accordance with the generatex through which the solution passes.}
\label{fig:nonautonomous_generatexes_aperiodic}
\end{figure*}

The {\sc BraMAH} cell complex $K_3$, illustrated in Figure~\ref{fig:complex_aperiodic}, is simplified into $\bar{K}_3$ by merging a subset of its cells. The components of the templex $T_3 = (\bar{K}_3, G_3)$ are shown in Fig.~\ref{fig:aperiodic_templex}. The simplified cell complex $\bar{K}_3$ consists of six 3-cells, $\{\gamma_i : i = 1, \ldots, 6\}$, and sixteen 2-cells, $\{\sigma_j : j = 1, \ldots, 16 \}$.
In the digraph $G_3$, the ingoing cells—$\gamma_4$, $\sigma_2$, $\sigma_5$, $\sigma_7$, $\sigma_{10}$, and $\sigma_{16}$— are represented by shaded nodes, while the outgoing cell $\gamma_1$ has a heavy solid contour.

The pink line in $\bar{K}_3$ represents the joining locus $\langle 0,1 \rangle$. In the 1-cell $\langle 0,1,2 \rangle$, the 0-cell $\langle 1 \rangle$ is a splitting 0-cell. The flow emanating from $\gamma_4$ bifurcates into two branches: $\gamma_1$ and $\sigma_{14}$.

There are two additional splitting loci (in blue): $\langle 3,4 \rangle$ and $\langle 5,6 \rangle$. At the first, trajectories can follow two distinct paths—one from $\gamma_5$ to $\gamma_6$, the other one from $\gamma_5$ to $\sigma_1$. At the second splitting locus, the trajectories can follow three distinct paths: from $\gamma_6$ to $\sigma_3$, to $\sigma_8$, or to $\sigma_{11}$.

Compared to the periodic case, a three-dimensional branch—$(\gamma_5, \gamma_6)$—emerges from the filled torus, giving rise to additional branches. In the digraph $G_3$, there are six non-equivalent cycles, which define a generatex set composed of six elements of order 1:
\begin{subequations}
\begin{align}
    \mathcal{G}_1(T_3) & = \{\gamma_1 \rightarrow \gamma_2 \rightarrow \gamma_3 \rightarrow \gamma_4 \rightarrow \gamma_1 \}, \\
    \mathcal{G}_2(T_3) & = \{\gamma_1 \rightarrow \gamma_{2} \rightarrow \gamma_{3} \rightarrow \gamma_5 \rightarrow \gamma_6 \rightarrow \sigma_{11}  \notag \\
    & \hspace{1cm}\rightarrow \sigma_{12} \rightarrow \sigma_{13} \rightarrow \sigma_{14} \rightarrow \sigma_{15}
     \notag \\
    & \hspace{1cm}\rightarrow \sigma_{16} \rightarrow \gamma_{1}\}, \\
    \mathcal{G}_3(T_3) & = \{\gamma_1 \rightarrow \gamma_{2} \rightarrow \gamma_{3} \rightarrow \gamma_{5} \rightarrow \gamma_{6} \rightarrow \sigma_6 \notag \\
    & \hspace{1cm} \rightarrow \sigma_7 \rightarrow \gamma_{1} \}, \\
    \mathcal{G}_4(T_3) & = \{\gamma_1 \rightarrow \gamma_{2} \rightarrow \gamma_{3} \rightarrow \gamma_{5} \rightarrow \gamma_{6} \rightarrow \sigma_1 \notag \\
    & \hspace{1cm} \rightarrow \sigma_2 \rightarrow \gamma_{1} \}, \\
    \mathcal{G}_5(T_3) & = \{\gamma_1 \rightarrow \gamma_{2} \rightarrow \gamma_{3} \rightarrow \gamma_5 \rightarrow \gamma_{6} \rightarrow \sigma_3 \notag \\
    & \hspace{1cm} \rightarrow \sigma_4 \rightarrow \sigma_5 \rightarrow \gamma_{1} \}, \\
    \mathcal{G}_6(T_3) & = \{\gamma_1 \rightarrow \gamma_{2} \rightarrow \gamma_{3} \rightarrow \gamma_{5}
    \rightarrow \gamma_{6}
    \rightarrow \sigma_{8}
    \notag \\
    & \hspace{1cm}\rightarrow \sigma_{9}
    \rightarrow \sigma_{10}
    \rightarrow \gamma_1 \}.
\end{align}
\end{subequations}
The generatexes $\mathcal{G}_2(T_3)$ and $\mathcal{G}_6(T_3)$ exhibit a torsion. 
In this case, the set of 3-cells shared by all six generatexes is $\{\gamma_1, \gamma_2, \gamma_3\}$. These common elements play a crucial role in \emph{plaiting} the generatexes together, and thus ensuring continuity and structural coherence within the templex.

Figure~\ref{fig:nonautonomous_generatexes_aperiodic} shows the $\Psi_3(t)$-component of a solution, along with the six generatexes $\{\mathcal{G}_i(T_3) : i=1,\dots,6 \}$; the latter are depicted in distinct colors within the cell complex $\bar{K}_3$. These colors are used consistently to indicate the corresponding segments of $\Psi_3(t),$ according to the generatex traversed by the solution. The six generatexes define six TMVs. Their estimated durations, in years, are:  6.5 for TMV-1, 27.4 for TMV-2, 13.0 for TMV-3, 14.9 for TMV-4, 21.3 for TMV-5, and 21.5 for TMV-6. 

\begin{figure*}
\centering
\begin{subfigure}[b]{0.4\textwidth}
\includegraphics[width=\textwidth]{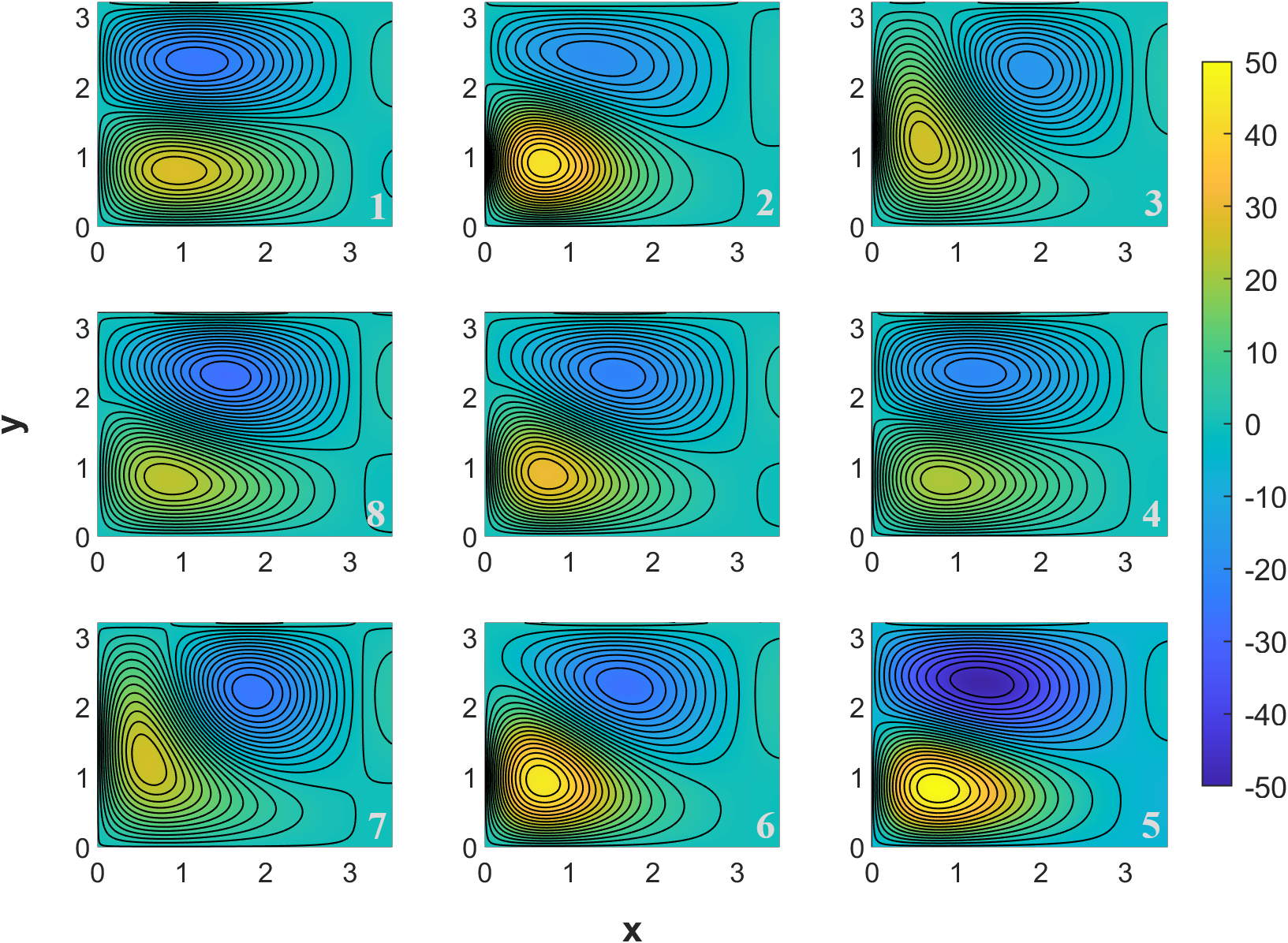}
	\caption{TMV-1}
	\end{subfigure}
~	
\begin{subfigure}[b]{0.4\textwidth}
	\includegraphics[width=\textwidth]{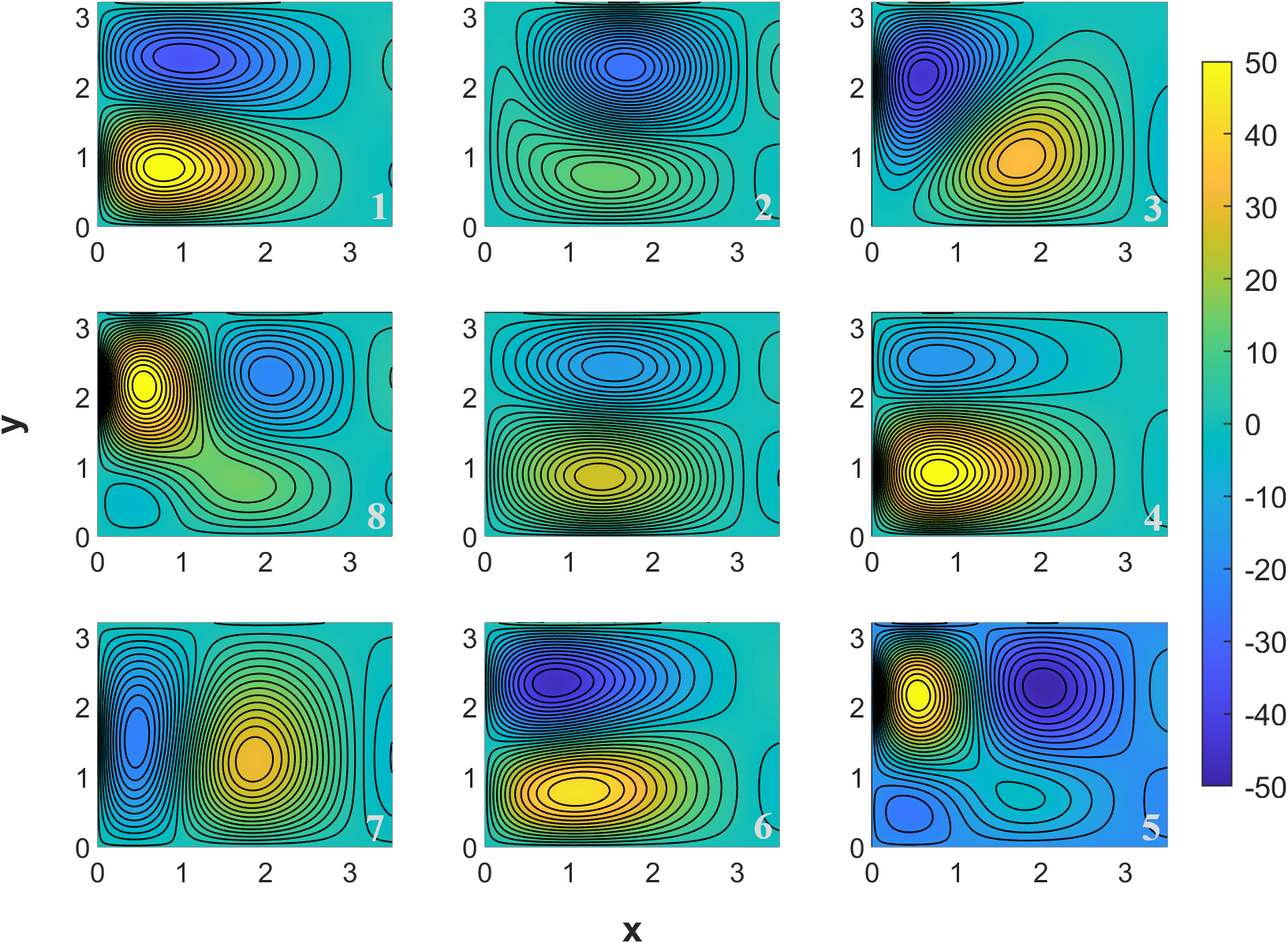}
	\caption{TMV-2}
	\end{subfigure}

\begin{subfigure}[b]{0.4\textwidth}
\includegraphics[width=\textwidth]{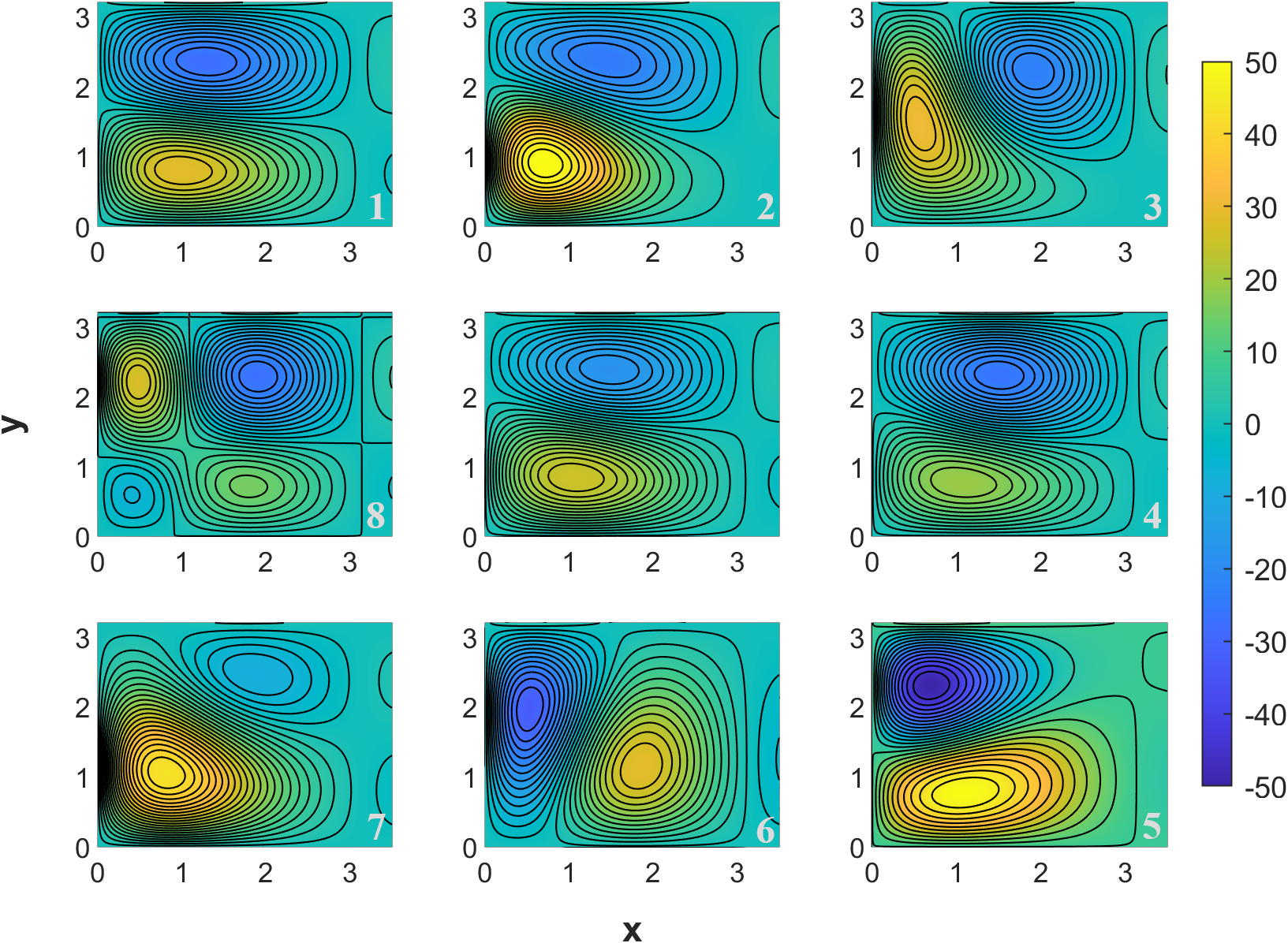}
	\caption{TMV-3}
	\end{subfigure}
   ~	
\begin{subfigure}[b]{0.4\textwidth}
	\includegraphics[width=\textwidth]{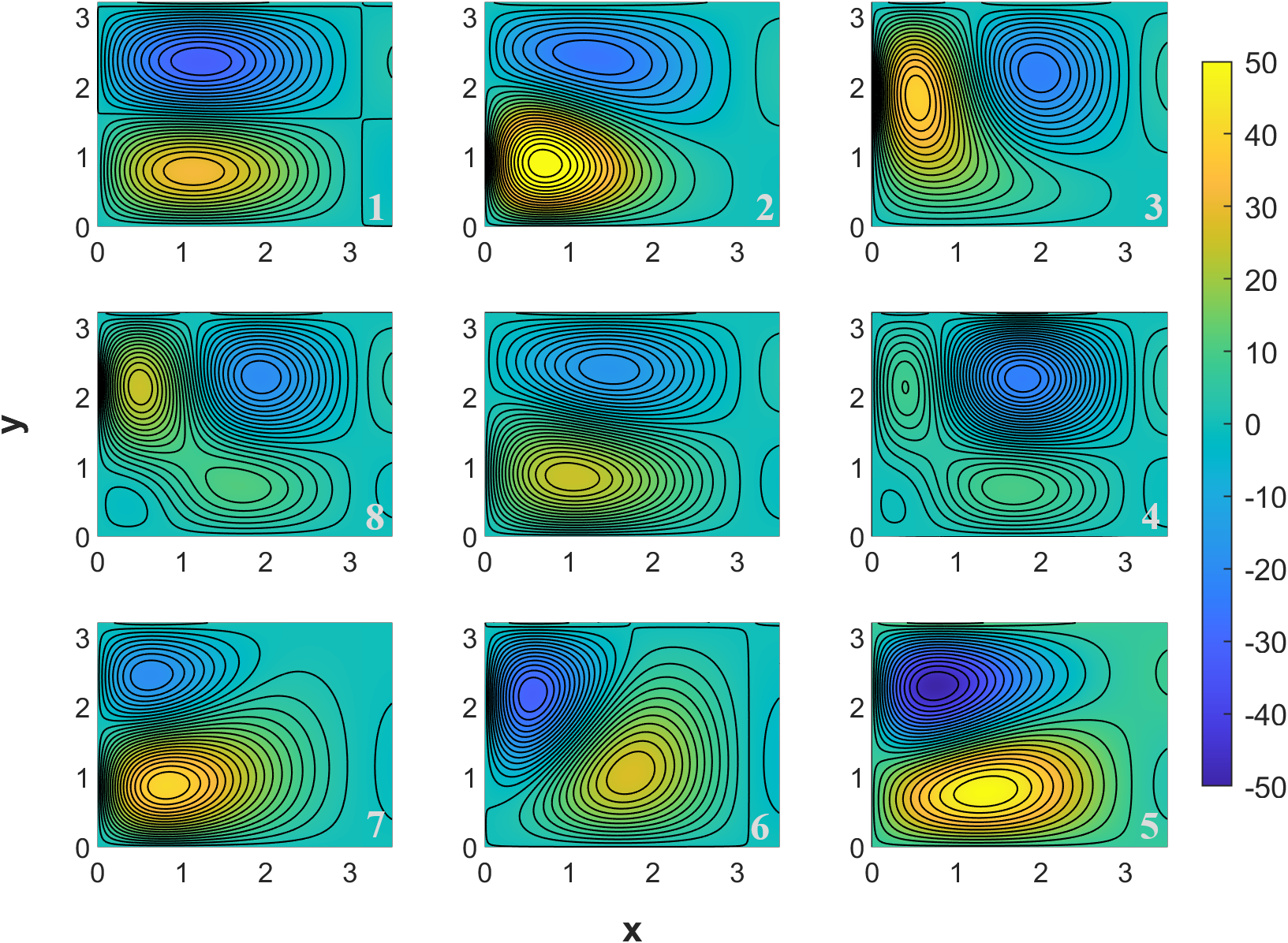}
	\caption{TMV-4}
	\end{subfigure} 
 	
\begin{subfigure}[b]{0.4\textwidth}
	\includegraphics[width=\textwidth]{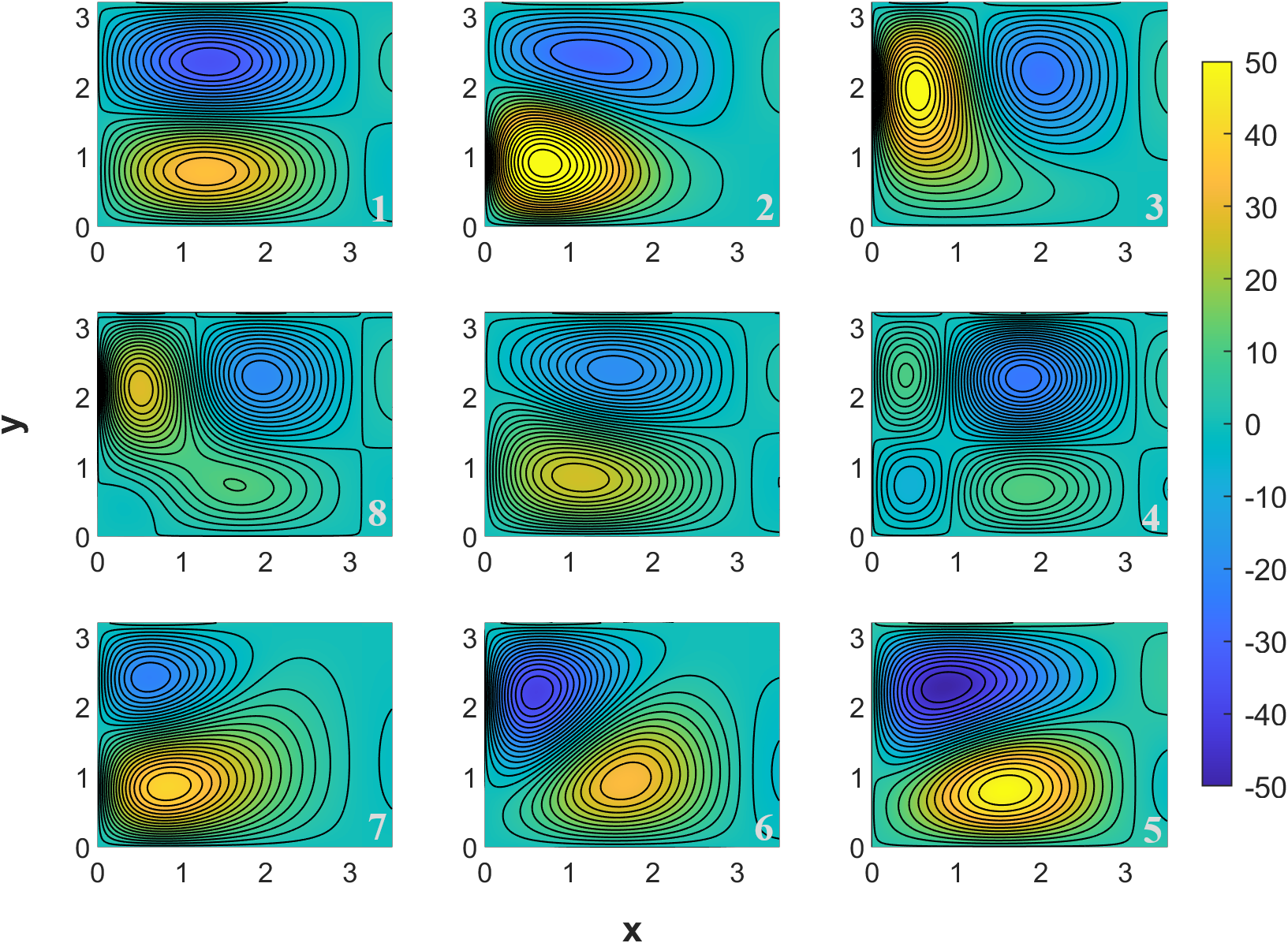}
	\caption{TMV-5}
	\end{subfigure} 
    ~	
\begin{subfigure}[b]{0.4\textwidth}
	\includegraphics[width=\textwidth]{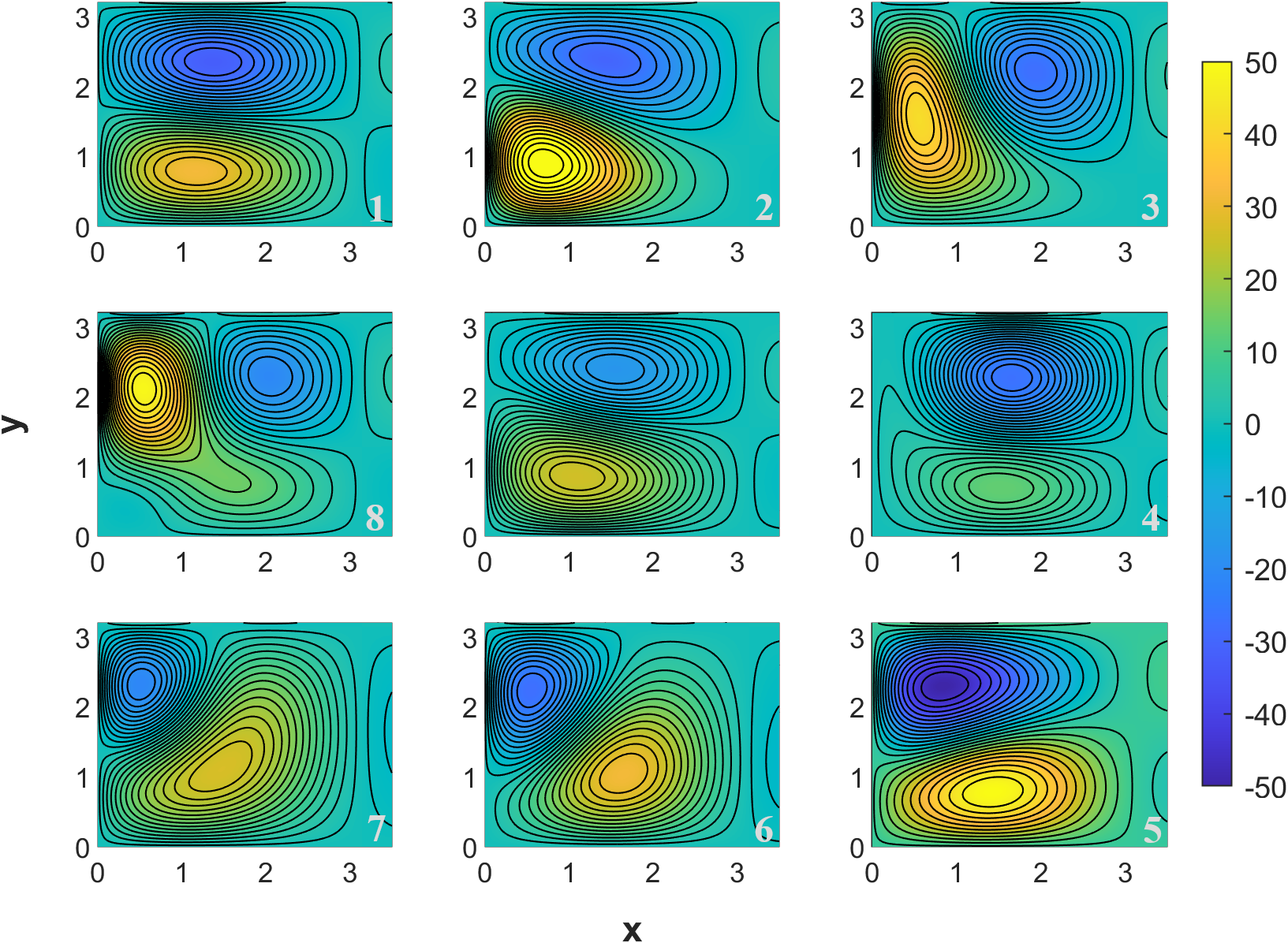}
	\caption{TMV-6}
	\end{subfigure} 
\caption{TMVs plotted in physical space for the aperiodically forced system;  same arrangement and
		labeling of subpanels as in Fig.~\ref{fig:tmv_autonomo}. The correspondence between the numbering of the subpanels and the cells of the generatex is as follows:(a) TMV-1: $\{1,5\} \to\gamma_1$, $\{2,6\} \to \gamma_2$, $\{3,7\} \to \gamma_3$, and $\{4,8\} \to \gamma_4$. (b) TMV-2: $\{1\} \to\gamma_2$, $\{2\} \to \gamma_5$, $\{3\} \to \sigma_{11}$, $\{4\} \to \sigma_{12}$,$\{5\} \to \sigma_{13}$, $\{6\} \to \sigma_{14}$, $\{7\} \to \sigma_{15}$, and $\{8\} \to \sigma_{16}$.(c) TMV-3:$\{1\} \to\gamma_1$, $\{2\} \to \gamma_2$, $\{3\} \to \gamma_3$, $\{4\} \to \gamma_5$, $\{5\} \to \gamma_6$, $\{6\} \to \sigma_6$, and $\{7,8\} \to \sigma_7$.(d) TMV-4:$\{1\} \to\gamma_1$, $\{2\} \to \gamma_2$, $\{3\} \to \gamma_3$, $\{4\} \to \gamma_5$, $\{5\} \to \sigma_1$, and $\{6,7,8\} \to \sigma_2$. (d) TMV-5:$\{1\} \to\gamma_1$, $\{2\} \to \gamma_2$, $\{3\} \to \gamma_3$, $\{4\} \to \gamma_5$, $\{5\} \to \gamma_6$, $\{6\} \to \sigma_3$, $\{7\} \to \sigma_4$, and $\{8\} \to \sigma_5$. (d) TMV-6:$\{1\} \to\gamma_1$, $\{2\} \to \gamma_2$, $\{3\} \to \gamma_3$, $\{4\} \to \gamma_5$, $\{5\} \to \gamma_6$, $\{6\} \to \sigma_8$, $\{7\} \to \sigma_9$, and $\{8\} \to \sigma_{10}$.}
  \label{fig:tmv_aperiodic}
\end{figure*}

The plots of the TMVs in physical space are shown in Figure~\ref{fig:tmv_aperiodic}. Note that, in panels~(d) and~(f), snapshots 1 through 4 of $\psi$ appear pairwise fairly similar, as they correspond to trajectories passing through the four 3-cells shared by $\mathcal{G}_4$ and $\mathcal{G}_6$, namely $\gamma_1$, $\gamma_2$, $\gamma_3$, and $\gamma_5$. In contrast, snapshots 5 through 8 exhibit marked differences between TMV-4 and TMV-6, reflecting the distinct sets of cells involved in the subentities of the two generatexes that are not mutually shared.

\section{TMV{\sc s} and nonautonomous attractors }
\label{sec:snapshots}

As already mentioned in the introduction, NDS theory differs from the older and better known theory of autonomous attractors in an essential way. Classical theory applies to systems in which the forcing and coefficients are constant in time, and in which solutions exist for all time, from $t \to - \infty$ to $t \to + \infty.$ In NDS theory, the coefficients or forcing or both may depend on time, and solutions do not just depend on $t - t_0,$ where $t_0$ is the initial time, but on both the initial time $s \equiv t_0$ and $t,$ the running time.

\begin{figure*}[h]
\centering
\begin{subfigure}[b]{0.3\textwidth}
\includegraphics[width=\textwidth]{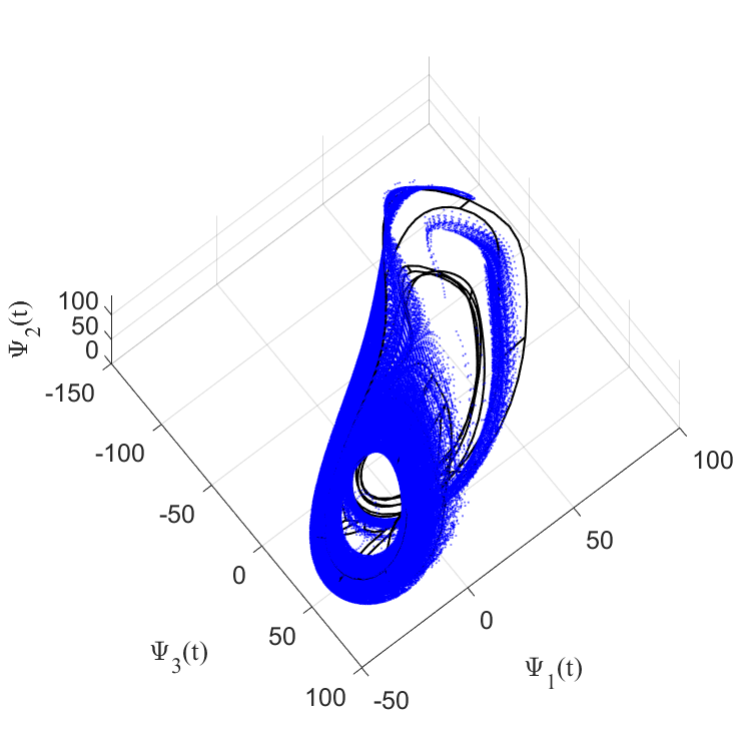}
	\caption{$\tau_1$: TMV-(1+3), $|\tau_1| = 6.74$ yrs.}
	\end{subfigure}
~	
\begin{subfigure}[b]{0.3\textwidth}
	\includegraphics[width=\textwidth]{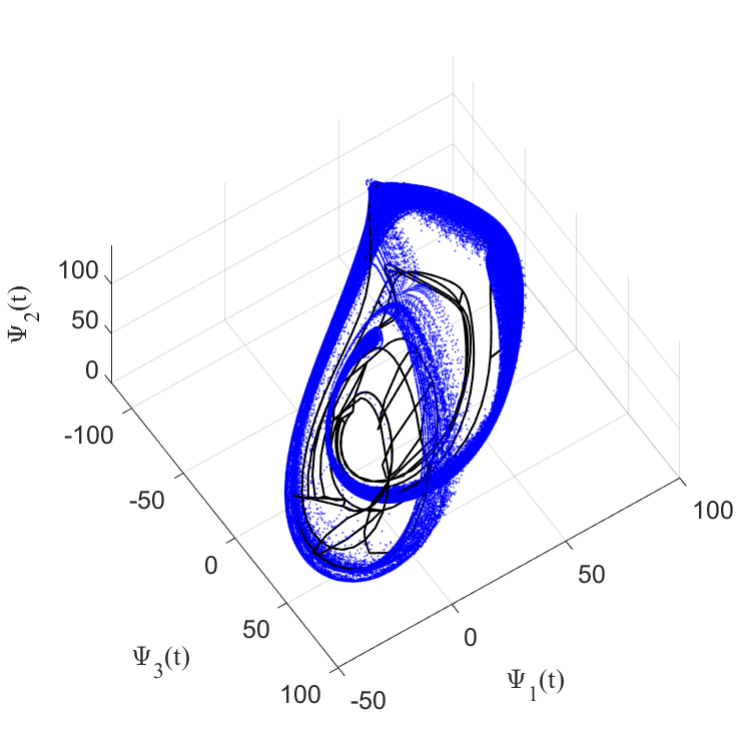}
	\caption{$\tau_2$: TMV-3, $|\tau_2| = 6.23$ yrs.}
	\end{subfigure} 

    \begin{subfigure}[b]{0.8\textwidth}
\includegraphics[width=\textwidth]{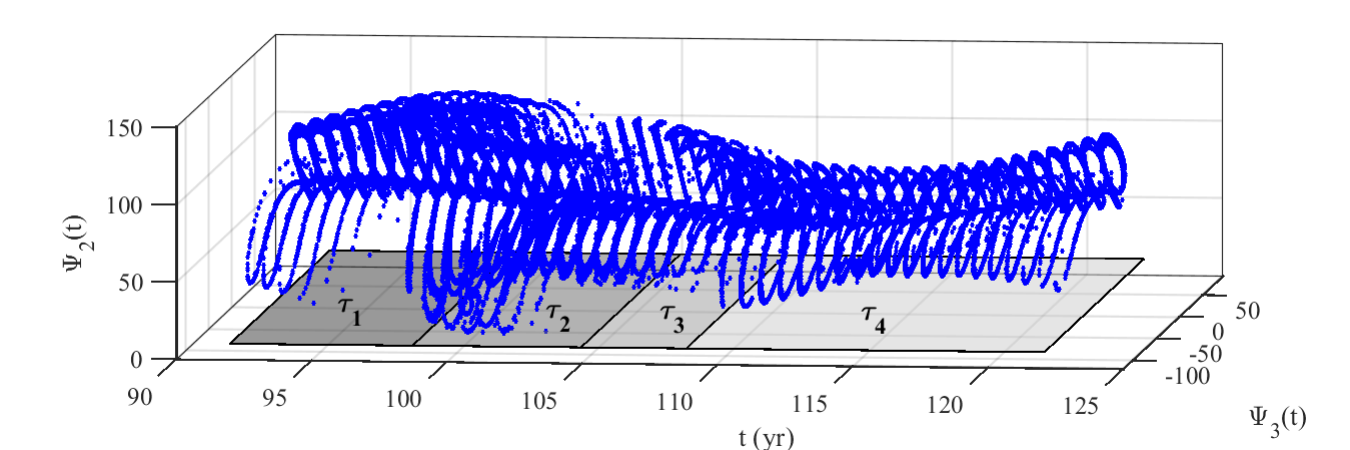}
	\caption{Evolution of the snapshot attractor over a time interval of length $T_p$.}
	\end{subfigure}
    
\begin{subfigure}[b]{0.3\textwidth}
\includegraphics[width=\textwidth]{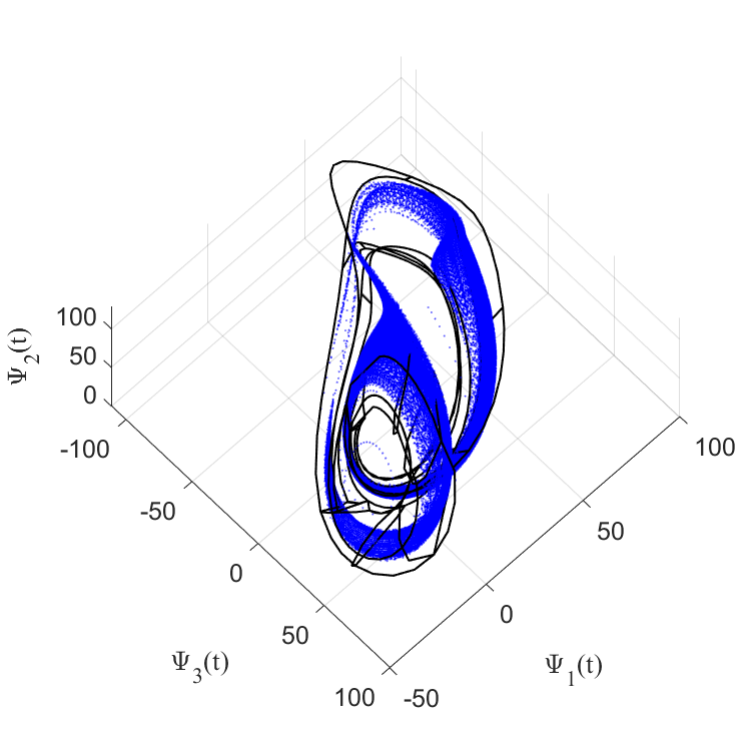}
	\caption{$\tau_3$: TMV-(3+4), $|\tau_3| = 3.89$ yrs.}
	\end{subfigure}
~	
\begin{subfigure}[b]{0.3\textwidth}
	\includegraphics[width=\textwidth]{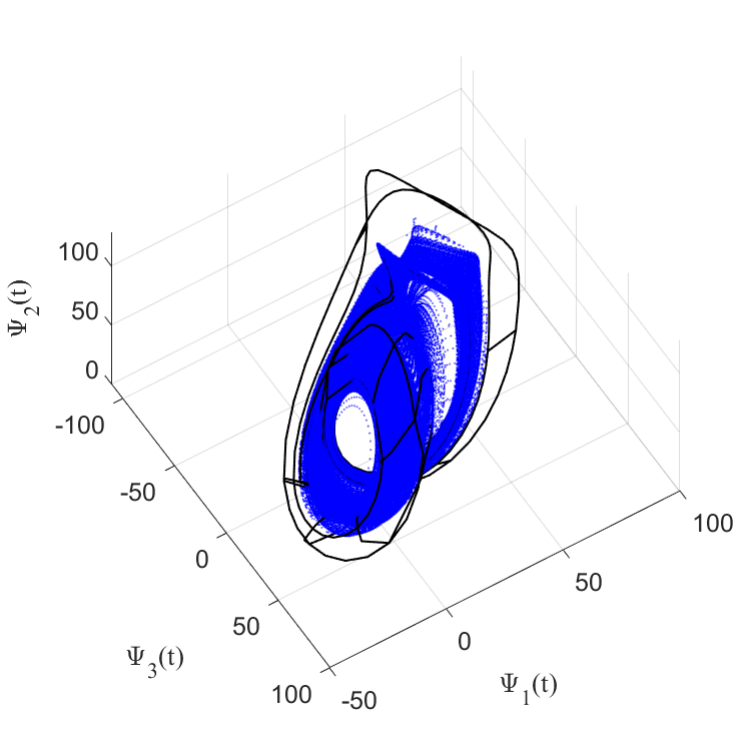}
	\caption{$\tau_4$: TMV-(4+2), $|\tau_4| = 13.23$ yrs.}
	\end{subfigure}  
\caption{Evolution in time of the fibers ({\em more geometrico}) or snapshots ({\em more physico}) in 
		the periodic-forcing case; the time interval equals $T_p = 30$ years. The central subfigure shows the evolution of the snapshots over a 30-year interval, within which all TMVs are visited at least once. The surrounding subfigures show an ensemble of snapshots (in blue) juxtaposed with the cell complex (in black) for four different time windows $\tau_i$ of length $|\tau_i|$, for $i = 1, \ldots, 4$. The TMVs active in each subinterval are:
		(a) TMV-(1+3): TMV-1+ TMV-3, (b) TMV-3, (d) TMV-(3+4): TMV-3+ TMV-4, and (e)TMV-(4+2): TMV-4+TMV-2.} 
\label{fig:snapshot_g_periodic}       
\end{figure*}

Accordingly, there are two kinds of attractors that can be defined for NDSs: {\it pullback  attractors (PBAs)}, obtained by letting $s \to - \infty$ for $t$ fixed \cite{crauel1994attractors, arnold1998random, Ghil.ea.2008}, and {\em forward attractors(FWAs)}, obtained by finding a special solution that does exist all the way to $+ \infty$ and such that all solutions converge to it. In terms of conditions for existence and uniqueness, much more is known for PBAs than for the FWAs \cite{Car.Han.2017, Kloeden.Yang.2020}.  In the physical literature, {\it snapshot attractors} have been introduced for time-dependent forcing by \citet{namenson1996fractal} without awareness of the existence of a mathematical NDS literature or the use of precise definitions of behavior at $\pm \infty.$ Tamás Tél and associates have treated climate-related problems of this type in the snapshot attractor context. \cite{Drotos.ea.2015, tel2020theory,Bodai.Tel.2012, janosi2024overview}

In numerical practice, calculating a PBA does not require pulling back all the way to $- \infty$, only to a distance in time that equals a finite multiple of dissipation times, which depends on the precision with which we want to calculate the attractor, and is actually rather small \cite{chekroun2011stochastic, charo2021noise, Drotos.ea.2015, Maraldi.ea.2025}. Still, it is more straightforward computationally to calculate an FWA or snapshot attractor, when one exists, even when it may not be unique \cite{Pierini.ea.2016}, and hence we do so here.

An FWA or snapshot attractor at time $t = T$ — called a {\em fiber} in the mathematical literature or a {\em snapshot} in the physical one — is the result of integrating a substantial array of initial conditions that coalesce, at least to within a good enough approximation, at time $T$. To obtain the attractor of system \eqref{eq:ODEs} at time $T,$ an ensemble of $1.5 \times 10^5$ random initial conditions lying in a hypercube $D \subset \mathbb{R}^4$ was integrated until the time $t=T$. In order to establish a basis for comparison between the templex approach and the dynamical approach in both nonautonomous cases — namely the periodic and aperiodic one — the parameter values of the time-dependent forcing were set to the values $\gamma=1.10$ and $\varepsilon=0.20$ in both Secs.~\ref{subsec:periodic} and \ref{sec:aper}.

As indicated in Sec.~\ref{sec:aper}, the temporal evolution of the attractor in the aperiodic-forcing case was investigated over a 400-year time interval.
The transition time for satisfactory convergence to the attractor was of 66~yrs, after which the attractor became periodic, with a period of $T_p=30$~yr  in  Sec.~\ref{subsec:periodic}.

\begin{figure*}[h]
\centering
\begin{subfigure}[b]{0.25\textwidth}
\includegraphics[width=\textwidth]{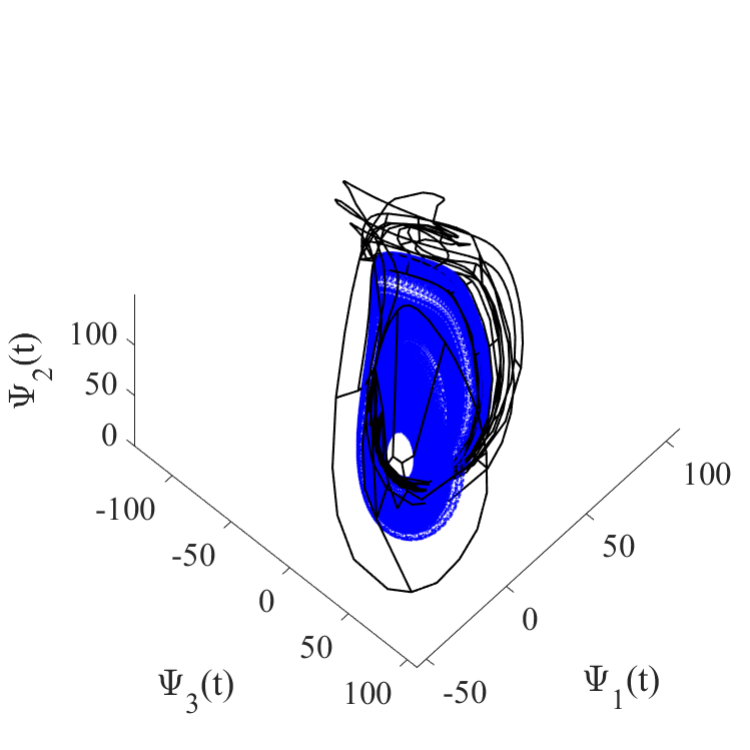}
	\caption{$\tau_1$: TMV-4, $|\tau_1|=22.05 $ yrs.}
	\end{subfigure}
~	
\begin{subfigure}[b]{0.25\textwidth}
	\includegraphics[width=\textwidth]{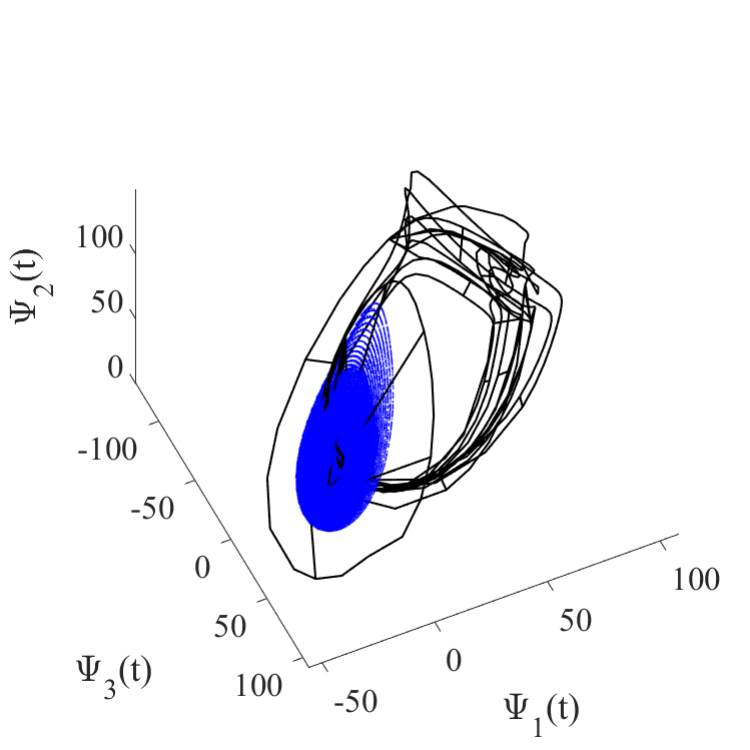}
	\caption{$\tau_2$: TMV-1, $|\tau_2|=22.05$ yrs.}
	\end{subfigure} 
~    
\begin{subfigure}[b]{0.25\textwidth}
\includegraphics[width=\textwidth]{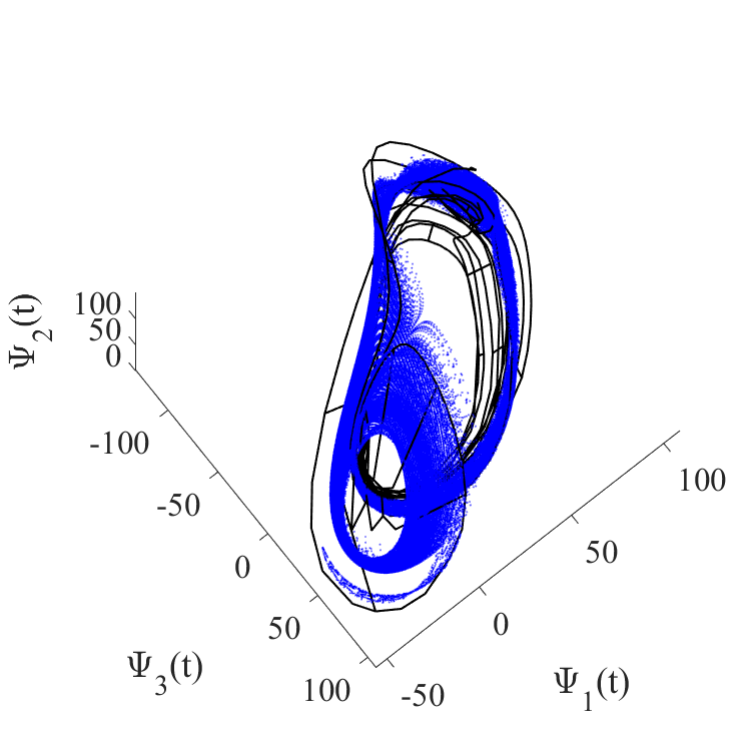}
	\caption{$\tau_3$: TMV-(3+2), $|\tau_3|=10.3$ yrs.}
	\end{subfigure}
    
	\begin{subfigure}[b]{0.8\textwidth}
	\includegraphics[width=\textwidth]{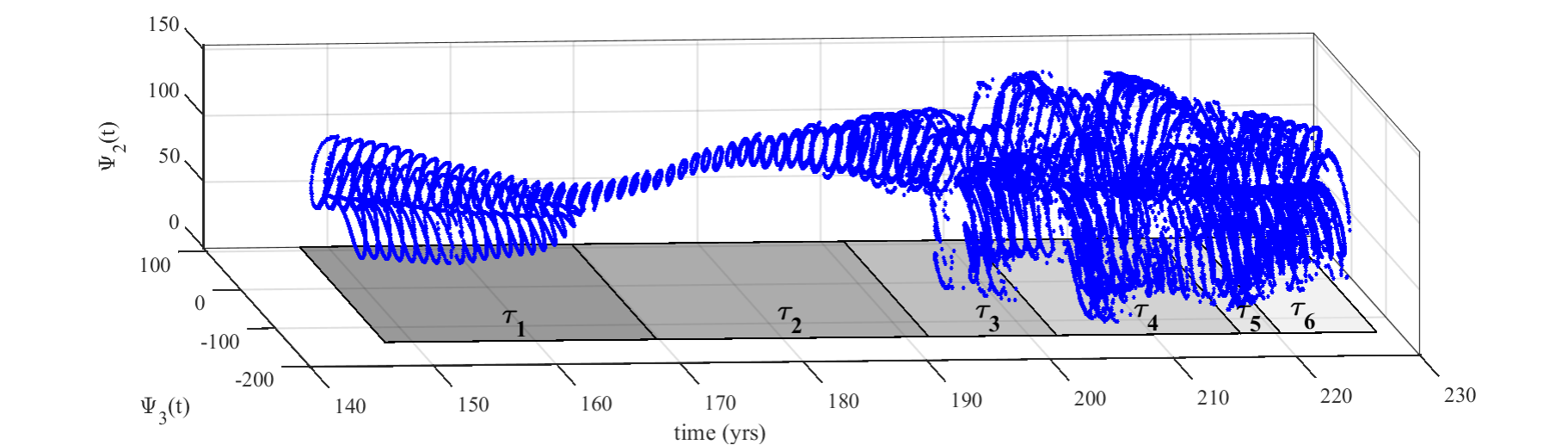}
	\caption{Evolution of the snapshot attractor within the time window [147, 228] years}
	\end{subfigure} 

\begin{subfigure}[b]{0.25\textwidth}
	\includegraphics[width=\textwidth]{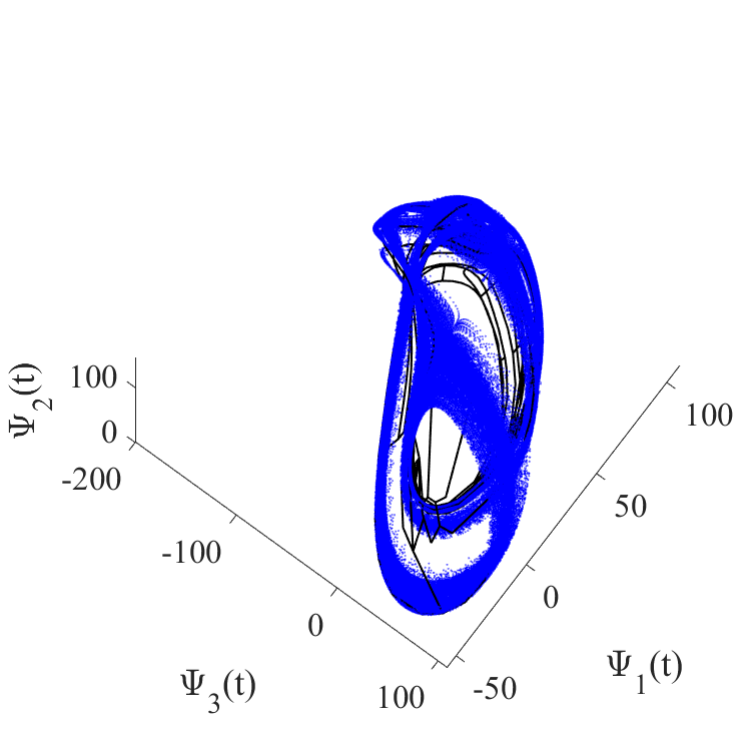}
	\caption{$\tau_4$:TMV-(2+6), $|\tau_4| =14.9 $ yrs.}
	\end{subfigure}  
    ~	
\begin{subfigure}[b]{0.25\textwidth}
	\includegraphics[width=\textwidth]{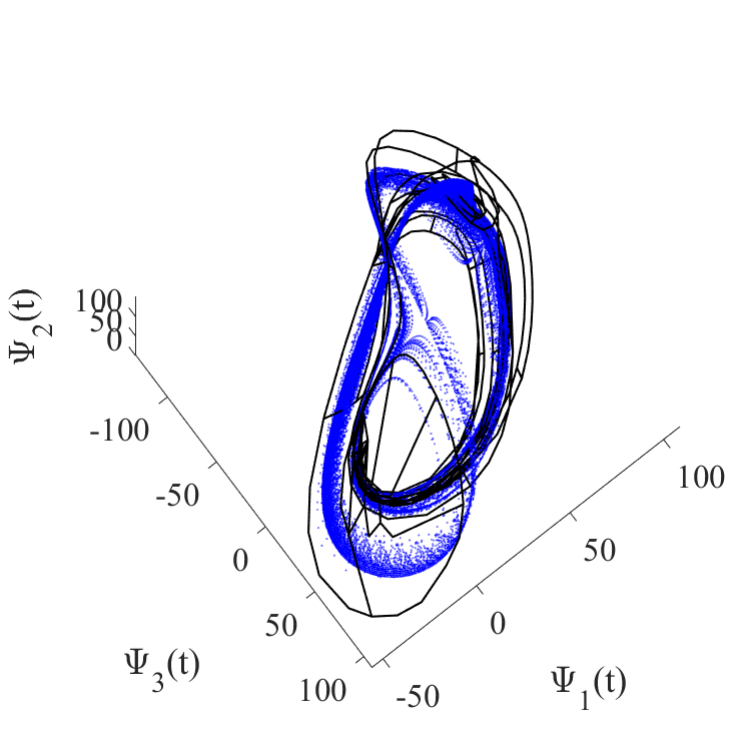}
	\caption{$\tau_5$: TMV-(6+3),  $|\tau_5|=3.2$ yrs.}
	\end{subfigure} 
    ~	
\begin{subfigure}[b]{0.25\textwidth}
	\includegraphics[width=\textwidth]{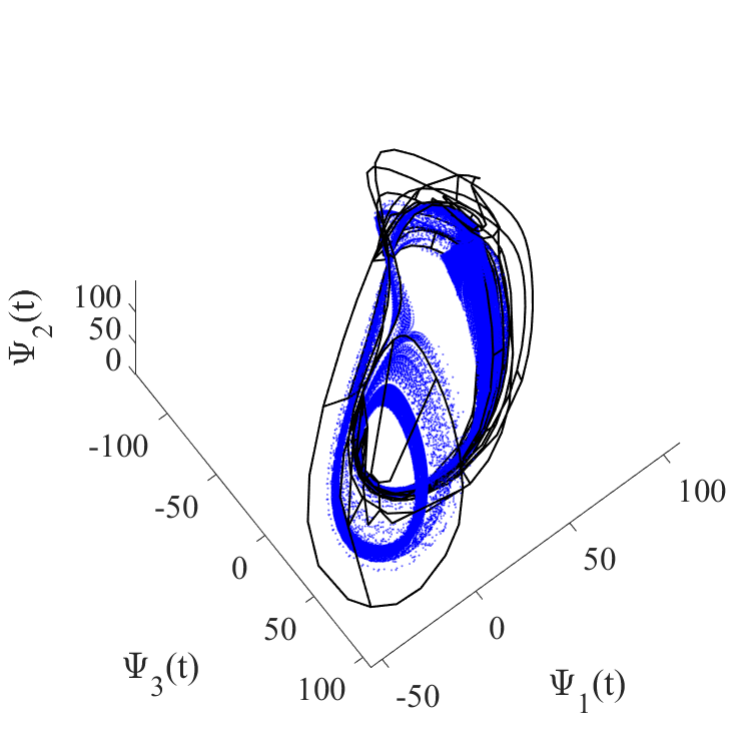}
	\caption{$\tau_6$, TMV-(3+5), $|\tau_6| =7.8$ yrs.}
	\end{subfigure} 
\caption{Evolution in time of the fibers or snapshots in the central subfigure over a time window of 80 years, within which all TMVs are visited at least once (d). The surrounding subfigures show an ensemble of snapshots (in blue) juxtaposed with the cell complex $\bar{K}_3$ (in black) for six different time windows $\tau_i$ of length $|\tau_i|$, for $i = 1, \ldots, 6$. The TMVs active in each subinterval are:
		(a) TMV-4, (b) TMV-1, (c) TMV-(3+2): TMV-3+TMV-2,(e) TMV-(2+6): TMV-2+TMV-6, (f) TMV-(6+3): TMV-6+TMV-3, and (g) TMV-(3+5): TMV-3+TMV-5.} 
\label{fig:snapshot_g_aperiodic}       
\end{figure*}

At a given time $T$, the point cloud corresponding to a fiber, or snapshot, does not represent a single trajectory in phase space. Instead, it captures the collection of states reached by an ensemble of initial conditions at time $T$. This explains why multiple TMVs can coexist at a given time. 

Figures~\ref{fig:snapshot_g_periodic} and~\ref{fig:snapshot_g_aperiodic} display the time evolution under periodic and aperiodic forcing, respectively. The central panel illustrates the dynamics of the attractor over a time window, during which all TMVs are visited at least once. 
The time window is divided into a sequence of time subintervals $\tau_i$.
The transition from one subinterval \( \tau_i \) to the next one \( \tau_{i+1} \) represents a \emph{tipping point (TP)} in the attractor, as it involves a shift from one set of active TMVs to another. The surrounding subfigures show ensembles of snapshots paired with the cell complex for each time interval $\tau_i$.

In the case of periodic forcing, Figure~\ref{fig:snapshot_g_periodic}~(c) shows the evolution of the set of fibers, or snapshots, over a full period $T_p=30$ years. The time window is divided into a sequence of four subintervals $\tau_i$, for $i = 1, \ldots, 4$. This window was selected to ensure that all TMVs are visited at least once during the total time interval. During $\tau_1$, the trajectories evolve through the generatexes $\mathcal{G}_1$ and $\mathcal{G}_3$; in $\tau_2$, through the generatex $\mathcal{G}_3$; in $\tau_3$, through the generatexes $\mathcal{G}_3$ and $\mathcal{G}_4$; and in $\tau_4$, through the generatexes $\mathcal{G}_4$ and $\mathcal{G}_2$.

In the case of aperiodic forcing, Figure~\ref{fig:snapshot_g_aperiodic}(d) shows the evolution of a set of snapshots over an 80-year time window.  The window is divided into a sequence of six subintervals $\tau_i$, for $i = 1, \ldots, 6$. The snapshot set evolves through the set of six generatexes as follows: $\tau_1$: $\{\mathcal{G}_4\}$, $\tau_2$: $\{\mathcal{G}_1\}$, $\tau_3$: $\{\mathcal{G}_3, \mathcal{G}_2\}$, $\tau_4$: $\{\mathcal{G}_2, \mathcal{G}_6\}$, $\tau_5$: $\{\mathcal{G}_6, \mathcal{G}_3\}$, and $\tau_6$:$\{\mathcal{G}_3, \mathcal{G}_5\}$. Note that, in this case, another 80-year interval might or might not contain all the TMVs identified in the 400 years of forcing analyzed here, or if a different 400-year window is considered, it might include additional ones.

\begin{figure}[ht]
  \centering
  \includegraphics[width=0.49\textwidth]{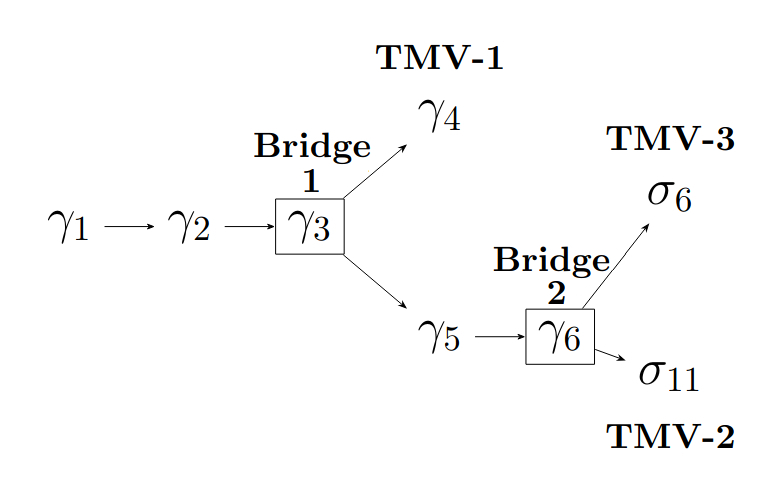}
   \caption{Schematic diagram illustrating two topological bridges between TMVs via 3-cells $\gamma_3$ and $\gamma_6$, corresponding to the transitions in Figs.~\ref{fig:snapshot_g_aperiodic}~(b,c).}
  \label{fig:tmv-bridges}
\end{figure}

The attractor's evolution in time can be followed through the structure of the associated digraph. 
While transitions between sets of TMVs may appear abrupt—for instance, Figs.~\ref{fig:snapshot_g_aperiodic}~(b,c) illustrate a shift from TMV-1 alone to TMV-3 and on to TMV-2—the underlying topology ensures continuity. Specifically, generatexes 1 and 3 are connected via the 3-cell $\gamma_3$, and generatexes 3 and 2 via the 3-cell $\gamma_6$; see Fig.~\ref{fig:tmv-bridges}.
These cells provide a topological bridge that renders such instantaneous changes consistent with a continuous deformation of the snapshots.
\phantomsection

\section*{Concluding remarks}

The present work introduces a general topological framework for analyzing the behavior of dynamical systems subject to  time-dependent external forcing. This framework is based on the extension to nonautonomous dynamics of the \emph{templex} $T$ introduced by \citet{charo2022templex} for autonomous ones. This extension is still comprised of a cell complex $K$ and a directed graph $G$, $T = (K, G)$, which encodes the organization of trajectories in phase space. 

To illustrate the application of this nonautonomous templex, we examined a prototypical model of the wind-driven ocean circulation in midlatitudes. 
The model chosen is based on a well-accepted set of closely related low-order models of the wind-driven ocean circulation. In the North Atlantic and the North Pacific, in particular, this circulation takes the form of a pair of counterrotating gyres, subtropical and subpolar. These low-order models \cite{jiang1995multiple, Pierini.ea.2016,Pierini.ea.2018} are governed by four ordinary differential equations, as opposed to the three to which much of the applied chaos literature --- like the Lorenz convection model \cite{lorenz1963lorenz} or the R\"ossler model \cite{Ros76c} --- is limited, on the one hand.  

On the other hand, models that are higher in the usual hierarchy of models \cite{Ghil.2001, Held.2005, Schneider.Dick.1974} in the climate sciences -- such as shallow-water \cite{jiang1995multiple, Speich.ea.1995, Pierini.ea.2016,Pierini.ea.2018} or quasi-geostrophic \cite{simonnet2005homoclinic,pierini2014ensemble} models governed by sets of partial differential equations --- have reproduced the predictions of the low-order ones or vice-versa. Hence we feel comfortable in asserting the usefulness of the present results in the broader context of the climate sciences and beyond. 

This model provides a meaningful test case thanks to its moderate complexity and the availability of numerically computed attractors under both periodic and aperiodic forcing. These features enable a systematic topological analysis grounded in physically relevant regimes.

The study shows that {\em generatexes}—directed cycles within the digraph of the templex—can be used to define modes of variability rooted in the topological organization of the attractor. This construction yields a genuine \emph{nonlinear modal decomposition} of the system’s behavior, where each mode is constrained by the topology of the flow imposed by deterministic dynamics. This representation puts topological results at the service of metric details and allows for a compact yet informative decomposition of the system's temporal evolution. Note that these {\em topological modes of variability (TMVs)} — unlike the types of modal decomposition summarized in Table~\ref{tab:lin_modes} — are concatenated in time, rather than superposed, like the modes in the table, for all times.

Under periodic forcing, we showed that the set of generatexes extracted from the templex can be considered complete, provided that the construction requirements are met—specifically, that the time window is sufficiently long to explore the full attractor. The system then evolves within a fixed topological scaffold defined by a stable generatex set, while the sequence of TMVs and their activation—that is, the effective visits of a trajectory to specific generatexes—are not fixed, but remain inherently sensitive to initial conditions due to the presence of {\em joining and splitting loci}, i.e., of alternative paths within the templex.

In the aperiodically forced case, we enter a genuinely nonautonomous regime. The generatex set obtained reflects only the finite segment of the forcing used to construct the templex. In this regime, the set of TMVs is no longer stable and may vary over time, as new generatexes can emerge or disappear beyond the analysis window. This highlights the necessarily partial and time-dependent nature of the topological analysis under aperiodic forcing.

Finally, the visualization of TMVs in physical space and their temporal association with the FWA or snapshot attractor provides a novel link between abstract topological structures and concrete dynamical patterns. In this way, seemingly abrupt dynamical changes acquire a topological interpretation as smooth transitions within the templex, mediated by bridges between TMVs; see again Fig.~\ref{fig:tmv-bridges}.

The ensemble dynamics—understood as the time-varying activation of specific TMVs—emerges from the interplay between the templex structure and the evolving attractor. The templex encodes the set of topological configurations accessible to the system, whereas the attractor captures the ensemble’s actual distribution across different TMVs at a given time. This complementarity
allows us to identify which generatexes are active and how transitions among them unfold. In this sense, the templex provides a structural scaffold for the dynamics, and the attractor traces the ensemble’s trajectory through this scaffold. The complementarity between these two aspects of the dynamics enables a precise description of the system’s temporal evolution in terms of activation of alternative TMVs and transitions from one TMV to another.

Beyond individual trajectories, our results also reveal a TMV organization in the attractor itself. By analyzing ensembles of states at successive times $T$, we identify sets of active TMVs that change across subintervals $\tau_i$. These changes correspond to tipping points (TPs) in the attractor, providing a topological reading of how ensemble simulations explore distinct sets of nonlinear modes at different times. These TPs definitely require further study, as they might furnish insights into possible early warnings for a transition from one TMV to another. Both \citet{evans2004rise} and \citet{mukhin2015predicting, mukhin2015predict_II} provide interesting examples of the application of rather abstract methods to improve prediction. \citet{mukhin2015predicting, mukhin2015predict_II}, in particular, also indicate how such methods can be successfully extended from a lower to a higher rung in the model hierarchy.

This work sets the stage for future applications of templex analysis to more realistic ocean and other climate models, as well as to observational datasets, where the identification of coherent, interpretable TMVs could enhance our understanding of regime transitions, tipping points, and predictability in climate dynamics.

\section*{Acknowledgments}
This work has received funding from the ANR project TeMPlex ANR-23-CE56-0002 (DS). GDC gratefully acknowledges her postdoctoral scholarship within CONICET and the ANR project. This is ClimTip contribution \#84; the ClimTip project has received funding from the European Union's Horizon Europe research and innovation program under grant agreement No. 101137601 (MG).

\appendix

\section{Homologies for a {\sc BraMAH} complex}
\label{ap:BraMAH}

\begin{figure}[ht]
\center
(a) \includegraphics[width=0.2\textwidth]{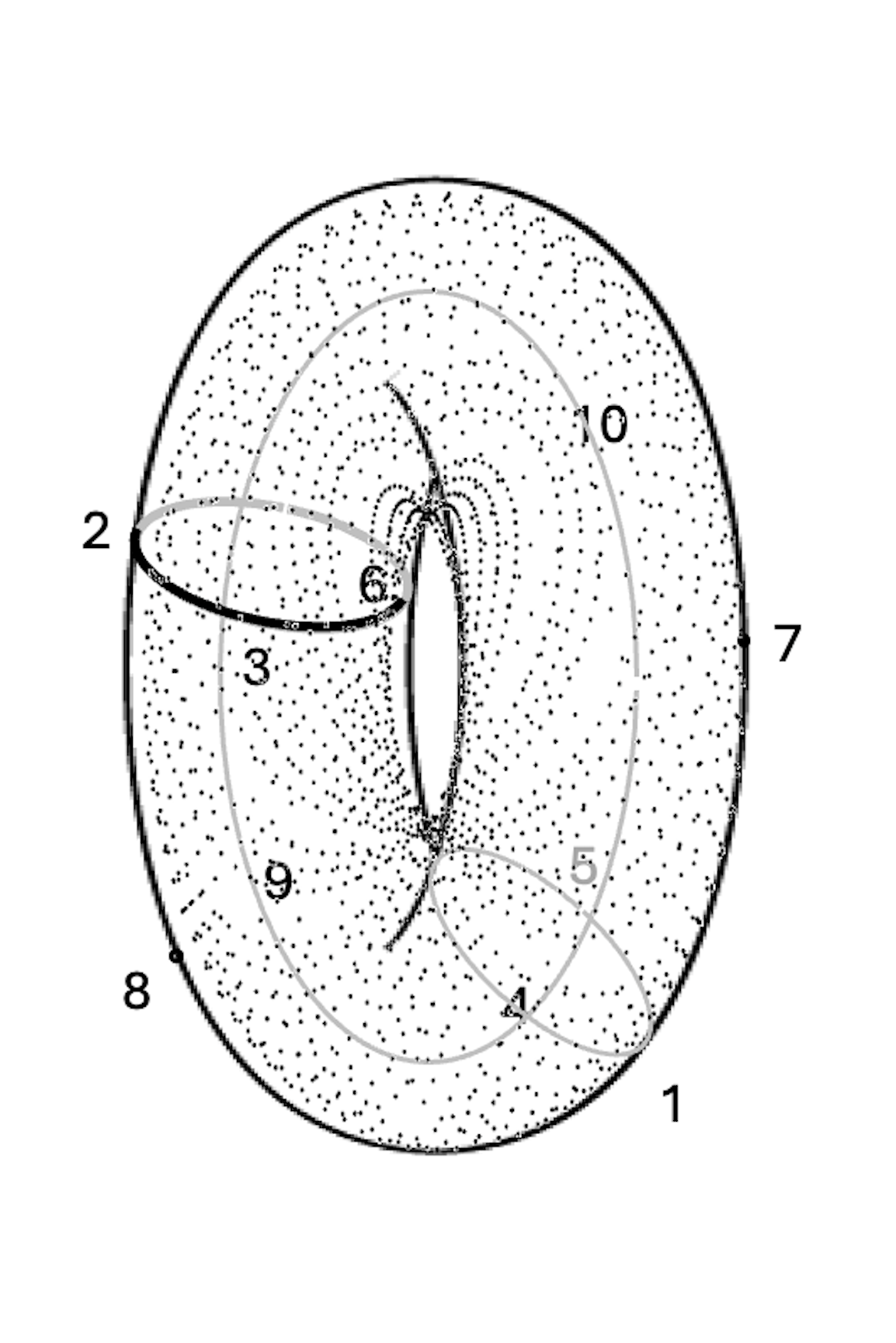}
(b) 
\includegraphics[width=0.2
\textwidth]{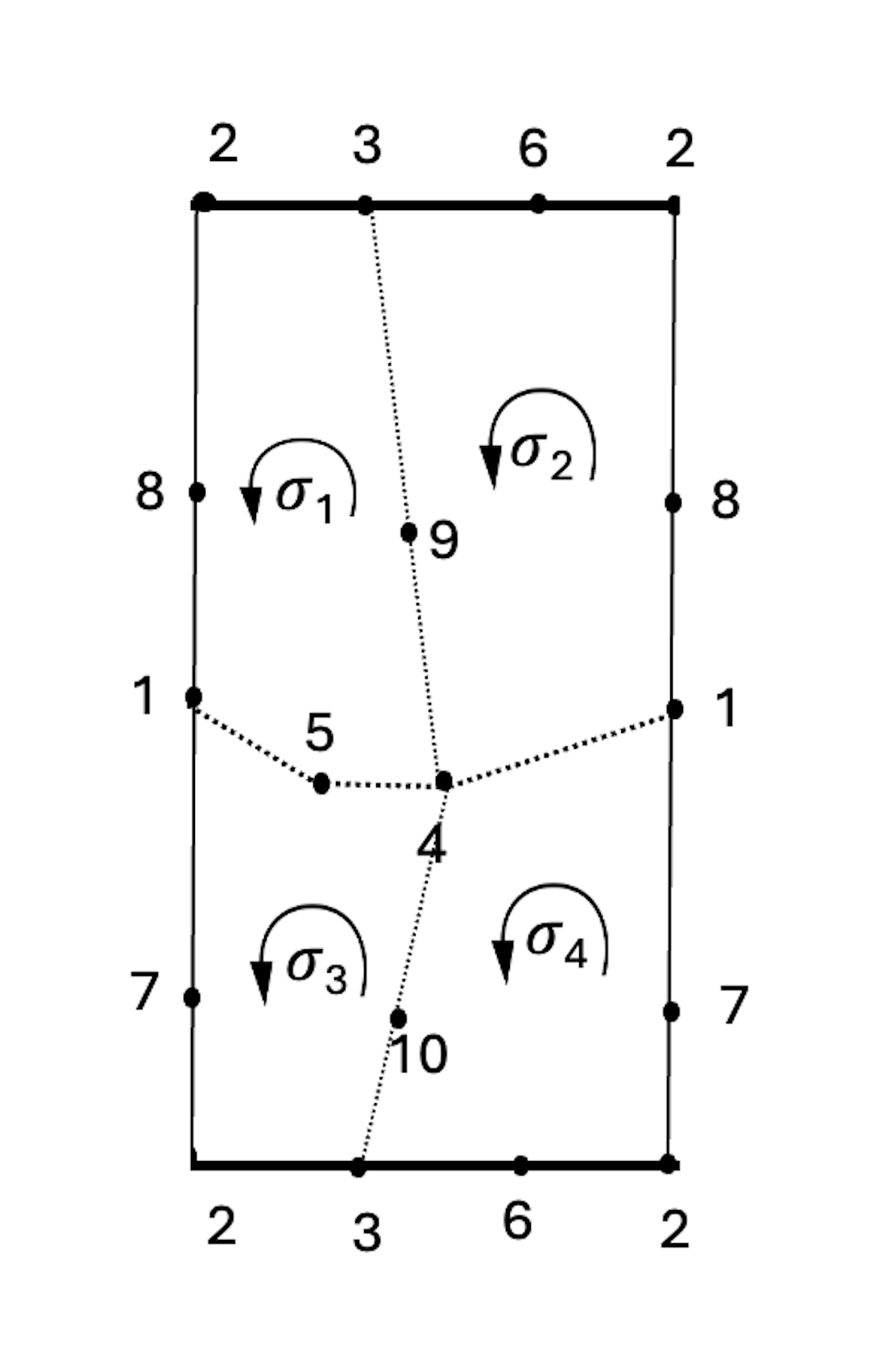}
\caption{{\sc BraMAH} cell complex for a torus. (a) A point cloud (gray dots) corresponding to a regular flow on a toroidal surface $\mathbb{T}^2$, with a {\sc BraMAH} cell complex $K(\mathbb{T}^2)$ sketched on it; and (b) a planar diagram of $K(\mathbb{T}^2)$, where repeated labels indicate gluing. The anticlockwise orientation of $\sigma_1$ is propagated to the other 2-cells so as to make $K(\mathbb{T}^2)$ be uniformly oriented.}
\label{fig:EgTorus}       
\end{figure}

This appendix illustrates the construction of a {\sc BraMAH} complex. Consider as an example a {\sc BraMAH} complex $K(\mathbb{T}^2)$ of dimension $2$ for a three-dimensional point cloud on a toroidal surface $\mathbb{T}^2$, as shown in Figure~\ref{fig:EgTorus}. We follow the definitions provided by \citet{kinsey2012topology} and the code in \citet{sciamarella2023code} to show how the homologies are computed. 

{\em Chain groups} are algebraic structures that encode the \(k\)-dimensional features of a cell complex. The \(k\)-th chain group is denoted by \(C_k(K)\) and it consists of all linear combinations of \(k\)-cells in the complex with integer coefficients. The chain groups of the cell complex $K(\mathbb{T}^2)$ are as follows:
\begin{enumerate}[$\bullet$, nosep]
    \item {\bf $C_2$}: The group of 2-chains, generated by the 2-cells $\{\sigma_1, \sigma_2, \sigma_3, \sigma_4\}$ with the orientations indicated in the planar diagram of Fig.~\ref{fig:EgTorus}(b). 
    \item {\bf $C_1$}: The group of 1-chains, generated by the edges of the 2-cells in the complex. The edges, oriented from the lower-numbered to the higher-numbered vertices, include:
    \[
    \langle 1,4 \rangle, \langle 1,5 \rangle, \langle 1,7 \rangle, \langle 1,8 \rangle, \langle 2,3 \rangle, \langle 2,7 \rangle, \langle 2,8 \rangle, \dots
    \]
    \item {\bf $C_0$}: The group of 0-chains, generated by the vertices $\{\langle 1 \rangle, \langle 2 \rangle, \langle 3 \rangle, \langle 4 \rangle, \dots\}$.
\end{enumerate}

These groups represent collections of \(k\)-dimensional cells, and their relationships are described by the boundary operators. A {\em boundary operator} $\partial_k$ maps $k$-chains to $(k-1)$-chains, so that: 
\begin{enumerate}[$\bullet$, nosep]
    \item For a 1-cell $\langle 1,4 \rangle$, the operator $\partial_1(\langle 1,4 \rangle)$ is the difference of its terminal and initial vertices:
    \[
    \partial_1(\langle 1,4 \rangle) = \langle 4 \rangle - \langle 1 \rangle.
    \]
    \item For the 2-cell $\sigma_2$:
     \begin{multline*}
    \partial_2(\sigma_2) = 
    \langle 1,8 \rangle - \langle 2,8 \rangle \\
    + \langle 2,6 \rangle - \langle 3,6 \rangle + \langle 3,9 \rangle - \langle 4,9 \rangle - \langle 1,4 \rangle.
    \end{multline*}
\end{enumerate}

Note that the orientations of 1-cells and 2-cells are essential for applying the boundary operator. Consistent orientations can be assigned arbitrarily, as long as they are respected throughout the computation. 

{\em Homology groups} are defined as:
\[
H_k(K) = \ker(\partial_k) / \operatorname{im}(\partial_{k+1});
\]
here \(\ker(\partial_k)\) is the group of cycles, i.e., of chains whose boundaries are zero, and \(\operatorname{im}(\partial_{k+1})\) is the group of boundaries, i.e., of chains that are boundaries of \((k+1)\)-chains.

The  $k^{{\rm{th}}}$ {\em Betti number}, denoted by \(\beta_k\), is the rank of the $k^{{\rm{th}}}$ homology group. In other words, \(\beta_k\) corresponds to the number of independent \(k\)-dimensional cycles in the complex:
\[
\beta_k = \text{rank}(H_k(K)).
\]

For the cell complex $K(\mathbb{T}^2)$ in Fig.~\ref{fig:EgTorus}, we compute the homology groups and Betti numbers, and explicitly describe the generators as follows:
\begin{enumerate}[$\bullet$, nosep]
    \item The $0^{{\rm{th}}}$ homology group represents the connected components of the manifold underlying the point cloud. The generator of \(H_0(K(\mathbb{T}^2))\) corresponds to any of the 0-cells, since all of them are homologous:
    $$ g_{0_1} = \langle 1 \rangle. $$
    We thus write:
    \[
    H_0(K(\mathbb{T}^2)) = [[ g_{0_1} ]] \simeq \mathbb{Z}, \quad \beta_0 = 1.
    \]
    \item The 1st homology group captures homologically independent nontrivial loops. The torus has two:
   \begin{enumerate}[$-$, nosep]
        \item A poloidal loop, aligned with the shortest circle on the surface, represented by:      
        \[
        g_{1_1} = \langle 2,3 \rangle + \langle 3,6 \rangle - \langle 2,6 \rangle.
        \]
        \item A toroidal loop around the central axis of the torus, represented by: 
        \[
        g_{1_2} = \langle 1,7 \rangle - \langle 2,7 \rangle + \langle 2,8 \rangle - \langle 1,8 \rangle.
        \]
    \end{enumerate}
    We thus write:
    \[
    H_1(K(\mathbb{T}^2)) = [[ g_{1_1}, g_{1_2} ]] \simeq \mathbb{Z}^2, \quad \beta_1 = 2.
    \]

    \item The 2nd homology group captures enclosed cavities. The torus encloses a single cavity, and the 2-generator is:
    \[
    g_{2_1} = \sigma_1 + \sigma_2 + \sigma_3 + \sigma_4.
    \]
    We thus write:
    \[
    H_2(K(\mathbb{T}^2)) = [[ g_{2_1} ]] \simeq \mathbb{Z}, \quad \beta_2 = 1.
    \]
\end{enumerate}

If the 0-cells along the bottom line of the planar diagram of Fig.~\ref{fig:EgTorus}(b) that are labeled $2,3,6,2$ are relabeled $2,6,3,2$, modifying thus the gluing instructions, the torus becomes a Klein bottle $\mathbb{K}^2$. While the 2-cells $\sigma_1$ and $\sigma_2$ remain the same, the 2-cells below them become $\sigma_3=\langle 2,6,10,4,5,1,7 \rangle$ and $\sigma_4=\langle 1,4,10,6,3,2,7 \rangle$. Computing homologies for this new complex, we still have $\beta_0 = 1$, but $\beta_1 = 1$ and $\beta_2 = 0$, i.e., there is no longer an enclosed cavity. The torsion group is given by the 1-chain forming the nonorientable cycle:
\[
-2 \langle 2,3 \rangle - 2 \langle 3,6 \rangle + 2 \langle 2,6 \rangle.
\]
The Klein bottle is said to have a weak boundary because its nonorientable structure implies that 1-cycles fail to bound 2-chains in a consistent manner. Specifically, the torsion group reflects this property.

We have thus shown how building a {\sc BraMAH} cell complex is useful in computing the topological properties of the manifold underlying a point cloud.

\section*{Data availability}
The Wolfram Mathematica code to compute homology groups and templex properties with examples is available at \url{https://git.cima.fcen.uba.ar/sciamarella/bramah_torus/} and \url{https://git.cima.fcen.uba.ar/sciamarella/templex-properties} or \url{https://community.wolfram.com/groups/-/m/t/3079776}.

\section*{References}
\bibliography{TMV_Chaos_bib}

\end{document}